\documentclass[12pt]{article}
\usepackage{amsmath,amssymb,amsthm,bbm,natbib,color,hyperref}
\usepackage[margin=1.3in]{geometry}
\usepackage{tikz}
\usetikzlibrary{decorations.pathreplacing}
\usepackage{caption}
\usepackage{graphicx}
\usepackage{subcaption}
\usepackage{afterpage}
\usepackage[hang,flushmargin]{footmisc}
\usepackage{multirow}
\usepackage{microtype}
\usepackage{ragged2e}
\usepackage{soul}

\newtheorem{theorem}{Theorem}
\newtheorem{lemma}{Lemma}

\newcommand{\R}{\mathbb{R}}
\newcommand{\eps}{\epsilon}
\newcommand{\EE}[1]{\mathbb{E}\left[{#1}\right]}
\newcommand{\EEst}[2]{\mathbb{E}\left[{#1}\  \middle| \ {#2}\right]}

\newcommand{\PP}[1]{\mathbb{P}\left\{{#1}\right\}}
\newcommand{\PPst}[2]{\mathbb{P}\left\{{#1}\  \middle| \ {#2}\right\}}

\newcommand{\eqd}{\stackrel{\textnormal{d}}{=}}
\newcommand{\One}[1]{{\mathbbm{1}}\left\{{#1}\right\}}
\newcommand{\quant}{\mathsf{Q}}
\newcommand{\alg}{\mathcal{A}}
\newcommand{\Ch}{\widehat{C}}
\newcommand{\muh}{\widehat{\mu}}

\newcommand*\samethanks[1][\value{footnote}]{\footnotemark[#1]}
\newcommand{\dtv}{\mathsf{d}_{\mathsf{TV}}}
\newcommand{\dmix}{\mathsf{d}_{\textnormal{mix}}}
\newcommand\independent{\protect\mathpalette{\protect\independenT}{\perp}}
\def\independenT#1#2{\mathrel{\rlap{$#1#2$}\mkern2mu{#1#2}}}
\newcommand{\iidsim}{\stackrel{\textnormal{iid}}{\sim}}

\newcommand{\Xcal}{\mathcal{X}}
\newcommand{\Scal}{\mathcal{S}}
\newcommand{\Tcal}{\mathcal{T}}
\newcommand{\Ycal}{\mathcal{Y}}

\graphicspath{{code/R_plots/}}

\title{Conformal Prediction Beyond Exchangeability}
\author{Rina Foygel Barber\thanks{Department of Statistics, University of Chicago} , 
Emmanuel J.~Cand{\`e}s\thanks{Departments of Statistics and Mathematics, Stanford University} ,
 \\ Aaditya Ramdas\thanks{Departments of Statistics and Machine Learning, Carnegie Mellon University} , 
 Ryan J.~Tibshirani\samethanks}
\date{}

\begin{document}
\maketitle

\begin{abstract}  
Conformal prediction is a popular, modern technique for providing valid
predictive inference for arbitrary machine learning models. Its validity relies
on the assumptions of exchangeability of the data, and symmetry of the given
model fitting algorithm as a function of the data. However, exchangeability is
often violated when predictive models are deployed in practice. For example, if
the data distribution drifts over time, then the data points are no longer
exchangeable; moreover, in such settings, we might want to use a nonsymmetric
algorithm that treats recent observations as more relevant. This paper
generalizes conformal prediction to deal with both aspects: we employ weighted
quantiles to introduce robustness against distribution drift, and design a new
randomization technique to allow for algorithms that do not treat data points
symmetrically. Our new methods are provably robust, with substantially less loss
of coverage when exchangeability is violated due to distribution drift or other
challenging features of real data, while also achieving the same coverage
guarantees as existing conformal prediction methods if the data points are in
fact exchangeable. We demonstrate the practical utility of these new tools with
simulations and real-data experiments on electricity and election forecasting. 
\end{abstract}

\newpage

\section{Introduction}\label{sec:intro}

The field of conformal prediction addresses a challenging modern problem: given
a ``black box'' algorithm that fits a predictive model to available training
data, how can we calibrate prediction intervals around the output of the model
so that these intervals are guaranteed to achieve some desired coverage level?

As an example, consider a holdout set approach. Suppose we have a pre-fitted
model \smash{$\muh$} mapping features $X$ to a prediction of a real-valued
variable $Y$ (e.g., \smash{$\muh$} is the output of some machine learning
algorithm trained on a prior data set), and a fresh holdout set of data
$(X_1,Y_1),\dots,(X_n,Y_n)$ not used for training. We can then use the empirical
quantiles of the errors \smash{$|Y_i - \muh(X_i)|$} on the holdout set to
compute a prediction interval around our prediction \smash{$\muh(X_{n+1})$} that
aims to cover the unseen response $Y_{n+1}$. Split conformal prediction
\citep{vovk2005algorithmic} formalizes this method, and gives guaranteed
predictive coverage when the data points $(X_i,Y_i)$ are drawn i.i.d.\ from
\emph{any} distribution (see Section~\ref{sec:background}). However, the
validity of this method hinges on the assumption that the data points are drawn
independently from the \emph{same} distribution, or more generally, that
$(X_1,Y_1),\dots,(X_{n+1},Y_{n+1})$ are exchangeable.

\begin{figure}[!h]
\centering
\includegraphics[width=0.775\textwidth]{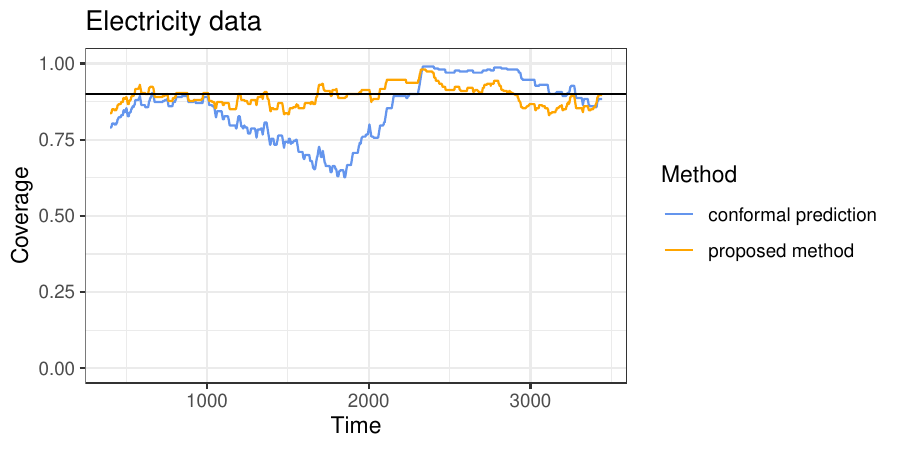}
\caption{Empirical results from a real data set (details will be given in
  Section~\ref{sec:electricity_data}).}  
\label{fig:intro}
\end{figure}

In many applied domains, this assumption is often substantially violated, due to
distribution drift, correlations between data points, or other phenomena. As an
example, Figure~\ref{fig:intro} shows results from an experiment on a real data
set monitoring electricity usage in Australia (the \texttt{ELEC2} data set
\citep{harries1999splice}, which we return to in
Section~\ref{sec:electricity_data}). We see that over a substantial stretch of
time, conformal prediction loses coverage, its intervals decreasing far below
the target 90\% coverage level, while our proposed method, \emph{nonexchangeable
conformal prediction}, is able to maintain approximately the desired coverage
level. In this paper, we will see how to quantify the loss of coverage due to
violations of exchangeability, and how we can modify the conformal prediction
methodology to regain predictive coverage even in the presence of distribution
drift or other violations of exchangeability.

\subsection{Beyond exchangeability}\label{sec:intro_preview}

In this paper, we will consider three important classes of
methods for distribution-free prediction: split conformal, full conformal, and
the jackknife+. (We give background on split and full conformal in
Section~\ref{sec:background}, and on jackknife+ in Appendix~\ref{app:jack}.)
These methods rely on exchangeability in two different ways: 

\begin{itemize} 
\item The data $Z_i = (X_i,Y_i)$ are assumed to be exchangeable (for example, i.i.d.).
\item The algorithm $\alg$, which maps data to a fitted model
  \smash{$\muh:\Xcal\rightarrow\R$}, is assumed to treat the data points
  symmetrically, to ensure that exchangeability of the data points $Z_i$ still
  holds even after we observe the fitted model(s).   
\end{itemize}

In this work, we aim to provide distribution-free prediction guarantees when we 
drop both of these assumptions:  

\begin{itemize}
\item We may have data points $Z_i$ that are not exchangeable---for instance, 
  they may be independent but nonidentically distributed (e.g., due to 
  distribution drift), or there may be dependence among them that creates 
  nonexchangeability (e.g., correlation over space or time). 
\item We may wish to use an algorithm $\alg$ that does not treat the input data 
  points symmetrically---for example, if $Z_i$ denotes data collected at time
  $i$, we may prefer to fit a model \smash{$\muh$} that places higher weight on
  more recent data points. 
\end{itemize}

\subsection{Our contributions} 

We generalize the split conformal, full conformal, and jackknife+ methods
(detailed later) to allow for both of these sources of nonexchangeability.
Our procedures can recover the original variants if a symmetric algorithm is
employed.  We will provide coverage guarantees that are identical to existing
guarantees if the data points are in fact exchangeable,  and only slightly lower
if the deviation from exchangeability is mild.

To elaborate, let us define the \emph{coverage gap} as the loss in coverage
compared to what is achieved under exchangeability. For example, in split
conformal prediction run with a desired coverage level $1-\alpha$,
we have
\[
\textnormal{Coverage gap} = (1 - \alpha) - \PP{Y_{n+1}\in\Ch_n(X_{n+1})}, 
\]
since, under exchangeability, the method guarantees coverage with probability
$1-\alpha$. To give an informal preview of our results, we write $Z_i=(X_i,Y_i)$
to denote the $i$th data point and
\begin{equation}\label{def:z}
Z=(Z_1,\dots,Z_{n+1})
\end{equation}
to denote the full (training and test) data sequence, and let $Z^i$ denote this
sequence after swapping the test point $(X_{n+1},Y_{n+1})$ with the $i$th
training point $(X_i,Y_i)$:  
\begin{equation}\label{eqn:z-swap-indices}
Z^i=(Z_1,\dots,Z_{i-1},Z_{n+1},Z_{i+1},\dots
,Z_{n}, Z_{i}).
\end{equation}
To enable robustness, our methods will allow for weights: let $w_i\in[0,1]$
denote a prespecified weight placed on data point $i$. We will see that the
coverage gap can be bounded as
\begin{equation}\label{eqn:informal_bound_Z}
\textnormal{Coverage gap} \leq \frac{\sum_{i=1}^n w_i \cdot 
\dtv(Z,Z^i)}{1 + \sum_{i=1}^n w_i},
\end{equation}
where \smash{$\dtv$} denotes the total variation distance between
distributions.  Notably, we do not make any assumption on the joint distribution
of the $n+1$ points.  Of course, the result will only be meaningful
if we are able to select fixed (non-data-dependent) weights $w_i$ such that this
upper bound is likely to be small. In particular, if we use these methods, we
are implicitly assuming that this weighted sum of total variation terms is
small. In contrast, past work on conformal prediction in a model-free setting
has relied on the assumption that the data are \emph{exactly} exchangeable,
since no prior results offered an analysis of the coverage gap under violations
of exchangeability. See Section~\ref{sec:remarks_on_theorem} for further
discussion.

Note that the upper bound in \eqref{eqn:informal_bound_Z} is a far stronger
result than simply asking whether the data is ``nearly exchangeable''. For instance,
in a time series, it might be the case that \smash{$\dtv(Z,Z^i)$} is quite small
but \smash{$\dtv(Z,Z_\pi) \approx 1$} for most permutations $\pi$. In words, if
the observations are noisy, then permuting only two data points might not be 
detectable---but if there is nonstationarity or dependence over time then $Z$ is 
likely to be far from exchangeable.

Several further remarks are in order.  First, for $w_i \equiv 1$ and a symmetric
algorithm, the proposed weighted methods will reduce to the usual conformal (or
jackknife+) methods. Thus, the result \eqref{eqn:informal_bound_Z} also
quantifies the degradation in coverage of standard methods in nonexchangeable
settings. Second, this result has new implications in exchangeable settings: if
the data points are in fact exchangeable (with i.i.d.\ as a special case), then
\smash{$Z \eqd Z^i$} and the coverage gap bound in~\eqref{eqn:informal_bound_Z}
is equal to zero (here we use \smash{$\eqd$} for equality in distribution).
Therefore, our use of a weighted conformal procedure (rather than choosing $w_i
\equiv 1$, which is the original unweighted procedure) does not hurt coverage if 
the data are exchangeable. Finally, the result provides insights on why one
might prefer to use our new weighted procedures in (possibly) nonexchangeable
settings: it can provide robustness in the case of distribution shift. To
elaborate, consider a setting where the data points $Z_i$ are independent, but
are not identically distributed due to distribution drift. The following result
relates \smash{$\dtv(Z,Z^i)$} to the distributions of the individual data
points.
\begin{lemma}\label{lem:dtv_swap}
If $Z_1,\dots,Z_{n+1}$ are independent, then
\[
\dtv(Z,Z^i) \leq 2\dtv(Z_i,Z_{n+1}) - \dtv(Z_i,Z_{n+1})^2 \leq 
2\dtv(Z_i,Z_{n+1}).
\]
\end{lemma}
\noindent 
Combining this lemma with~\eqref{eqn:informal_bound_Z}, we can see that if we
are able to place small weights $w_i$ on data points $Z_i$ with large total
variation distance \smash{$\dtv(Z_i,Z_{n+1})$}, then the coverage gap will be
low. For example, under distribution drift, we might have
\smash{$\dtv(Z_i,Z_{n+1})$} decreasing with $i$; we can achieve a low coverage
gap by using, say, weights $w_i = \rho^{n+1-i}$ for some $\rho<1$. We will
return to this example in Section~\ref{sec:examples_theory}.

We will also see that the result in~\eqref{eqn:informal_bound_Z} actually stems
from a stronger result: 
\begin{equation}\label{eqn:informal_bound_R}
\textnormal{Coverage gap} \leq \frac{\sum_{i=1}^n w_i \cdot 
\dtv(R(Z),R(Z^i))}{1+ \sum_{i=1}^n w_i}.
\end{equation}
Here $R(Z)$ denotes a vector of residuals: for split conformal prediction, this
is the vector with entries \smash{$R(Z)_i = |Y_i - \muh(X_i)|$}, where
\smash{$\muh$} is a pre-fitted model, while for full conformal the entries are
again given by \smash{$R(Z)_i = |Y_i - \muh(X_i)|$} but now \smash{$\muh$} is
the model obtained by running $\alg$ on the entire data sequence $Z$. Now
\smash{$R(Z^i)$} is simply the same function applied to the swapped data $Z^i$
instead of $Z$---that is, the residuals are computed after swapping data points
$i$ and $n+1$ in the data set.
  (We also later generalize to any
outcome space $\Ycal$ and to other definitions of residuals.)

Clearly, the bound in \eqref{eqn:informal_bound_R} is strictly stronger
than~\eqref{eqn:informal_bound_Z}, because the total variation distance between
any function applied to each of $Z$ and $Z^i$, cannot be larger than
\smash{$\dtv(Z,Z^i)$} itself---and in many cases, the bound
in~\eqref{eqn:informal_bound_R} may be substantially tighter. For example, if
the data is high dimensional, with $Z_i=(X_i,Y_i)\in\R^p\times\R$ for large $p$,
then the distance \smash{$\dtv(Z,Z^i)$} may be extremely large since $Z$ and
$Z^i$ each contain $p+1$ dimensions of information about each data point. On the
other hand, if we only observe the residuals (e.g., \smash{$R(Z)_i = |Y_i -
\muh(X_i)|$} for each $i$), then this reveals only a one-dimensional summary of
each data point; this typically reduces the distance between the two
distributions, and by a considerable amount if the distribution drift occurs in
features that happen to be irrelevant for prediction and are thus ignored by
\smash{$\muh$}. In Section~\ref{sec:examples_theory}, we will see a specific
example demonstrating the potentially large gap between these two upper bounds.

\section{Background and related work}\label{sec:background}

We briefly review several distribution-free prediction methods that offer
guarantees under an exchangeability assumption on the data and symmetry of the
underlying algorithm. We also set up notation that will be useful later in the
paper.

\paragraph{Split conformal prediction.}

Split conformal prediction \citep{vovk2005algorithmic} (also called inductive
conformal prediction) is a holdout method for constructing prediction intervals
around a pre-trained model. Specifically, given a model
\smash{$\muh:\Xcal\rightarrow\R$} that was fitted on an initial training data
set, and given $n$ additional data points $(X_1,Y_1),\dots,(X_n,Y_n)$ (the
holdout set), we define residuals 
\[
R_i = |Y_i - \muh(X_i)|, \ i=1,\dots,n,
\]
and then compute the prediction interval at the new feature vector $X_{n+1}$ as 
\[
\Ch_n(X_{n+1}) = \muh(X_{n+1}) \pm \textnormal{(the $\lceil
  (1-\alpha)(n+1)\rceil$-th smallest of $R_1,\dots,R_n$)}.
\] 
Equivalently, we can write
\begin{equation}\label{eqn:def_splitCP}
\Ch_n(X_{n+1}) = \muh(X_{n+1})\pm \quant_{1-\alpha}\left(\sum_{i=1}^n
  \tfrac{1}{n+1} \cdot \delta_{R_i} +
  \tfrac{1}{n+1}\cdot\delta_{+\infty}\right), 
\end{equation}
where \smash{$\quant_{\tau}(\cdot)$} denotes the $\tau$-quantile of its
argument,\footnote{If the quantile is not unique, then $\quant_\tau(\cdot)$ denotes the smallest
possible $\tau$-quantile, throughout this paper.}  and $\delta_a$ denotes the point mass at $a$.  This
method is well-known to guarantee distribution-free predictive coverage at the
target level $1-\alpha$.

A drawback of the split conformal method is the loss of accuracy due to sample
splitting, since the pre-trained model $\muh$ needs to be independent from the
holdout set---in practice, if only $n$ labeled data points are available in
total, we might use $n/2$ data points for training \smash{$\muh$}, and then the
procedure defined in~\eqref{eqn:def_splitCP} above would actually be run with a
holdout set of size $n/2$ in place of $n$. In this paper, however, we will
continue to write $n$ to denote the holdout set size for the split conformal
method, in order to allow for universal notation across different methods.

\paragraph{Full conformal prediction.}

To avoid the cost of data splitting, an alternative is the full conformal method
\citep{vovk2005algorithmic}, also referred to as transductive conformal
prediction. Fix any regression algorithm 
\[\alg: \ \cup_{n\geq 0} \left(\Xcal\times \R\right)^n \ \rightarrow \
  \left\{\textnormal{measurable functions $\muh: \Xcal\rightarrow\R$}\right\}, 
\] 
which maps a data set containing any number of pairs $(X_i,Y_i)$, to a fitted
regression function \smash{$\muh$}. The algorithm $\alg$ is required to treat
data points symmetrically, i.e.,\footnote{If $\alg$ is a randomized algorithm, 
then this equality is only required to hold in a distributional sense.}
\begin{equation}\label{eqn:alg_symmetric}
\alg\big((x_{\pi(1)},y_{\pi(1)}),\dots,(x_{\pi(n)},y_{\pi(n)})\big) =
\alg\big((x_1,y_1),\dots,(x_n,y_n)\big)
\end{equation}
for all $n \geq 1$, all permutations $\pi$ on $[n]:=\{1,\dots,n\}$, and all
$\{(x_i,y_i)\}_{i=1,\dots,n}$. Next, for each $y\in\R$, let \[
\muh^y = \alg\big((X_1,Y_1),\dots,(X_n,Y_n),(X_{n+1},y)\big)
\]
denote the trained model, fitted to the training data together with a
hypothesized test point $(X_{n+1},y)$, and let  
\begin{equation}\label{eqn:R_y_i}
R^y_i = 
\begin{cases}|Y_i - \muh^y(X_i)|, & i=1,\dots,n,\\ 
|y-\muh^y(X_{n+1})|, & i=n+1.
\end{cases}
\end{equation}
The prediction set (which might or might not be an interval) for feature vector
$X_{n+1}$ is then defined as 
\begin{equation}\label{eqn:def_fullCP}
\Ch_n(X_{n+1}) =\left\{y \in\R \ : \ R^y_{n+1}\leq
  \quant_{1-\alpha}\left(\sum_{i=1}^{n+1} \tfrac{1}{n+1} \cdot \delta_{R^y_i}
  \right)\right\}. 
\end{equation}
The full conformal method is known to guarantee distribution-free
predictive coverage at the target level $1-\alpha$: 

\begin{theorem}[Full conformal prediction
  \citep{vovk2005algorithmic}]\label{thm:background_fullCP} 
If the data points $(X_1,Y_1),\dots,(X_n,Y_n),(X_{n+1},Y_{n+1})$ are i.i.d.\ (or
more generally, exchangeable), and the algorithm $\alg$ treats the input data
points symmetrically as in~\eqref{eqn:alg_symmetric}, then the full conformal
prediction set defined in~\eqref{eqn:def_fullCP} satisfies 
\[
\PP{Y_{n+1}\in\Ch_n(X_{n+1})} \geq 1-\alpha.
\]
 The same result holds true for split conformal.
\end{theorem}
\noindent  Indeed, since split conformal can be viewed as a special case of
full conformal (by considering a trivial algorithm $\alg$ that returns the same
fixed \emph{pre-fitted} function \smash{$\muh$} regardless of the input data),
this theorem also implies that the same coverage result holds for the split
conformal method~\eqref{eqn:def_splitCP}. For completeness, and to set up our
proof strategy, we will give a succinct proof of this theorem in
Section~\ref{sec:proofs-thm123}.

By avoiding data splitting, full conformal  often (but not always)
yields more precise prediction intervals than split conformal. This potential
benefit comes at a steep computational cost, since in order to compute the
prediction set~\eqref{eqn:def_fullCP} we need to rerun the model training 
algorithm $\alg$ for each $y\in\R$ (or in practice, for each $y$ in a fine
grid). Luckily, in certain special cases such as ordinary least squares, kernel ridge
regression \citep{burnaev2014efficiency}, or the Lasso \citep{lei2017},
the prediction set~\eqref{eqn:def_fullCP}  can be computed more efficiently
using specialized techniques.

 As a compromise between the greater computational efficiency of
split conformal and the greater statistical efficiency of full conformal, the
jackknife+ and CV+ methods \citep{barber2019predictive} (closely related to
cross-conformal prediction~\citep{vovk2015cross}) use a cross-validation-style  
approach for distribution-free predictive inference. For example, for
jackknife+, the procedure requires fitting $n$ leave-one-out models
\smash{$\muh_{-i}$}. Later, in Appendix~\ref{app:jack}, we will give more 
detailed background on these methods, and provide a nonexchangeable version of
the jackknife+.

\paragraph{General nonconformity scores.}

In the exchangeable setting, conformal prediction (both split and full) was
initially proposed in terms of ``nonconformity scores''
\smash{$\widehat{S}(X_i,Y_i)$}, where \smash{$\widehat{S}$} is a fitted function
that measure the extent to which a data point $(X_i,Y_i)$ is unusual relative to
a training data set \citep{vovk2005algorithmic} (whose dependence we make
implicit in the notation). For simplicity, so far we have only presented the
most commonly used nonconformity score, which is the residual from the fitted
model
\begin{equation}\label{eqn:standard_nonconformity_score}
\widehat{S}(X_i,Y_i) := |Y_i - \muh(X_i)|
\end{equation}
(where \smash{$\muh$} is pre-trained for split conformal, and \smash{$\muh =
\alg((X_j,Y_j): j\in[n+1])$} for full conformal).  We will also present our new
methods with this particular form of score.  In many settings, other
nonconformity scores can be more effective---for example,
\cite{romano2019conformalized,kivaranovic2020adaptive} propose scores based on
quantile regression that often lead to tighter prediction intervals in
practice. Our proposed nonexchangeable conformal prediction procedures can also
be extended to allow for general nonconformity scores---we will return to this
generalization in Appendix~\ref{app:general_scores}.

\paragraph{Further related work.}

Conformal prediction was pioneered by Vladimir Vovk and various collaborators in
the early 2000s; the book by~\cite{vovk2005algorithmic} details their advances
and remains a critical resource. The recent spurt of interest in these ideas in
the field of statistics was catalyzed by Jing Lei, Larry Wasserman, and
colleagues (see,
e.g.,~\cite{lei2013distribution,lei2014distribution,lei2018distribution}). For
gentle introduction and more history, we refer to the tutorials
by~\cite{shafer2008tutorial} and~\cite{angelopoulos2021gentle}.

\cite{tibshirani2019conformal} extended conformal prediction to handle
nonexchangeable data under an assumption called \emph{covariate shift}, where
training and test data can have a different $X$ distribution, but are assumed to
have an identical distribution of $Y$ given $X$.  The data is reweighted using
the likelihood ratio to compare the test and training covariate distributions
(with this likelihood ratio assumed to be known or accurately approximable),
coverage can be guaranteed via an argument based on a concept that they called
\emph{weighted exchangeability}.

Our current work differs from~\cite{tibshirani2019conformal} in several
fundamental ways, such that neither work subsumes the other in terms of
methodology or theory.  In their work, the covariate shift assumption must hold,
and the aforementioned high-dimensional likelihood ratio must be known exactly
or well approximated for correct coverage. Furthermore, the weights on the data
points are then calculated as a function of the data point $(X_i,Y_i)$ to
compensate for the known distribution shift.  In the present work, on the other
hand, the weights are required to be \emph{fixed} rather than data-dependent,
and can compensate for \emph{unknown} violations of the exchangeability
assumption, as long as the violations are small (to ensure a low coverage
gap). Moreover, our theory can handle nonsymmetric algorithms that treat
different data points differently, and in particular, can depend on their
order. Finally, and importantly, if there was actually no distribution shift,
and the data happened to be exchangeable, their weighted algorithm does not have
any coverage guarantee, while ours retains exact coverage.

Since its publication, the ideas and methods from~\cite{tibshirani2019conformal}
have been applied and extended in several ways. For
example,~\cite{podkopaev2021distribution} demonstrate that reweighting can also
deal with \emph{label shift} (the marginal distribution of $Y$ changes from
training to test, but the conditional distribution of $X$ given $Y$ is assumed
unchanged). \cite{lei2021conformal} show how reweighting can be extended to
causal inference setups for predictive inference on individual treatment
effects, and~\cite{candes2021conformalized} show how to apply these ideas in the
context of censored outcomes in survival analysis.
\cite{fannjiang2022conformal} use reweighting in a setup where the test
covariate distribution is under the statistician's control.  A different
weighted approach is taken in \cite{guan2021localized}, called ``localized''
conformal prediction, where the weight on data point $i$ is determined as a
function of the distance $\|X_i - X_{n+1}\|_2$, to enable predictive coverage
that holds locally (in neighborhoods of $X$ space, i.e., an approximation of
prediction that holds conditional on the value of $X_{n+1}$). Each of these
works also contributes new ideas to problem-specific challenges (and differs
substantially from the work proposed here, both in terms of methods and the
nature of the resulting guarantees), but we omit the details for brevity.

Conformal methods have also be used for sequential tests for exchangeability of
the underlying data~\citep{vovk2021testing}, and these sequential tests can form
the basis of sequential algorithms for changepoint
detection~\citep{volkhonskiy2017inductive} or outlier
detection~\citep{bates2021testing}. This line of work is differs from ours in
that they employ conformal prediction for detecting nonexchangeability, but do
not provide algorithms or guarantees for the use of conformal methods for
predictive inference on nonexchangeable data. Several other recent works propose
conformal inference type methods for time
series~\citep{chernozhukov2018exact,xu2021conformal,stankeviciute2021conformal},
but these results require exchangeability assumptions or other distributional
conditions (e.g., assuming a strongly mixing time series), while in our present
work we aim to avoid these conditions.  

The recent work of~\cite{gibbs2021adaptive} takes a different approach towards
handling distribution drift in an online manner. Informally, they compare the
current attained coverage to the target $1-\alpha$ level, and if the former is
bigger (or smaller) than the latter, then they iteratively increase (or
decrease) the nominal level $\alpha_t$ to employ for the next prediction.
\cite{zaffran2022adaptive} build further on this approach, allowing for
adaptivity to the amount of dependence in the time series. An alternative
approach is that of~\cite{cauchois2020robust}, where robustness is introduced
under the assumption that the test distribution is bounded in $f$-divergence
from the distribution of the training data points.

For data that is instead drawn from a \emph{spatial} domain, the recent work of
\cite{mao2020valid} uses weighted conformal prediction with higher weights
assigned to data points drawn at spatial locations near that  of the test point
(or, as a special case, giving a weight of 1 to the nearest neighbors of the
test point, and weight 0 to all other points), but their theoretical guarantees
require distributional assumptions. 

Finally, we return full circle to the book of~\cite{vovk2005algorithmic}, which
has chapters that discuss moving beyond exchangeability, for example using
Mondrian conformal prediction (and its generalization, online compression
models). Mondrian methods informally divide the observations into groups, and
assume that the observations within each group are still exchangeable (e.g.,
class-conditional conformal classification). We also note the work
of~\cite{dunn2018distribution} that studies the case of two-layer hierarchical
models (like random effect models) that shares strength across groups. These
works involve very different ideas from those presented in the current paper. 

\section{Nonexchangeable conformal prediction}\label{sec:method_CP}

We now present our new nonexchangeable conformal prediction method, in both its
split and full versions, in this section. 
For clarity of the exposition, we will use \smash{$|y - \muh(x)|$} as the score
used to measure the nonconformity of a point $(x, y)$ in the data set, as
in~\eqref{eqn:standard_nonconformity_score}, but our methods and accompanying
theoretical guarantees can be extended in a straightforward way to arbitrary
nonconformity scores---we give details for this extension in
Appendix~\ref{app:general_scores}.

\subsection{Robust inference through weighted quantiles}
\label{sec:methods_w_only}

As described above, our new methodology moves beyond the exchangeable setting by
allowing both for nonexchangeable data, and for nonsymmetric algorithms. For
simplicity, we will first consider only the first extension---the data points
$Z_i=(X_i,Y_i)$ are no longer required to be exchangeable, but the model fitting
algorithm $\alg$ will still be assumed to be symmetric for now. The next
subsection generalizes the method to allow nonsymmetric algorithms as well.

For our nonexchangeable conformal methods, we choose weights
$w_1,\dots,w_n\in[0,1]$, with the intuition that a higher weight $w_i$ should 
be assigned to a data point $Z_i$ that is ``trusted'' more, i.e., that we believe
comes from (nearly) the same distribution as the test point $Z_{n+1}$. We assume
the weights $w_i$ are fixed (see Section~\ref{sec:extensions} for further
discussion on this point). For instance if data point $Z_i$ occurs at time $i$,
and we are concerned about distribution drift, we might choose weights $w_1 \leq
\dots \leq w_n$ so that our prediction interval relies mostly on recent data 
points and places little weight on data from the distant past.  Alternatively,
in a spatial setting, if data point $i$ is collected at a (prespecified)
location $L_i$, then the weight $w_i$ might be chosen as a function of the
distance $\textnormal{dist}(L_i,L_{n+1})$, with the intuition that data points
collected nearby in the spatial domain are more likely to have similar
distributions.

We now modify the split and full conformal predictive inference methods to use
weighted quantiles, rather than the original definitions where all data points
are implicitly given equal weight. To simplify notation, in what follows, given
$w_i \in [0,1]$, $i=1,\dots,n$, we will define normalized weights   
\begin{equation}\label{eqn:normalized_weights}
\tilde{w}_i = \frac{w_i}{w_1+\dots+w_n+1}, \ i=1,\dots,n, 
\textnormal{ and } \tilde{w}_{n+1}= \frac{1}{w_1+\dots+w_n+1}.
\end{equation}

\paragraph{Nonexchangeable split conformal with a symmetric algorithm.} 

The prediction interval is given by
\begin{equation}\label{eqn:def_splitCP_symm}
\Ch_n(X_{n+1}) = \muh(X_{n+1})\pm \quant_{1-\alpha}\left(\sum_{i=1}^n
  \tilde{w}_i \cdot \delta_{R_i} + \tilde{w}_{n+1}\cdot\delta_{+\infty}\right), 
\end{equation} 
where \smash{$R_i = |Y_i - \muh(X_i)|$} for the pre-trained model
\smash{$\muh$}, as before. 

\paragraph{Nonexchangeable full conformal with a symmetric algorithm.}

The prediction set is given by
\begin{equation}\label{eqn:def_fullCP_symm}
\Ch_n(X_{n+1}) = \left\{y : R^y_{n+1} \leq
  \quant_{1-\alpha}\left(\sum_{i=1}^{n+1} \tilde{w}_i \cdot
    \delta_{R^y_i}\right)\right\},
\end{equation} 
where as before, we define \smash{$\muh^y =
  \alg((X_1,Y_1),\dots,(X_n,Y_n),(X_{n+1},y))$} by running the algorithm $\alg$
on the training data together with the hypothesized test point $(X_{n+1},y)$,
and define \smash{$R^y_i$} as in~\eqref{eqn:R_y_i} from before.   

\medskip

Notice that for both methods, their original (unweighted) versions are recovered 
by choosing weights $w_1 = \dots = w_n=1$.  

The theoretical results for this section, which we previewed
in~\eqref{eqn:informal_bound_Z} and~\eqref{eqn:informal_bound_R}, will follow as
a corollary of more general results that also accommodate nonsymmetric
algorithms (introduced next); we avoid restating the results here for
brevity. In addition, the interested reader may already jump forward to 
Appendix~\ref{app:multiplicative_bound} to examine a different style of result
on the robustness of weighted (and unweighted) conformal methods---using
symmetric algorithms---under a Huber-style adversarial contamination model
 (which relies on stronger assumptions to allow for a
  tighter guarantee).

\subsection{Enhanced predictions with nonsymmetric algorithms} 
\label{sec:method_CP_general}    

Now, we will allow the algorithm $\alg$ to be an arbitrary function of the data
points, removing the requirement of a symmetric algorithm. This generalization
will require only a small modification to the previous conformal method to
ensure validity, and can result in more accurate predictors and boost efficiency
of the resulting prediction sets, as we will demonstrate in the experiments
(Section~\ref{sec:experiments}).  

To begin, let us give some examples of algorithms that do not treat data points
symmetrically, to see what types of settings we want to handle:
\begin{itemize}
\item {\bf Weighted regression.} 
The algorithm $\alg$ might fit a model \smash{$\muh(x) = x^\top\widehat\beta$}
where the parameter vector \smash{$\widehat\beta$} is fitted via a weighted
regression. Specifically, for nonnegative weights $t_i$, consider solving
\begin{equation}\label{eqn:weighted_regr_example}
\widehat\beta = \arg\min_{\beta\in\R^p}\left\{\sum_i t_i \cdot
  \ell(X_i^\top\beta, Y_i) + h(\beta)\right\},
\end{equation} 
for some loss function $\ell$ and penalty function $h$. For example, weighted
least squares would be obtained by taking the loss function $\ell(u,y)=(u-y)^2$.

\item {\bf Adapting to changepoints.} 
In a streaming data setting, if sudden changes may occur in the data
distribution, then the quality of our predictions will suffer if our models are
always trained on the full set of available training data without accounting for
possible changepoints. We might therefore aim to improve the model by building
in a changepoint detection step. Assume data points arrive in an ordered fashion
so that $i = 1$ is the first arrival, $i = 2$ the second, and so on. Then, we
might have  
\begin{equation}\label{eqn:automatic_changepoint_example}
\widehat\beta = \arg\min_{\beta\in\R^p}\left\{\sum_{i >\widehat{T}}
  \ell(X_i^\top\beta, Y_i) + h(\beta)\right\},
\end{equation} 
for some loss function $\ell$ and penalty function $h$, where
\smash{$\widehat{T}$} is the time of the most recent detected changepoint (or
\smash{$\widehat{T}=0$} if no changepoint is detected). To be clear, here the
algorithm $\alg$ incorporates estimation of both \smash{$\widehat{T}$} and of
\smash{$\widehat{\beta}$}. 

\item {\bf Autoregressive models.} 
Suppose that the response $Y_{n+1}$ is best predicted by combining information
from the features $X_{n+1}$ together with response $Y_n$ from the previous time
point---for example, we might solve for  
\begin{equation}\label{eqn:AR_example}
(\widehat\beta,\widehat\gamma) =
\arg\min_{(\beta,\gamma)\in\R^p\times\R}\left\{\sum_i \left(Y_i - (X_i^\top\beta
  + \gamma Y_{i-1})\right)^2\right\},
\end{equation} 
to return a fitted function of the form \smash{$\muh(x,y_{\textnormal{prev}}):=
  x^\top \widehat\beta + \widehat\gamma \cdot y_{\textnormal{prev}}$}.
\end{itemize}

To accommodate these and many other settings, we will now define $\alg$ as  
\begin{equation}\label{eqn:general-alg}
\alg: \ \cup_{n\geq 0} \left(\Xcal\times \R\times \Tcal\right)^n \ \rightarrow \
\left\{\textnormal{measurable functions $\muh: \Xcal\rightarrow\R$}\right\}, 
\end{equation}
mapping a data sequence containing any number of ``tagged'' data points
$(X_i,Y_i,t_i)\in \Xcal\times \R\times \Tcal$, to a fitted regression function
$\muh$. The tag $t_i$ associated with data point $(X_i,Y_i)$ can play a variety
of different roles, depending on the application: 
\begin{itemize}
\item $t_i$ can provide the weight for data point $i$ in a weighted regression; 
\item $t_i$ can indicate the time or spatial location at which data point $i$ is
  sampled; 
\item $t_i$ can simply indicate the order of the data points (i.e., setting
  $t_i=i$ for each $i$), so that $\alg$ is ``aware'' that data point $(X_i,Y_i)$
  is the $i$th data point, and is thus able to use the ordering of the data
  points when fitting the model. 
\end{itemize}
In particular, the algorithm $\alg$ is no longer required to treat the input
data points $(X_i,Y_i)$ symmetrically, because if we swap $(X_i,Y_i)$ with
$(X_j,Y_j)$ (and the algorithm receives tagged data points $(X_j,Y_j,t_i)$
and $(X_i,Y_i,t_j)$), the fitted model may indeed change.\footnote{For many
  common examples, the algorithm $\alg$ will instead be symmetric as a function
  of the \emph{tagged} data points $(X_i,Y_i,t_i)$, but we do not require this
  assumption in this work.} 
As for the weights $w_i$, we require the tags $t_1,\dots,t_{n+1}$ to be fixed. 

With the added flexibility of a nonsymmetric regression algorithm, we will need
a key modification to the methods defined earlier in
Section~\ref{sec:methods_w_only} to maintain predictive coverage. Our
modification requires that, before applying the model fitting algorithm $\alg$,
we first randomly swap the tags of two of the data points in the ordering.
First, draw a random index $K\in[n+1]$ from the multinomial distribution that
takes the value $i$ with probability \smash{$\tilde{w}_i$} (defined
in~\eqref{eqn:normalized_weights}):
\begin{equation}\label{eqn:draw_K}
K \sim \sum_{i=1}^{n+1} \tilde{w}_i \cdot\delta_i.
\end{equation}
Note that $K$ is drawn independently from the
data. We will apply our algorithm to the data $Z^K$ (defined
in~\eqref{eqn:z-swap-indices}) in place of $Z$. In particular, the tagged data
points are now $(X_{n+1},Y_{n+1},t_K)$ and $(X_K,Y_K,t_{n+1})$, i.e., these two 
data points have swapped tags. This modification is carried out as follows.

\paragraph{Nonexchangeable split conformal with a nonsymmetric algorithm.}

For split conformal, the model \smash{$\muh$} is pre-fitted on separate data,
and does not depend on the data points $(X_i,Y_i)$ of the holdout set---in other
words, \smash{$\muh$} is trivially a symmetric function of the $(X_i,Y_i)$
points. Thus, no modification is needed, and our prediction
interval~\eqref{eqn:def_splitCP_symm} is unaltered.  

\paragraph{Nonexchangeable full conformal with a nonsymmetric algorithm.}

First, for any $y\in\R$ and any $k\in[n+1]$, define 
\[
\muh^{y,k} = \alg\left((X_{\pi_k(i)},Y^y_{\pi_k(i)},t_i) : i\in[n+1]\right),
\] 
where $\pi_k$ is the permutation on $[n+1]$ swapping indices $k$ and $n+1$ 
(and $\pi_{n+1}$ is the identity permutation), and where we define 
\[
Y^y_i = \begin{cases}
Y_i , & i= 1,\ldots,n, \\ 
y, & i=n+1.\end{cases} 
\]
In other words, \smash{$\muh^{y,k}$} is fitted by applying the algorithm $\alg$
to the training data $(X_1,Y_1),\dots,(X_n,Y_n)$ together with the hypothesized
test point $(X_{n+1},y)$, but with the $k$th and $(n+1)$st data points swapped
(note that the tags $t_k$ and $t_{n+1}$ are now assigned to data points
$(X_{n+1},y)$ and $(X_k,Y_k)$, respectively, after this swap).

Define the residuals from this model, 
\[
R^{y,k}_i = 
\begin{cases}
|Y_i - \muh^{y,k}(X_i)|, & i=1,\dots,n, \\
|y - \muh^{y,k}(X_{n+1})|, & i = n+1.
\end{cases}
\]
Then, after drawing a random index $K$ as in~\eqref{eqn:draw_K}, the prediction
set is given by 
\begin{equation}\label{eqn:def_fullCP_general}
\Ch_n(X_{n+1}) = \left\{y : R^{y,K}_{n+1} \leq
  \quant_{1-\alpha}\left(\sum_{i=1}^{n+1} \tilde{w}_i \cdot
    \delta_{R^{y,K}_i}\right)\right\}.
\end{equation} 

\medskip

\paragraph{Symmetric algorithms as a special case.}

The symmetric setting, discussed in Section~\ref{sec:methods_w_only}, 
is actually a special case of the broader setting defined here. Specifically,
for any symmetric algorithm $\alg$ that acts on (untagged) data points
$(x_i,y_i)$, we can trivially regard it as an algorithm $\alg'$ that acts on
tagged data points $(x_i,y_i,t_i)$ by simply ignoring the tags. For this reason,
we will only give theoretical results for the general forms of the methods given
in this section, but our theorems apply also to the symmetric setting considered
in Section~\ref{sec:methods_w_only}.

\paragraph{The swap step.}

In the case of a nonsymmetric algorithm, our swap step requires that $\alg$ is
run on the swapped data set---that is, with data points $(X_{n+1},Y_{n+1},t_K)$
and $(X_K,Y_K,t_{n+1})$, rather than the original tagged data points
$(X_K,Y_K,t_K)$ and $(X_{n+1},Y_{n+1},t_{n+1})$.  This swap step is necessary
for our theoretical guarantees to hold (in fact, it plays a key role in the
theory even for \emph{symmetric} algorithms, with fixed weights $w_i$ as
in~\eqref{eqn:def_fullCP_symm}, even though the fitted models are unchanged in 
that case).

Of course, for nonsymmetric algorithm $\alg$, the swap step will alter the
fitted model \smash{$\muh$} produced by $\alg$ and thus may affect the precision
of the resulting prediction interval. The extent to which this swap will perturb
the output of the algorithm $\alg$, will undoubtedly depend on the nature of the 
algorithm itself. In many practical situations, we would not expect the random
swap to have a large impact on the output of the method, since many algorithms
$\alg$ applied to a large number of data points are often not very sensitive to perturbing the training set in this fashion, although interestingly, our
theoretical results do not rely on any stability conditions or any assumptions
of this type.  (For comparison, if we were to instead permute the data at random
before applying the algorithm $\alg$---that is, use a permutation $\pi$ chosen 
uniformly at random, rather than the single swap permutation $\pi_K$, so that
the algorithm is now trained on tagged data points
\smash{$(X_{\pi(i)},Y_{\pi(i)},t_i)$}---then this would restore the symmetric
algorithm assumption, but could potentially result in a highly inaccurate model
since the information carried by the tags is now meaningless.)

However, in certain settings it may be the case that the fitted model
\smash{$\muh$} returns predictions that are far less accurate due to the
swap. For instance, this may be the case in an autoregressive setting where the
tag $t_{n+1}$ indicates the most recent data point and thus plays a
disproportionately large role in the resulting predictions.  We leave the
important question of practical implementation for such settings, and the
question of how to choose algorithms $\alg$ that will not be too sensitive to
the swap, to future work.

\section{Theory}

 In this section, we establish theory on the coverage of our
proposed method. Since split conformal is a special case of full conformal (even
in this nonexchangeable setting), we only present theory for the nonexchangeable
full conformal method.

We first need to define the map from a data sequence $z=(z_1,\dots,z_{n+1}) \in
(\Xcal\times\R)^{n+1}$, with entries $z_i=(x_i,y_i)$, to a vector of residuals
$R(z)$. Given $z$, we first define the model 
\[
\muh =  \alg\big((x_i,y_i,t_i) : i\in[n+1]\big).
\]
Then define the residual vector \smash{$R(z)\in\R^{n+1}$}
with entries 
\[
\big(R(z)\big)_i = |y_i - \muh(x_i)|, \ i=1,\ldots,n+1.
\]

\subsection{Lower bounds on coverage}

Recall the notation $Z_i, Z, Z^i$ defined in~\eqref{def:z}
and~\eqref{eqn:z-swap-indices}.  We now present our coverage
guarantee for nonexchangeable full conformal (and consequently, the same bound
holds for nonexchangeable split conformal as a special case). This theorem can
be viewed as a generalization of Theorem~\ref{thm:background_fullCP}.

\begin{theorem}[Nonexchangeable full conformal
  prediction]\label{thm:fullCP} 
Let $\alg$ be an algorithm mapping a sequence of  triplets $(X_i,Y_i,t_i)$ to a 
fitted function as in~\eqref{eqn:general-alg}. Then the nonexchangeable  
full conformal method defined in~\eqref{eqn:def_fullCP_general} satisfies 
\[
\PP{Y_{n+1}\in\Ch_n(X_{n+1})}\geq 1-\alpha - \sum_{i=1}^n\tilde{w}_i \cdot
\dtv\big(R(Z),R(Z^i)\big).
\] 
 The same result holds true for nonexchangeable split conformal.
\end{theorem}

To summarize, we see that the coverage gap is bounded by
\smash{$\sum_{i=1}^n\tilde{w}_i\cdot \dtv(R(Z),R(Z^i))$}. 
Since it holds that
\[
\dtv\big(R(Z),R(Z^i)\big) \leq \dtv(Z,Z^i)
\]
for each $i$, we therefore also see that 
\[
\textnormal{Coverage gap} \leq \sum_i \tilde{w}_i\cdot\dtv(Z,Z^i).
\] This last bound is arguably more interpretable,
but could also be significantly more loose, and we consider it an important
point that the coverage gap depends on the total variation between swapped
residual vectors, and not the swapped raw data vectors. Finally, recalling
Lemma~\ref{lem:dtv_swap}, we see that in the case of independent data points, we
have
\[
\textnormal{Coverage gap} \leq 2\sum_i \tilde{w}_i\cdot 
\dtv\big((X_i,Y_i),(X_{n+1},Y_{n+1})\big).
\]

\subsection{Upper bounds on coverage}\label{sec:overcover}

To complement the results in the last subsection, it is also possible to verify,
for the nonexchangeable conformal method, that the procedure does
not substantially overcover---that is, under mild deviations from
exchangeability, our method is not overly conservative.

For the exchangeable setting, \citet[Theorem 2.1]{lei2018distribution} show
that, in a setting where the residuals $R_i$ (for split conformal) or
\smash{$R^y_i$} (for full conformal) are distinct with probability 1, conformal
prediction satisfies 
\[
1-\alpha \leq \PP{Y_{n+1}\in\Ch_n(X_{n+1})} < 1- \alpha + \frac{1}{n+1}. 
\]  
Here we give the analogous results for our nonexchangeable methods.

\begin{theorem}\label{thm:overcover}
For any algorithm $\alg$ 
as in~\eqref{eqn:general-alg},  
if
\smash{$R^{Y_{n+1},K}_1,\dots,R^{Y_{n+1},K}_n,R^{Y_{n+1},K}_{n+1}$} are distinct
with probability 1, then the nonexchangeable full conformal
method~\eqref{eqn:def_fullCP_general} satisfies 
\[
\PP{Y_{n+1}\in\Ch_n(X_{n+1})} <  1 -  \alpha + \tilde{w}_{n+1} +
\sum_{i=1}^n\tilde{w}_i \cdot \dtv\big(R(Z), 
R(Z^i)\big).
\] 
 The same result holds true for nonexchangeable split conformal.
\end{theorem}
\noindent (In the split conformal context, since \smash{$R^{Y_{n+1},K}_i
= |Y_i -\muh(X_i)|$} for the pre-fitted function \smash{$\muh$}, the statement
becomes simpler---we simply require that the residuals \smash{$|Y_i
-\muh(X_i)|$} are distinct with probability 1.)

From this result, we see that if \smash{$\tilde{w}_{n+1} =
  \frac{1}{w_1+\dots+w_n+1}$} is small (which corresponds to the effective
sample size of our weighted method being large), then mild violations of
exchangeability can only lead to mild undercoverage (as in
Theorem~\ref{thm:fullCP}) or to mild overcoverage.  

Of course, when we use these methods in practice, it would be useful to know 
whether overcoverage or undercoverage is to be expected; however, without
further assumptions, this cannot be determined in advance. As a simple example, 
if the data exhibits mild violations of exchangeability due to the conditional
variance of $Y \mid X$ changing over time, then we might see undercoverage if 
$\textnormal{Var}(Y \mid X)$ increases over time (and thus the residual of the 
test point $(X_{n+1},Y_{n+1})$ is larger than typical training residuals), or
overcoverage if $\textnormal{Var}(Y \mid X)$ is instead decreasing over time.

\subsection{Remarks on the theorems}\label{sec:remarks_on_theorem}

A few comments are in order to help us further understand the implications of
these theoretical results. 

\paragraph{New results in the exchangeable setting.}

We point out that  when the data happen to be exchangeable, that is,
$\dtv(Z,Z^i)=0$ for all $i$, then the above results are new and cannot be
inferred from the existing conformal literature. In particular, existing
conformal methods are not able to handle nonsymmetric algorithms, which limits
their applicability in many practical settings (e.g., streaming data, as
described above). In addition, our results show that, under exchangeability,
there is no coverage lost by introducing fixed weights $w_i$ into the quantile
calculations used for constructing the prediction interval; this means that we
are free to use these weights to help ensure robustness against
nonexchangeability without sacrificing any guarantees if indeed exchangeability
happens to hold. 

\paragraph{Robustness results for the original algorithms.} 

Another interesting implication of these new bounds is that they yield
robustness results for the original algorithms. In more detail, the original
split conformal~\eqref{eqn:def_splitCP} and full
conformal~\eqref{eqn:def_fullCP} algorithms presented in
Section~\ref{sec:background} can be viewed as special cases of our proposed
nonexchangeable methods~\eqref{eqn:def_splitCP_symm}
and~\eqref{eqn:def_fullCP_general}, respectively, by taking weights $w_1 = \dots
= w_n = 1$ and using a symmetric $\alg$, i.e., without tags.
(As we will see in Appendix~\ref{app:jack} below, the same is true
for viewing jackknife+ as a special case of the nonexchangeable jackknife+.)
In this setting, our theorems establish a new robustness result,
\[
\textnormal{Coverage gap} \leq \frac{\sum_{i=1}^n \dtv(R(Z),R(Z^i))}{n+1} \leq
\frac{\sum_{i=1}^n \dtv(Z,Z^i)}{n+1}.
\] 
For example, in the case of independent data points, applying
Lemma~\ref{lem:dtv_swap} we obtain 
\[
\textnormal{Coverage gap} \leq \frac{2\sum_{i=1}^n
  \dtv\big((X_i,Y_i),(X_{n+1},Y_{n+1})\big))}{n+1}. 
\] 
These new bounds ensure robustness of existing methods against mild violations
of the exchangeability (or i.i.d.) assumption, and thus help explain the success
of these methods on real data, where the exchangeability assumption may not
hold. 

\paragraph{Choosing the weights.}

Our theoretical results above confirm the intuition that we should give higher
weights $w_i$ to data points $(X_i,Y_i)$ that we believe are drawn from a
similar distribution as $(X_{n+1},Y_{n+1})$, and lower weights to those that are
less reliable. As is always the case with inference methods, we are faced with a
tradeoff: if many weights $w_i$ are chosen to be quite low, then this reduces
the effective sample size of the method (e.g., for split conformal prediction,
we are reducing the effective sample size for estimating the empirical quantile
of the residual distribution). Thus, overly low weights will often lead to
wider prediction intervals---at the extreme, if we choose $w_1= \dots = w_n=0$,
this yields a coverage gap of zero but results in \smash{$\Ch_n(X_{n+1}) \equiv
\R$}, a completely uninformative prediction interval. How to choose weights
optimally (and, even how to quantify optimality) is an interesting and important
question that we leave for future work.

\paragraph{Is the guarantee useful?} 

While the upper bound on the coverage gap holds with no assumptions on the distribution
of the data, the result is meaningless if this upper bound is extremely
large. Thus we would ideally use these methods in settings where we have some
\emph{a priori} knowledge about the properties of the data distribution, so that
the weights $w_i$ can be chosen in advance in such a way that we believe the
resulting coverage gap is likely to be small. We emphasize in practice we likely
only need qualitative (not quantitative) knowledge of the likely deviations from
exchangeability---for example, under gradual distribution drift, a geometric decay as in
$w_i = \rho^{n+1-i}$ will likely lead to a low coverage gap, without requiring
knowledge of the exact rate or nature of the distribution drift.  On the other
hand, if the test point comes from a new distribution that bears no resemblance
to the training data, neither our bound nor any other method would be able
to guarantee valid coverage without further assumptions.  An
important open question is whether it may be possible to determine, in an
adaptive way, whether coverage will likely hold for a particular data set, or
whether that data set exhibits high deviations from exchangeability such that
the coverage gap may be large.

\subsection{Examples}\label{sec:examples_theory}

Before turning to our empirical results, we pause to give several examples of
settings where the coverage gap bound is favorable. 

\paragraph{Bounded distribution drift.}

First, consider a setting where the data points $(X_i,Y_i)$ are independent, but
experience distribution drift over time.  In this type of setting, we would want
to choose weights $w_i$ that decay as we move into the distant past, for
example, $w_i = \rho^{n+1-i}$ for some decay parameter $\rho\in(0,1)$.  If we
assume that the distribution drift is bounded with a Lipschitz-type condition,
\[
\dtv(Z_i,Z_{n+1})\leq \eps \cdot (n+1-i), \ i=1,\dots,n+1,
\] 
for some $\eps>0$, then the coverage gap for our proposed methods is bounded as 
\begin{multline*}
\textnormal{Coverage gap} \leq \sum_i \tilde{w}_i\cdot\dtv(Z,Z^i) \leq \sum_i
\tilde{w}_i \cdot 2\dtv(Z_i,Z_{n+1}) \\ 
\leq \sum_{i=1}^n \frac{\rho^{n+1-i}}{1 + \sum_{j=1}^n \rho^{n+1-j}}\cdot
2\eps\cdot (n+1-i)\leq \frac{2\eps}{1-\rho},
\end{multline*} 
which is small as long as the distribution drift parameter $\eps$ is sufficiently small.

\paragraph{Changepoints.} 

In other settings with independent data points $(X_i,Y_i)$, we might have
periodic large changes in the distribution rather than the gradual drift studied
above---that is, we may be faced with a changepoint. Suppose that the most
recent changepoint occurred $k$ time steps ago, so that
\smash{$\dtv(Z_i,Z_{n+1}) = 0$} for $i>n-k$  (but, before that time, the
distribution might be arbitrarily different from the test point, so we might
even have \smash{$\dtv(Z_i,Z_{n+1}) = 1$} for $i\leq n-k$). In this setting,
again taking weights $w_i = \rho^{n+1-i}$ that decay as we move into the past, 
we have
\[
\textnormal{Coverage gap} \leq \sum_{i=1}^n \tilde{w}_i\cdot\dtv(Z,Z^i) 
\leq \sum_{i=1}^{n-k}\tilde{w}_i 
= \frac{\sum_{i=1}^{n-k} \rho^{n+1-i}}{1 + \sum_{i=1}^n \rho^{n+1-i}} 
\leq \rho^k.  
\]
This yields a small coverage gap as long as $k$ is large, i.e., as long as we
have plenty of data observed after the most recent changepoint. 

\paragraph{Covariate time series.}

Next, to highlight the distinction between \smash{$\dtv(Z,Z^i)$} and
\smash{$\dtv(R(Z),R(Z^i))$}, we will consider a setting where the data points
$(X_i,Y_i)$ are no longer independent. Suppose that $Y_i = X_i^\top \beta +
\eps_i$ where \smash{$\eps_i\iidsim \mathcal{N}(0,\sigma^2)$} but where the
covariates $X_i$ are not i.i.d. For example, the covariates may be dependent 
due to a time series structure, or may be independent but not identically
distributed. Writing $X\in\R^{(n+1)\times p}$ to denote the covariate matrix (with
$i$th row $X_i$), we will assume that $\textnormal{vec}(X) \sim
\mathcal{N}(0,\Sigma)$ for some $\Sigma\in\R^{(n+1)p\times (n+1)p}$, allowing
for both nonindependent and/or nonidentically distributed rows $X_i$. Now
consider running full conformal with least squares regression as the base
algorithm, so that we have residuals
\[
R(Z) = Y - X(X^\top X)^{-1}X^\top Y =
\mathcal{P}^\perp_X(Y) =  \mathcal{P}^\perp_X(\eps), 
\]
where $Y=(Y_1,\dots,Y_{n+1})$, $\eps = (\eps_1,\dots,\eps_{n+1})$, and
$\mathcal{P}^\perp_X$ denotes projection to the orthogonal complement of 
the column span of $X$. In
Appendix~\ref{app:calculations_eqn:example_covariate_timeseries} we prove that 
\begin{equation}\label{eqn:example_covariate_timeseries}
\dtv(R(Z),R(Z^i)) \leq  
\sqrt{8} \kappa_\Sigma \cdot \frac{p}{\sqrt{n+1-p}},
\end{equation}
where $\kappa_\Sigma$ is the condition number of $\Sigma$; if $n \gg p^2$ while $\kappa_\Sigma$ is bounded,
then
this total variation distance is very small. 

On the other hand, it is likely that \smash{$\dtv(Z,Z^i)$} is very large (it may
even be close to the largest possible value of 1), unless the covariates are
essentially exchangeable. For example, in dimension $p=1$, we can consider the 
autoregressive model $X_i = \gamma  X_{i-1} + \mathcal{N}(0,1-\gamma^2)$,
with $X_1\sim \mathcal{N}(0,1)$, so $X_1,\dots,X_{n+1}$ are identically
distributed. Then, for $2\leq i\leq n$ we have   
\begin{multline*}
\dtv(Z,Z^i) \geq \dtv(X_i - \gamma X_{i-1},X_{n+1} - \gamma X_{i-1} ) \\
= \dtv\big(\mathcal{N}(0,1-\gamma^2), \mathcal{N}(0, 1+ \gamma^2
-2\gamma^{n+3-i}) \big), 
\end{multline*}
which is proportional to $\gamma^2$. This shows that
\smash{$\dtv(R(Z),R(Z^i))$} can be 
vanishingly small even when \smash{$\dtv(Z,Z^i)$} is bounded away from zero. 

\subsection{Extensions and explorations}\label{sec:extensions}

We now briefly describe several extensions of our general framework.

\paragraph{Additive versus multiplicative bounds.}

In our theoretical results above, the reduction in coverage is
additive---that is, the probability
\smash{$\mathbb{P}\{Y_{n+1}\not\in\Ch_n(X_{n+1})\}$} has the form $\alpha +
\Delta$, where the term $\Delta$ reflects the extent to which the
exchangeability assumption is violated (as measured by total variation
distance). If the target non-coverage level $\alpha$ is extremely low, then this
additive bound may represent a substantial increase in the probability of
error. In Appendix~\ref{app:multiplicative_bound}, we give an alternative bound
under a Huber contamination model, which is multiplicative rather than additive,
but holds only for the symmetric algorithm case.

\paragraph{Fixed versus data-dependent weights.}

Throughout this paper, we have worked under the assumption that the weights
$w_i$ on the conformal residuals, as well as the tags $t_i$ used in model
fitting in the nonsymmetric case, are fixed a priori. In contrast, when weighted
versions of
conformal prediction are used for addressing problems such as covariate shift 
\citep{tibshirani2019conformal}, data censoring
\citep{candes2021conformalized},  or local coverage 
\citep{guan2021localized}, the weights are data-dependent, i.e., $w_i = w(X_i)$
or $w_i = w(X_i,X_{n+1})$, in each of these settings. We pause here to comment
on this distinction.

In practical applications of our proposed methods, it may be the case that we
would like to use weights and/or tags that are somehow random---for example, if
each data point $(X_i,Y_i)$ is gathered at a random time $T_i$, the weight $w_i$
and tag $t_i$ might then need to depend on $T_i$. In the setting where the weights $w_i$
and/or tags $t_i$ may be random or data-dependent, our results
will still apply if the terms \smash{$\dtv(Z,Z^i)$} appearing in our bounds on
the coverage gap are replaced with suitable conditional versions, 
\[
\textnormal{Coverage gap} \leq \EE{\sum_{i=1}^n \tilde{w}_i\cdot \dtv\big(Z,Z_i
  \mid w_1,\dots,w_n,t_1,\dots,t_{n+1}\big)},
\] 
where now the $i$th term on the right-hand side is the total variation distance
between the \emph{conditional} distributions of $Z$ and $Z^i$, conditioning on
the weights and tags.
 We might therefore consider a possible extension that unifies the
proposed framework with the weighted conformal prediction
\citep{tibshirani2019conformal} and/or localized conformal prediction (LCP)
\citep{guan2021localized} methods, where the weight $w_i$ (and, potentially, the tag $t_i$) placed on data point
$i$ might now additionally incorporate data-dependent information---e.g., the weight $w_i$ 
might depend on both the index $i$, as in our framework, and on $\|X_i -
X_{n+1}\|_2$, as in the LCP framework. We leave a more detailed investigation
of data dependent weights for future work.

\paragraph{Are these results assuming the data is approximately
    exchangeable?}

Finally, we point out that these coverage gap bounds are very different in
flavor than simply assuming that $Z$ is ``nearly exchangeable''. In particular, 
in a setting where \smash{$\dtv(Z,\tilde{Z})$} is small for some exchangeable
\smash{$\tilde{Z}$}, it follows immediately that the coverage gap is bounded by 
\smash{$\dtv(Z,\tilde{Z})$} for (unweighted) split or full conformal, since these methods are guaranteed to have coverage
$1-\alpha$ with exchangeable data
\smash{$\tilde{Z}$}. By comparison, our coverage gap bound \smash{$\sum_i
\tilde{w}_i\cdot\dtv(Z,Z^i)$} is substantially stronger.

To see this through an example, consider a distribution where the covariates
$X_i$ are i.i.d., and where $Y_i\sim \textnormal{Bernoulli}(0.5 + (-1)^i \cdot
\eps)$, for some small constant $\eps>0$. Suppose that we run conformal
prediction without weights, $w_i \equiv 1$. Then we have
\smash{$\dtv(Z_i,Z_{n+1}) \leq 2 \eps$} for all $i$, and so our coverage gap
bound ensures that conformal prediction has coverage at least
$1-\alpha-4\eps$. On the other hand, we have   
\[
\dtv(Z,\tilde{Z}) \approx 1 \textnormal{ for any exchangeable $\tilde{Z}$}.
\]
To verify the above claim, note that under the distribution of $Z$, we have 
\[
\sum_{i=1}^{\lfloor \frac{n+1}{2}\rfloor} \One{Y_{2i-1} < Y_{2i}} +
B_i\One{Y_{2i-1}=Y_{2i}} \sim\textnormal{Binomial}\left(\lfloor
  \tfrac{n+1}{2}\rfloor, 0.5+\eps \right),
\] 
where $B_i\iidsim \textnormal{Bernoulli}(0.5)$,
while under any exchangeable distribution, the left-hand side above is
distributed as \smash{$\textnormal{Binomial}\left(\lfloor \tfrac{n+1}{2}\rfloor,
    0.5 \right)$}, and these two binomial distributions have total variation
distance $\approx 1$, for large $n$. Thus, in this example, we see that our coverage
gap is low even though it is not the case that $Z$ is ``nearly exchangeable''.  

\section{Experiments}\label{sec:experiments}

In this section, we examine the empirical performance of nonexchangeable
full conformal prediction, with residual weights and allowing for a nonsymmetric 
algorithm, against the original full conformal method. (Additional experiments
that implement split conformal and jackknife+ can be found in
Appendix~\ref{app:additional_simulations}.) We will see that adding weights
enables robustness against changes in the data distribution (i.e., better
coverage), while moving to a nonsymmetric algorithm enables shorter
prediction intervals.\footnote{Code for reproducing the experiments in Sections~\ref{sec:simulations}
and~\ref{sec:electricity_data} is available at 
\url{https://rinafb.github.io/code/nonexchangeable_conformal.zip}.}

\subsection{Simulations}\label{sec:simulations}

We consider three simulated data distributions: 
\begin{itemize}
\item {\bf Setting 1: i.i.d.\ data.} We generate $N=2000$ i.i.d.\ data points
  $(X_i,Y_i)$, with \smash{$X_i \iidsim \mathcal{N}(0,\mathbf{I}_4)$} and $Y_i
  \sim X_i^\top \beta + \mathcal{N}(0,1)$ for a coefficient vector
  $\beta=(2,1,0,0)$.  

\item {\bf Setting 2: changepoints.} We generate $N=2000$ data points
  $(X_i,Y_i)$, with \smash{$X_i \iidsim \mathcal{N}(0,\mathbf{I}_4)$} and
  \smash{$Y_i \sim X_i^\top \beta^{(i)} + \mathcal{N}(0,1)$}. Here
  \smash{$\beta^{(i)}$} is the coefficient vector at time $i$, and changes two
  times over the duration of data collection:
\begin{align*}
&\beta^{(1)}=\dots = \beta^{(500)} = (2,1,0,0), \\
&\beta^{(501)}=\dots = \beta^{(1500)} = (0,-2,-1,0), \\
&\beta^{(1501)}=\dots = \beta^{(2000)}=(0,0,2,1).
\end{align*}

\item {\bf Setting 3: distribution drift.} We generate $N=2000$ data points
  $(X_i,Y_i)$, with \smash{$X_i \iidsim \mathcal{N}(0,\mathbf{I}_4)$} and
  \smash{$Y_i \sim X_i^\top \beta^{(i)} + \mathcal{N}(0,1)$}. As before,
  \smash{$\beta^{(i)}$} is the coefficient vector at time $i$; but now we set
  \smash{$\beta^{(1)} = (2,1,0,0)$}, \smash{$\beta^{(N)}=(0,0,2,1)$}, and then
  compute each intermediate \smash{$\beta^{(i)}$} by linear interpolation.  
\end{itemize}

For each task, we implement the following three methods, with target coverage
level $1-\alpha = 0.9$. 

\begin{itemize}
\item {\bf CP+LS: full conformal prediction with least squares.} We consider the
  original definition of full conformal prediction~\eqref{eqn:def_fullCP}, with 
  \smash{$\muh$} the least squares fit, i.e., $\alg$ is the least squares
  regression algorithm.\footnote{ In principle, full conformal run
    with $\alg$ given by least squares may return a prediction set
    \smash{$\Ch(X_{n+1})$} that is a disjoint union of intervals, but this is
    rare in most typical settings. The same is true for NexCP and for weighted
    least squares. For interpretability, we implement each method to always
    return an interval (i.e., if \smash{$\Ch(X_{n+1})$} happens to be a disjoint
    union of intervals, then we return its convex hull).}  

\item {\bf NexCP+LS: nonexchangeable full conformal with least squares.}
  We also run nonexchangeable full conformal
  prediction~\eqref{eqn:def_fullCP_symm} using weights $w_i = 0.99^{n+1-i}$,  
  and with the same algorithm $\alg$ (least squares regression). 

\item {\bf NexCP+WLS: nonexchangeable full conformal with weighted 
    least squares.} Lastly we use nonexchangeable full conformal
  prediction~\eqref{eqn:def_fullCP_symm} but now with a nonsymmetric algorithm,
  weighted least squares regression. Specifically, to fit \smash{$\muh$} given
  tagged data points $(x_i,y_i,t_i)$, the algorithm $\alg$ will run weighted
  least squares regression placing weight $t_i$ on data point $(x_i,y_i)$. We
  implement the algorithm with $t_i = 0.99^{n+1-i}$, and again use weights $w_i
  = 0.99^{n+1-i}$.  
\end{itemize}

After a burn-in period of the first 100 time points, at each time
$n=100,\dots,N-1$ we run the methods with training data $i=1,\dots,n$ and test
point $n+1$. The results shown are averaged over 200 independent replications of
the simulation.

\begin{figure}[htb]
\includegraphics[width=\textwidth]{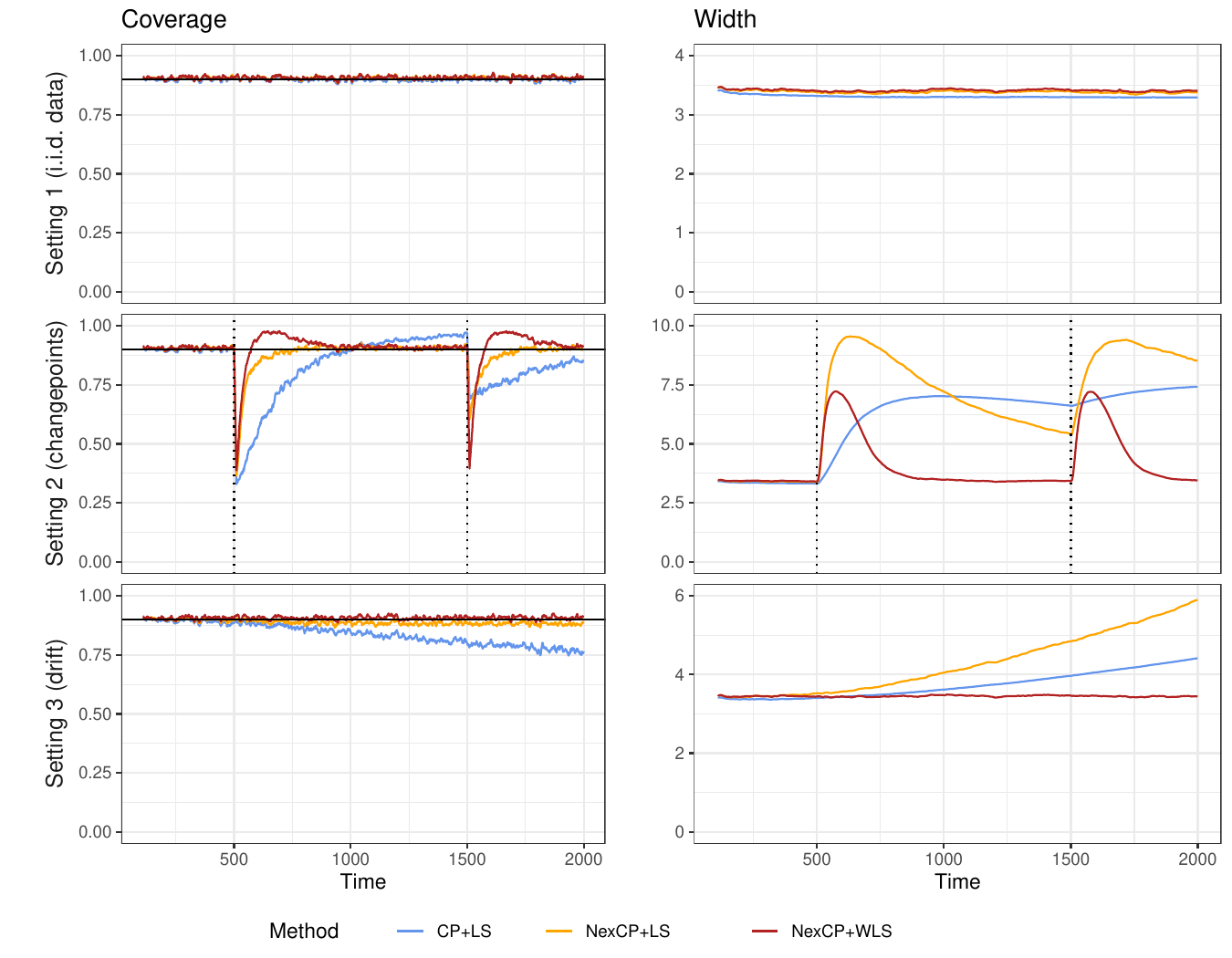}
\caption{Simulation results showing mean prediction interval coverage and width,
  averaged over 200 independent trials. The displayed curves are smoothed by
  taking a rolling average with a window of 10 time points.}
\label{fig:simulation}
\end{figure}

Our results are shown in Figure~\ref{fig:simulation}, and are summarized in
Table~\ref{tab:simulation}. In terms of coverage, we see that all three methods
have coverage $\approx 90\%$ across the time range of the experiment for the
i.i.d.\ data setting (Setting 1), while for the changepoint (Setting 2) and
distribution drift (Setting 3) experiments, the two proposed methods achieve
approximately the desired coverage level, but the original full conformal method
CP+LS undercovers. In particular, as expected, CP+LS shows steep drops in
coverage in Setting 2 after changepoints, while in Setting 3 the coverage for
CP+LS declines gradually over time as the distribution drift grows.  The
NexCP+LS and NexCP+WLS methods are better able to maintain coverage in
these settings. (In fact, in Setting 2, we see that NexCP+WLS overcovers for
a period of time after each changepoint---this is because, a short period of
time after the changepoint, the fitted weighted least squares model is already
quite accurate for the new data distribution, but the weights
\smash{$\tilde{w}_i$} are still placing some weight on residuals from data
points from before the changepoint, leading briefly to an overestimate of our
model error.)

{\begin{table}[htb]\small\centering
\begin{tabular}{l|cc|cc|cc}
&\multicolumn{2}{c|}{Setting 1 (i.i.d.\ data)}
&\multicolumn{2}{c|}{Setting 2 (changepoints)}
&\multicolumn{2}{c}{Setting 3 (drift)}\\
&Coverage& Width &Coverage& Width&Coverage& Width\\\hline
CP+LS & 0.900 &3.31&0.835 &5.99 &0.838 &3.73\\
NexCP+LS &0.907 &3.39&0.884 &6.83&0.888&4.29\\
NexCP+WLS &0.907 &3.42& 0.906&4.13&0.907&3.45
\end{tabular}
\caption{Simulation results showing mean prediction interval coverage and width,
  averaged over all time points and over 200 trials.}
\label{tab:simulation}
\end{table}}


Turning to the prediction interval width, for the i.i.d.\ data setting (Setting
1), the three methods show similar mean widths, although the widths for
NexCP+LS and NexCP+WLS are very slightly higher than for CP+LS; in
addition, variability is higher for NexCP+LS and NexCP+WLS than for CP+LS,
which is to be expected since using decaying weights $w_i$ for computing the
prediction intervals leads to a lower effective sample size.  For the
changepoint (Setting 2) and distribution drift (Setting 3) experiments, we see
that NexCP+LS leads to wider prediction intervals than the original method
CP+LS, which is to be expected since NexCP+LS is using the same model fitting
algorithm but avoiding the undercoverage issue of CP+LS. More importantly,
NexCP+WLS is able to construct narrower prediction intervals than CP+LS,
while avoiding undercoverage. This is due to the fact that weighted least
squares leads to more accurate fitted models. This highlights the utility of
nonsymmetric algorithms for settings where data are not exchangeable.

\subsection{Electricity data set}\label{sec:electricity_data}

We now compare the three methods on a real data set. The \texttt{ELEC2} data
set\footnote{Data was obtained from
  \url{https://www.kaggle.com/yashsharan/the-elec2-data set}.}  
 \citep{harries1999splice} tracks electricity usage and pricing
in the states of New South Wales and Victoria in Australia, every 30 minutes
over a 2.5 year period in 1996--1999. (This data set was previously analyzed by
\cite{vovk2021protected} in the context of conformal prediction, finding
distribution drift that violated exchangeability.) 

For our experiment, we use four covariates: \texttt{nswprice} and
\texttt{vicprice}, the price of electricity in each of the two states, and
\texttt{nswdemand} and \texttt{vicdemand}, the usage demand in each of the two
states. Our response variable is \texttt{transfer}, the quantity of
electricity transferred between the two states. We work with a subset of the
data, keeping only those observations in the time range 9:00am--12:00pm (aiming
to remove daily fluctuation effects), and discarding an initial stretch of time
during which the value \texttt{transfer} is constant. After these steps, we have
$N=3444$ time points. We then implement the same three methods as in the
simulations (CP+LS, NexCP+LS, and NexCP+WLS), using the exact same
definitions and settings as before.  

\begin{figure}[tb]
\includegraphics[width=\textwidth]{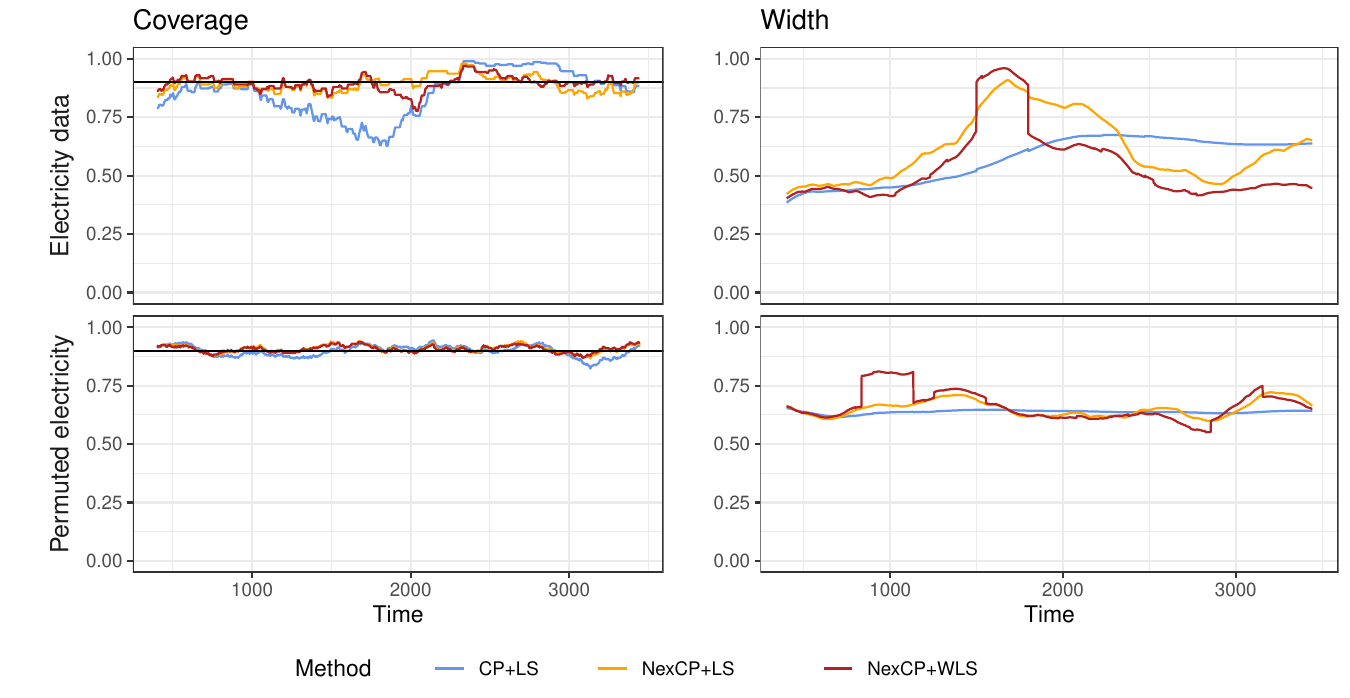}
\caption{Electricity data results showing coverage and prediction interval width
  on the original data and the permuted data. The displayed curves are smoothed
  by taking a rolling average with a window of 300 time points.} 
\label{fig:electricity}
\end{figure}

Our goal is to examine how distribution drift over the duration of this 2.5 year
period will affect each of the three methods. As a sort of ``control group'', we
also perform the experiment with a permuted version of this same data set---we
draw a permutation $\pi$ on $[N]$ uniformly at random, and then repeat the same
experiment on the permuted data set
\smash{$(X_{\pi(1)},Y_{\pi(1)}),\dots,(X_{\pi(N)},Y_{\pi(N)})$}.  The random
permutation ensures that the distribution of this data set now satisfies
exchangeability.

Our results are shown in Figure~\ref{fig:electricity}, and summarized in
Table~\ref{tab:electricity}.  On the original data set, we see that the
unweighted method CP+LS shows some undercoverage, while both NexCP+LS and 
NexCP+WLS achieve nearly the desired $90\%$ coverage level.  In particular,
CP+LS shows undercoverage during a long range of time around the middle of the
duration of the experiment, and then recovers, showing the effects of
distribution drift in this data set---this occurs as the response variable
\texttt{transfer} is more noisy during the middle of the time range, as compared
to the beginning and end of the time range.  On the permuted data set, on the
other hand, all three methods show coverage that is close to $90\%$ throughout
the time range, which is expected since the permuted data set is exchangeable.

{\begin{table}[htb]\small\centering
\begin{tabular}{l|cc|cc}
&\multicolumn{2}{c|}{Electricity data}&\multicolumn{2}{c}{Permuted electricity data}\\
& Coverage & Width& Coverage & Width\\\hline
CP+LS & 0.852 &0.565&0.899&0.639\\
NexCP+LS &0.890 &0.606&0.908&0.652\\
NexCP+WLS &0.893 &0.527&0.908&0.663
\end{tabular}
\caption{Electricity data results showing coverage and prediction interval width
  on the original data and the permuted data, averaged over all time points.} 
\label{tab:electricity}
\end{table}}


Turning now to prediction interval width, on the original data set we see that
the interval width of NexCP+LS is generally larger than that of NexCP+WLS, again
demonstrating the advantage of a nonsymmetric algorithm. For the permuted data
set, on the other hand, the interval widths are similar, although NexCP+LS and
NexCP+WLS show higher variability; this is explained by the lower effective
sample size that is introduced by weighting the data points, combined with the
heavy-tailed nature of the data.

\subsection{Election data set}\label{sec:election_data}

Finally, we apply our weighted methods to predict how Americans voted in the
2020 U.S.\ presidential election. Our experiments in this subsection are
inspired by the work of \cite{Cherian2020How} for The Washington Post.    

\begin{figure}
\includegraphics[width=0.53\textwidth]{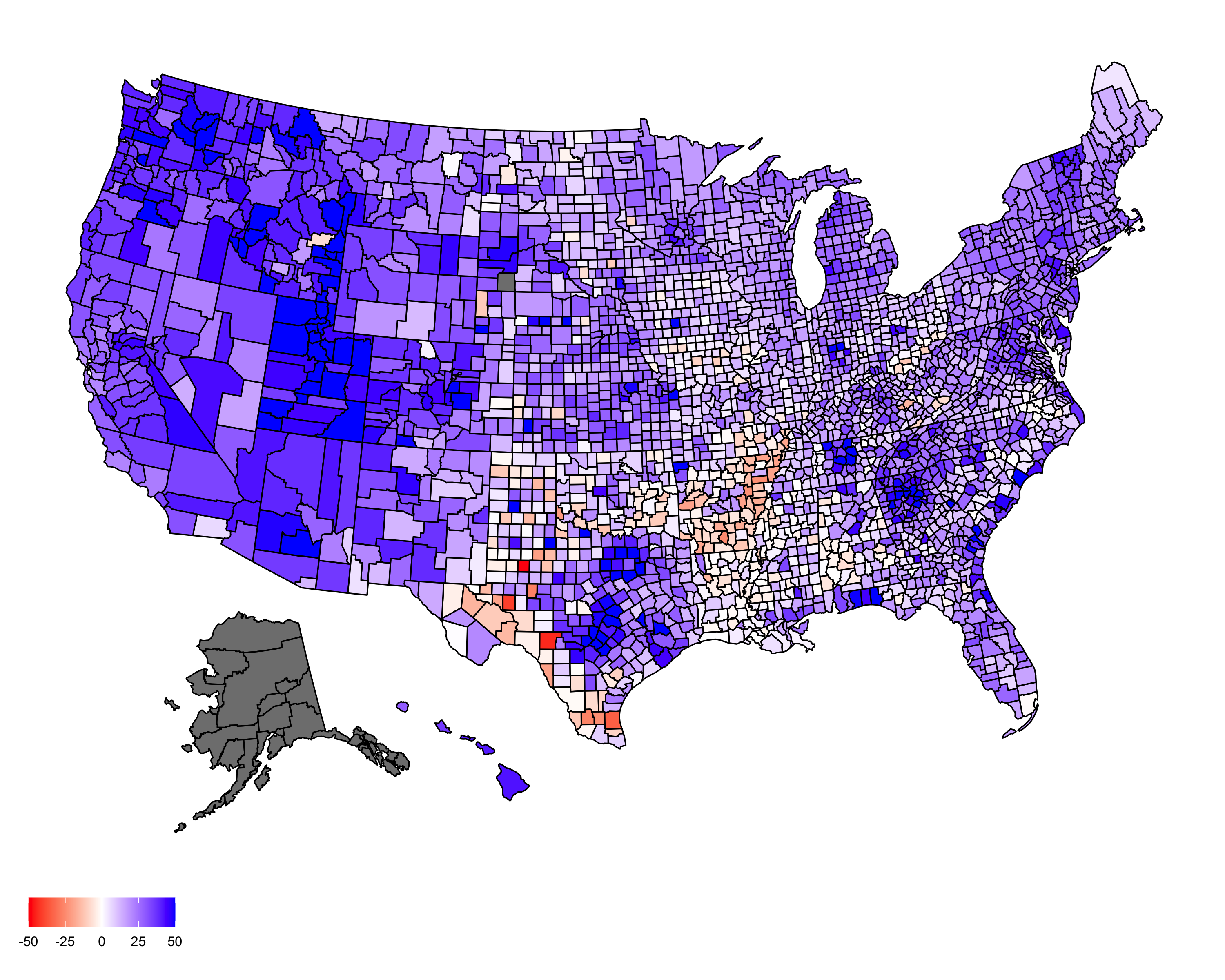}
\hspace{-20pt}
\includegraphics[width=0.53\textwidth]{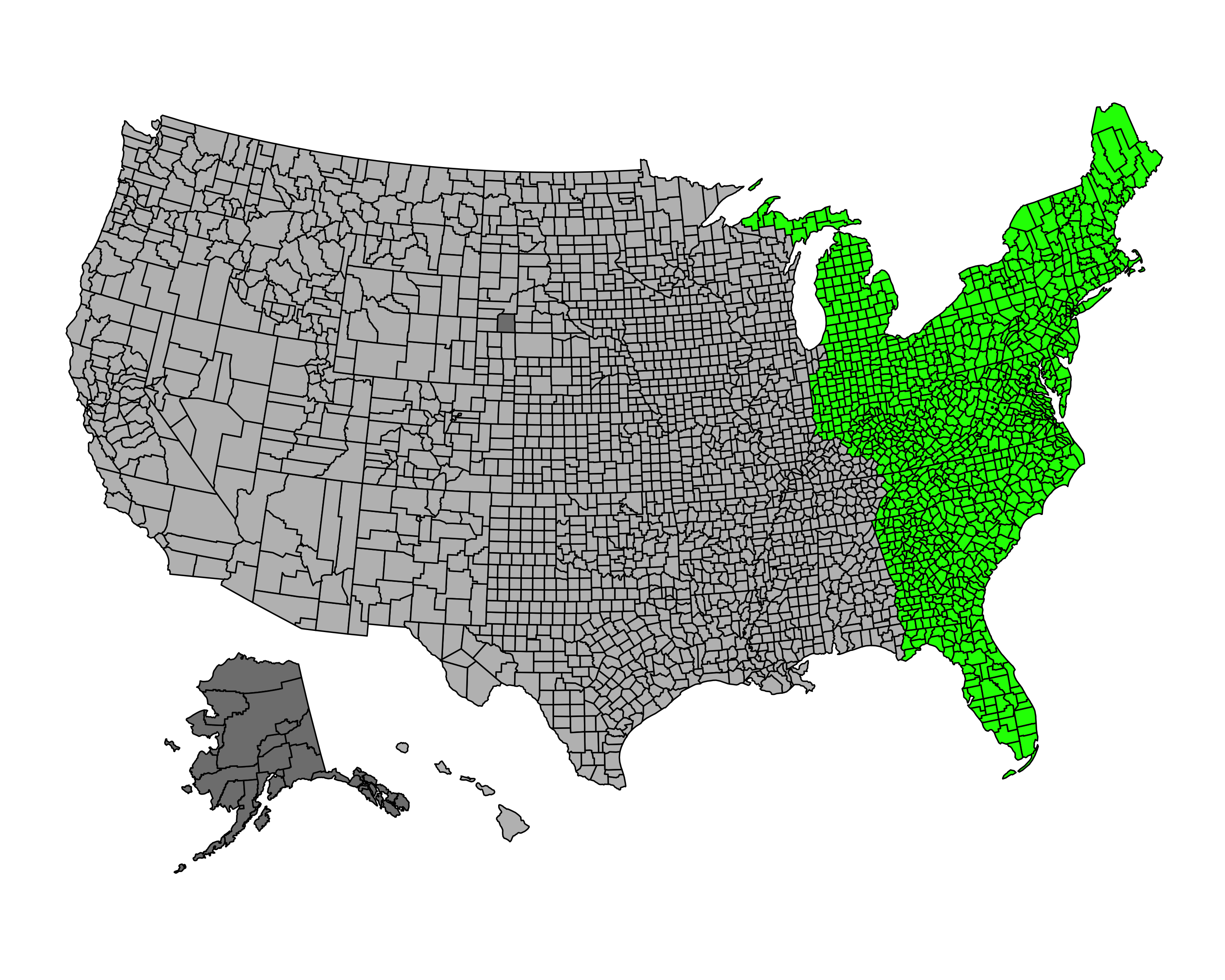}
\caption{Left map: relative change in the number of votes for the Democratic 
  presidential candidate from 2016 to 2020. Blue colors indicate an increase in  
  Democratic votes and red colors indicate a decrease, with the darkest shade of
  blue (respectively, red) corresponding to a 50\% increase (respectively,
  decrease). Right map: the counties that form the training set (in green), and
  all remaining counties are in the test set.}  
\label{fig:maps}
\end{figure}

The left map in Figure~\ref{fig:maps} shows, county by county, the relative
change in the number of votes for the Democratic Candidate between 2016 and
2020, defined as:
\[
Y = \frac{\textnormal{Dem}_{2020} - \text{Dem}_{2016}}{\textnormal{Dem}_{2016}}, 
\]
where $\textnormal{Dem}_{2020}$ is the number of Democratic votes in a given
county in 2020 (and similarly for 2016). In our experiments, the covariate
vector $X$ includes information on the makeup of the county population by
ethnicity, age, sex, median income and education (see Appendix
\ref{app:election} for details and for information about the data sources),
given the data that was available in 2020.

During real-time election forecasting, after observing the response $Y$ for a
subset of the counties (those counties that have reported), the problem is to
predict the vote change $Y$ in each of the counties where vote counts are not
yet available.  If the order in which counties report their vote totals were
drawn uniformly at random, then the exchangeability of the resulting training
and test sets would mean that conformal prediction can be applied in a
straightforward manner to obtain valid predictive intervals for the unobserved
counties. In practice, however, the time at which a county reports its votes may
depend on various factor such as the time zone of the county, the size of the
county, and so on. Therefore, if at any point in time we were to train on
counties whose votes have already been reported, then this can create a division
of training and test sets that violates exchangeability, and can thus lead to a
failure of the predictive coverage guarantee.

To mimic this type of biased split, for the current experiment, we use counties
that fall under the Eastern time zone as our training set, and the remaining
counties as the test set, as highlighted in the right-hand map in
Figure~\ref{fig:maps}. This results in 1119 training points and 1957 test 
points. 

To run the experiment, we implement the same three full conformal methods as
before (CP+LS, NexCP+LS, and NexCP+WLS). To define weights $w_i$ for
NexCP, we will use some available side information---namely,
$X^{\textnormal{prev}}\in\R^p$, which gives the 2016 measurements for the same
set of demographic and socioeconomic variables as contained in $X$ for
2020. The weights are then defined as \smash{$w_i = e^{-\gamma 
  \|X^{\textnormal{prev}}_i - X^{\textnormal{prev}}_{n+1}\|_2}$}, where we
choose $\gamma$ to satisfy 
\[
\frac{(\sum_{i=1}^n w_i + 1)^2}{\sum_{i=1}^n w_i^2 + 1} = 100,
\]
essentially corresponding to an effective training sample size of $100$ once we
use the weighted training sample. We note that, since these weights depend only
on data from 2016, we can treat these weights as fixed (i.e., these weights were
determined ``earlier'' than gathering the data set
\smash{$\{(X_i,Y_i)\}_{i=1}^{3076}$} in 2020). By using these weights within
nonexchangeable conformal, we are implicitly invoking a hypothesis that
counties which had similar demographics in 2016 will generate approximately
exchangeable data in 2020. Finally, for NexCP+WLS, we use the same choice
for the tags used for running the weighted least squares regression, i.e.,
setting $t_i = w_i$.   

In addition, we also repeat the entire experiment with quantile regression in
place of linear regression, and use a corresponding choice of the nonconformity
score function---specifically, after fitting a lower 5\% percentile function
\smash{$\hat{q}_{0.05}(\cdot)$}, and an upper 95\% percentile function
\smash{$\hat{q}_{0.95}(\cdot)$} to the data, the nonconformity score is given by
\smash{$\widehat{S}(X_i,Y_i) = \max\{\hat{q}_{0.05}(X_i) - Y_i, Y_i -
  \hat{q}_{0.95}(X_i)\}$}, as in \cite{romano2019conformalized}. This yields
three additional methods: conformal prediction with quantile regression (CP+QR),
nonexchangeable conformal with quantile regression (NexCP+QR), and
nonexchangeable conformal with weighted quantile regression (NexCP+WQR), where
the weights $w_i$ and the tags $t_i$ are defined the same way as in linear
regression.

Table~\ref{tab:election} shows the resulting predictive coverage, averaged over
the test set, for each of the three methods, when they are run with target
coverage level $1-\alpha=0.9$. We can see that CP undercovers substantially,
particularly when combined with least squares, due to the construction of
nonexchangeable training and test counties. In contrast, NexCP (with or
without the nonsymmetric algorithm) is able to achieve a coverage level that is
much closer to the target level $90\%$.  

{\begin{table}[t]\small\centering
    \begin{tabular}{l|c}
        & Coverage\\ \hline 
        CP+LS  &  0.743 \\
        NexCP+LS  &  0.820\\
        NexCP+WLS & 0.840 \\
        \end{tabular}
        \quad \quad \quad
        \begin{tabular}{l|c}
        & Coverage \\ \hline 
        CP+QR   &  0.782\\
        NexCP+QR   &  0.836\\
        NexCP+WQR & 0.835 \\
        \end{tabular}
\caption{Election data results showing coverage, averaged over all test
  counties.} 
\label{tab:election}
\end{table}}

\section{Proofs}

In this section, we give proofs of all theorems presented so far.

\subsection{Background: proof of Theorem~\ref{thm:background_fullCP}}
\label{sec:proofs-thm123}

To help build intuition for the proof techniques we will use later on, we
reformulate \cite{vovk2005algorithmic}'s proofs of these results, and we then
explain some of the challenges in extending these existing results to our new
setting. 

Let \smash{$R_i = R^{Y_{n+1}}_i$} denote the $i$th residual, at the hypothesized
value $y=Y_{n+1}$. By our assumptions, the data points
$(X_1,Y_1),\dots,(X_{n+1},Y_{n+1})$ are i.i.d. (or exchangeable), and the fitted
model \smash{$\muh = \muh^{Y_{n+1}} = \alg((X_1,Y_1),\dots,(X_{n+1},Y_{n+1}))$}
is constructed via an algorithm $\alg$ that treats these $n+1$ data points
symmetrically. The residuals \smash{$R_i = |Y_i - \muh(X_i)|$} are thus
exchangeable. 

Now define the set of ``strange'' points
\[
\Scal(R) = \left\{i\in [n+1] \ : \ R_i > \quant_{1-\alpha}\left(\sum_{j=1}^{n+1}
\frac{1}{n+1} \cdot \delta_{R_j}\right)\right\}.
\] 
That is, an index $i$ corresponds to a ``strange'' point if its residual $R_i$
is one of the \smash{$\lfloor\alpha(n+1)\rfloor$} largest elements of the list 
$R_1,\dots,R_{n+1}$. By definition, this can include at most $\alpha(n+1)$
entries of the list, i.e.,  
\[
|\Scal(R)|\leq \alpha(n+1).
\]
Next, by definition of the full conformal prediction set, we see that
\smash{$Y_{n+1}\not\in\Ch_n(X_{n+1})$} (i.e., coverage fails) if and only if
\smash{$R_{n+1} > \quant_{1-\alpha}\left(\sum_{i=1}^{n+1} \frac{1}{n+1} \cdot
  \delta_{R_i}\right)$}, or equivalently, if and only if the test point $n+1$ is
``strange'', i.e., $n+1\in\Scal(R)$. Therefore, we have
\begin{multline*}
\PP{Y_{n+1}\not\in\Ch_n(X_{n+1})} = \PP{n+1\in\Scal(R)} =
\frac{1}{n+1}\sum_{i=1}^{n+1}\PP{i\in\Scal(R)} \\
= \frac{1}{n+1}\EE{\sum_{i=1}^{n+1} \One{i\in\Scal(R)}} =
\frac{1}{n+1}\EE{|\Scal(R)|} \leq \frac{1}{n+1}\cdot \alpha(n+1) 
=\alpha, 
\end{multline*} 
where the second equality holds due to the exchangeability of
$R_1,\dots,R_{n+1}$. 

\paragraph{Challenges for the new algorithms.}

We will now see why the above proof does not
obviously extend to our nonexchangeable conformal method,
even if we were to assume that the data points $(X_i,Y_i)$ are exchangeable. First, suppose that
$\alg$ is symmetric (i.e., we do not use tags $t_i$). For the original full
conformal prediction method, in the proof of
Theorem~\ref{thm:background_fullCP}, exchangeability of the data points is used
to verify that $\PP{n+1\in\Scal(R)}=\PP{i\in\Scal(R)}$ for each $i\in[n]$, or
equivalently,   
\[
\PP{R_{n+1}>\quant_{1-\alpha}\left(\sum_{j=1}^{n+1}
\tfrac{1}{n+1}\cdot\delta_{R_j}\right)}  
=\PP{R_i>\quant_{1-\alpha}\left(\sum_{j=1}^{n+1}
\tfrac{1}{n+1}\cdot\delta_{R_j}\right)}. 
\]
This equality holds since the residuals $R_i$ are exchangeable (by assumption on
the data) and since
\smash{$\quant_{1-\alpha}(\sum_{j=1}^{n+1}\frac{1}{n+1}\cdot\delta_{R_j})$} is a
symmetric function of $R_1,\dots,R_{n+1}$. For the nonexchangeable full
conformal algorithm proposed in~\eqref{eqn:def_fullCP_symm}, on the other hand,
we would need to check whether
\[
\PP{R_{n+1}>\quant_{1-\alpha}\left(\sum_{j=1}^{n+1}
\tilde{w}_j\cdot\delta_{R_j}\right)} 
\stackrel{\textnormal{\normalsize ?}}{=}\PP{R_i>\quant_{1-\alpha}
\left(\sum_{j=1}^{n+1}\tilde{w}_j\cdot\delta_{R_j}\right)}. 
\]
Even when the residuals $R_i$ are exchangeable (i.e., when the data points are
exchangeable and the algorithm is symmetric), the weighted quantile
\smash{$\quant_{1-\alpha}(\sum_{j=1}^{n+1}\tilde{w}_j\cdot\delta_{R_j})$} is no
longer a symmetric function of $R_1,\dots,R_{n+1}$ if the weights
\smash{$\tilde{w}_j$} are not all equal, and therefore, the equality will no
longer be true in general. 

Next, if we use nonsymmetric algorithms that take tagged data points
$(X_i,Y_i,t_i)$ as input, the situation becomes even more complex---even if the
data points $(X_i,Y_i)$ are exchangeable, the residuals $R_1,\dots,R_{n+1}$ may
no longer be exchangeable as they depend on a fitted model \smash{$\muh$}
that treats the training data points nonsymmetrically. 

Finally, in this paper we are of course primarily interested in the setting
where the data points are no longer exchangeable, and in bounding the resulting
coverage gap. This leads to additional challenges, all of which we address in
the proofs below.  

\subsection{Proof of Theorem~\ref{thm:fullCP}}\label{sec:proofs_CP} 
Since nonexchangeable split conformal is simply a special case of
nonexchangeable full conformal, we only need to prove the result for full
conformal.

For each $k\in[n+1]$, denote
\[
\muh^k = \muh^{Y_{n+1},k} = \alg\Big((X_{\pi_k(1)},Y_{\pi_k(1)},t_1),\dots,
(X_{\pi_k(n+1)},Y_{\pi_k(n+1)},t_{n+1})\Big),
\]
where for any $k\in[n]$, as before $\pi_k$ denotes the permutation on $[n+1]$
that swaps indices $k$ and $n+1$, while $\pi_{n+1}$ is the identity
permutation. Then, for any $k\in[n+1]$, we can calculate
\[
\big(R(Z^k)\big)_i=|Y_{\pi_k(i)} - \muh^k(X_{\pi_k(i)})|,  
\]
and therefore,\begin{equation}\label{eqn:quantile_step_1}
\big(R(Z^K)\big)_i =
\begin{cases}
R^{Y_{n+1},K}_i, & \textnormal {if } i \not= K \textnormal{ and } i \not= n+1, \\ 
R^{Y_{n+1},K}_{n+1}, & \textnormal {if } i = K, \\
R^{Y_{n+1},K}_K, & \textnormal {if } i = n+1.
\end{cases}
\end{equation}
The definition of the nonexchangeable full conformal prediction
set~\eqref{eqn:def_fullCP_general} reveals 
\[
Y_{n+1}\not\in\Ch_n(X_{n+1}) \ \iff \ R^{Y_{n+1},K}_{n+1} >
\quant_{1-\alpha}\left(\sum_{i=1}^{n+1} \tilde{w}_i \cdot
\delta_{R^{Y_{n+1},K}_i}\right),
\]
and we can equivalently write this as
\begin{equation}\label{eqn:quantile_noncoverage_1}
Y_{n+1}\not\in\Ch_n(X_{n+1}) \ \iff \
R^{Y_{n+1},K}_{n+1} > \quant_{1-\alpha}\left(\sum_{i=1}^n \tilde{w}_i \cdot
  \delta_{R^{Y_{n+1},K}_i} + \tilde{w}_{n+1}\cdot\delta_{+\infty}\right).
\end{equation} 
Next, we verify that deterministically \eqref{eqn:quantile_step_1} implies 
\begin{equation}\label{eqn:quantile_step_2}
\quant_{1-\alpha}\left(\sum_{i=1}^n \tilde{w}_i \cdot \delta_{R^{Y_{n+1},K}_i} +
  \tilde{w}_{n+1}\cdot\delta_{+\infty}\right) \geq
\quant_{1-\alpha}\left(\sum_{i=1}^{n+1} \tilde{w}_i \cdot
  \delta_{(R(Z^K))_i}\right).
\end{equation} 
Indeed, if $K=n+1$, then \smash{$R(Z^K)=R^{Y_{n+1},K}$}
by~\eqref{eqn:quantile_step_1}, and so the bound holds trivially. If instead
$K\leq n$, then the distribution on the left-hand side
of~\eqref{eqn:quantile_step_2} equals
\begin{multline*}
\sum_{i=1}^n \tilde{w}_i \cdot \delta_{R^{Y_{n+1},K}_i} +
\tilde{w}_{n+1}\cdot\delta_{+\infty} \\
= \sum_{i=1,\dots,n; i\neq K} \tilde{w}_i \cdot \delta_{R^{Y_{n+1},K}_i} +
\tilde{w}_K (\delta_{R^{Y_{n+1},K}_K} + \delta_{+\infty}) + (\tilde{w}_{n+1} -
\tilde{w}_K)\delta_{+\infty},
\end{multline*}
while the distribution on the right-hand side of~\eqref{eqn:quantile_step_2} can 
be rewritten as 
\begin{multline*}
\sum_{i=1}^{n+1} \tilde{w}_i \cdot \delta_{(R(Z^K))_i} =
\sum_{i=1,\dots,n; i\neq K} \tilde{w}_i \cdot \delta_{R^{Y_{n+1},K}_i} +
\tilde{w}_K  \delta_{R^{Y_{n+1},K}_{n+1}} +
\tilde{w}_{n+1}\delta_{R^{Y_{n+1},K}_K} \\ 
=\sum_{i=1,\dots,n; i\neq K} \tilde{w}_i \cdot \delta_{R^{Y_{n+1},K}_i} +
\tilde{w}_K (\delta_{R^{Y_{n+1},K}_K} + \delta_{R^{Y_{n+1},K}_{n+1}}) +
(\tilde{w}_{n+1} - \tilde{w}_K)\delta_{R^{Y_{n+1},K}_K},
\end{multline*}
by applying~\eqref{eqn:quantile_step_1}. Since $w_K\in[0,1]$ by assumption, we
have \smash{$\tilde{w}_{n+1}\geq \tilde{w}_K$}, which from the last two displays 
verifies that~\eqref{eqn:quantile_step_2} must hold.
Combining~\eqref{eqn:quantile_noncoverage_1} and~\eqref{eqn:quantile_step_2}, we
have 
\[
Y_{n+1}\not\in\Ch_n(X_{n+1}) \ \implies \ 
R^{Y_{n+1},K}_{n+1} > \quant_{1-\alpha}\left(\sum_{i=1}^{n+1} \tilde{w}_i \cdot
  \delta_{(R(Z^K))_i}\right),
\] 
or equivalently by~\eqref{eqn:quantile_step_1},
\begin{equation}\label{eqn:quantile_noncoverage_2}
Y_{n+1}\not\in\Ch_n(X_{n+1}) \ \implies \ 
(R(Z^K))_K >  \quant_{1-\alpha}\left(\sum_{i=1}^{n+1}
  \tilde{w}_i \cdot \delta_{(R(Z^K))_i}\right). 
\end{equation}
Next define a function $\Scal$ from $\R^{n+1}$ to subsets of $[n+1]$, as
follows: for any $r\in\R^{n+1}$, 
\begin{equation}\label{eq:strange} 
\Scal(r) = \left\{i\in[n+1] \ : \ r_i > \quant_{1-\alpha}\left(\sum_{j=1}^{n+1}
    \tilde{w}_j \cdot\delta_{r_j}\right)\right\}. 
\end{equation}
These are the ``strange'' points---indices $i$ for which $r_i$ is unusually
large, relative to the (weighted) empirical distribution of $r_1,\dots,r_{n+1}$.
A direct argument (see, e.g., the deterministic inequality in \cite[Lemma
A.1]{harrison2012conservative}) shows that  
\begin{equation}\label{eqn:S_r_alpha}
\sum_{i\in\Scal(r)} \tilde{w}_i \leq \alpha \textnormal{ for all
  $r\in\R^{n+1}$}, 
\end{equation}
 that is, the (weighted) fraction of ``strange'' points cannot
  exceed $\alpha$. From~\eqref{eqn:quantile_noncoverage_2}, we have that
noncoverage of $Y_{n+1}$ implies strangeness of point $K$: 
\begin{equation}\label{eq:noncoverage-implies-strangeness}
Y_{n+1}\not\in\Ch_n(X_{n+1}) \ \implies \ K
\in\Scal\big(R(Z^K)\big). 
\end{equation}
Finally,
\allowdisplaybreaks
\begin{align}
\PP{K\in\Scal\big(R(Z^K)\big)} 
\notag
&=\sum_{i=1}^{n+1}\PP{K=i\textnormal{ and }
i\in\Scal\big(R(Z^i)\big)}\\ 
\label{eqn:R_i_indep_step}
&=\sum_{i=1}^{n+1}\tilde{w}_i\cdot
\PP{i\in\Scal\big(R(Z^i)\big)}\\ 
\notag
&\leq \sum_{i=1}^{n+1}\tilde{w}_i\cdot
\left(\PP{i\in\Scal\big(R(Z)\big)} +
\dtv\big(R(Z),R(Z^i)\big)\right)\\ 
\notag
&=\EE{\sum_{i\in\Scal(R(Z))} \tilde{w}_i} +
\sum_{i=1}^n\tilde{w}_i \cdot
\dtv\big(R(Z),R(Z^i)\big)\\ 
\notag
&\leq\alpha + \sum_{i=1}^n\tilde{w}_i\cdot
\dtv\big(R(Z),R(Z^i)\big), 
\end{align}
where the last step holds by~\eqref{eqn:S_r_alpha}, 
whereas step~\eqref{eqn:R_i_indep_step} holds because $K \independent Z$ and  
$Z^i=\pi_i(Z)$ is a function of the data $Z$, and therefore, $K \independent
Z^i$.  

\section{Discussion}

Our main contribution in this paper was to demonstrate how conformal
prediction, which has crucially relied on exchangeability, can be modified to
handle nonsymmetric regression algorithms, and utilize weighted residual
distributions in order to provide robustness against deviations from
exchangeability in the data. With no assumptions whatsoever on the underlying joint
distribution of the data, it is possible to give a coverage guarantee for both
existing conformal methods, and our new proposed nonexchangeable conformal
procedures. The coverage gap, expressing the extent to which the guaranteed
coverage level is lower than what would be guaranteed under exchangeability, is
bounded by a weighted sum of total variation distances between the residual
vectors obtained by swapping the $i$th point with the $(n+1)$st point.

Our work opens the door to applying conformal prediction in applications where
the data is globally likely far from exchangeable but locally deviates mildly
from exchangeability. Tags and weights can be prudently used to downweight
``far away'' points during training and calibration, and recover reasonable
coverage in practice. 
 We hope our work will lead to more targeted methods that
focus on custom design of nonsymmetric algorithms and weighting schemes to
improve efficiency and robustness in specific applications, through the lens of
nonexchangeable conformal prediction.

\section*{Acknowledgements}

The authors are grateful to the American Institute of Mathematics for supporting
and hosting our collaboration.  R.F.B.\ was supported by the National Science
Foundation via grants DMS-1654076 and DMS-2023109, and by the Office of Naval
Research via grant N00014-20-1-2337.  E.J.C.\ was supported by the Office of
Naval Research grant N00014-20-1-2157, the National Science Foundation grant
DMS-2032014, the Simons Foundation under award 814641, and the ARO grant
2003514594. R.J.T.\ was supported by ONR grant N00014-20-1-2787. The authors 
are grateful to Vladimir Vovk for helpful feedback on an earlier draft of this
paper. E.J.C.\ would like to thank John Cherian and Isaac Gibbs for their help
with the presidential election data.  

\bibliographystyle{plainnat}
\bibliography{bib}

\begin{thebibliography}{45}
\providecommand{\natexlab}[1]{#1}
\providecommand{\url}[1]{\texttt{#1}}
\expandafter\ifx\csname urlstyle\endcsname\relax
  \providecommand{\doi}[1]{doi: #1}\else
  \providecommand{\doi}{doi: \begingroup \urlstyle{rm}\Url}\fi

\bibitem[Angelopoulos and Bates(2021)]{angelopoulos2021gentle}
Anastasios~N. Angelopoulos and Stephen Bates.
\newblock A gentle introduction to conformal prediction and distribution-free
  uncertainty quantification.
\newblock \emph{arXiv preprint arXiv:2107.07511}, 2021.

\bibitem[Barber et~al.(2021)Barber, Cand{\`e}s, Ramdas, and
  Tibshirani]{barber2019predictive}
Rina~Foygel Barber, Emmanuel~J. Cand{\`e}s, Aaditya Ramdas, and Ryan~J.
  Tibshirani.
\newblock Predictive inference with the jackknife+.
\newblock \emph{Annals of Statistics}, 49\penalty0 (1):\penalty0 486--507,
  2021.

\bibitem[Bates et~al.(2021)Bates, Cand{\`e}s, Lei, Romano, and
  Sesia]{bates2021testing}
Stephen Bates, Emmanuel~J. Cand{\`e}s, Lihua Lei, Yaniv Romano, and Matteo
  Sesia.
\newblock Testing for outliers with conformal p-values.
\newblock \emph{arXiv preprint arXiv:2104.08279}, 2021.

\bibitem[Burnaev and Vovk(2014)]{burnaev2014efficiency}
Evgeny Burnaev and Vladimir Vovk.
\newblock Efficiency of conformalized ridge regression.
\newblock In \emph{Conference on Learning Theory}, pages 605--622, 2014.

\bibitem[Cand{\`e}s et~al.(2021)Cand{\`e}s, Lei, and
  Ren]{candes2021conformalized}
Emmanuel~J. Cand{\`e}s, Lihua Lei, and Zhimei Ren.
\newblock Conformalized survival analysis.
\newblock \emph{arXiv preprint arXiv:2103.09763}, 2021.

\bibitem[Cauchois et~al.(2020)Cauchois, Gupta, Ali, and
  Duchi]{cauchois2020robust}
Maxime Cauchois, Suyash Gupta, Alnur Ali, and John~C. Duchi.
\newblock Robust validation: {Confident} predictions even when distributions
  shift.
\newblock \emph{arXiv preprint arXiv:2008.04267}, 2020.

\bibitem[Cherian and Bronner(2020)]{Cherian2020How}
John Cherian and Leonard Bronner.
\newblock How {T}he {W}ashington {P}ost estimates outstanding votes for the
  2020 presidential election, 2020.
\newblock URL
  \url{https://s3.us-east-1.amazonaws.com/elex-models-prod/2020-general/write-up/election_model_writeup.pdf}.

\bibitem[Chernozhukov et~al.(2018)Chernozhukov, W{\"u}thrich, and
  Yinchu]{chernozhukov2018exact}
Victor Chernozhukov, Kaspar W{\"u}thrich, and Zhu Yinchu.
\newblock Exact and robust conformal inference methods for predictive machine
  learning with dependent data.
\newblock In \emph{Conference On learning theory}, pages 732--749. PMLR, 2018.

\bibitem[Dunn et~al.(2022)Dunn, Wasserman, and Ramdas]{dunn2018distribution}
Robin Dunn, Larry Wasserman, and Aaditya Ramdas.
\newblock Distribution-free prediction sets for two-layer hierarchical models.
\newblock \emph{Journal of the American Statistical Association (to appear)},
  2022.

\bibitem[Fannjiang et~al.(2022)Fannjiang, Bates, Angelopoulos, Listgarten, and
  Jordan]{fannjiang2022conformal}
Clara Fannjiang, Stephen Bates, Anastasios~N. Angelopoulos, Jennifer
  Listgarten, and Michael~I. Jordan.
\newblock Conformal prediction for the design problem.
\newblock \emph{arXiv preprint arXiv:2202.03613}, 2022.

\bibitem[Gibbs and Cand{\`e}s(2021)]{gibbs2021adaptive}
Isaac Gibbs and Emmanuel~J. Cand{\`e}s.
\newblock Adaptive conformal inference under distribution shift.
\newblock \emph{Advances in Neural Information Processing Systems}, 34, 2021.

\bibitem[Guan(2021)]{guan2021localized}
Leying Guan.
\newblock Localized conformal prediction: {A} generalized inference framework
  for conformal prediction.
\newblock \emph{arXiv preprint arXiv:2106.08460}, 2021.

\bibitem[Harries(1999)]{harries1999splice}
Michael Harries.
\newblock Splice-2 comparative evaluation: {Electricity} pricing.
\newblock Technical report, University of New South Wales, 1999.

\bibitem[Harrison(2012)]{harrison2012conservative}
Matthew~T. Harrison.
\newblock Conservative hypothesis tests and confidence intervals using
  importance sampling.
\newblock \emph{Biometrika}, 99\penalty0 (1):\penalty0 57--69, 2012.

\bibitem[Kivaranovic et~al.(2020)Kivaranovic, Johnson, and
  Leeb]{kivaranovic2020adaptive}
Danijel Kivaranovic, Kory~D. Johnson, and Hannes Leeb.
\newblock Adaptive, distribution-free prediction intervals for deep networks.
\newblock In \emph{International Conference on Artificial Intelligence and
  Statistics}. PMLR, 2020.

\bibitem[Landau(1953)]{landau1953dominance}
Hyman~G. Landau.
\newblock On dominance relations and the structure of animal societies: {III.
  The} condition for a score structure.
\newblock \emph{The Bulletin of Mathematical Biophysics}, 15\penalty0
  (2):\penalty0 143--148, 1953.

\bibitem[Lei(2019)]{lei2017}
Jing Lei.
\newblock Fast exact conformalization of the lasso using piecewise linear
  homotopy.
\newblock \emph{Biometrika}, 106\penalty0 (4):\penalty0 749--764, 2019.

\bibitem[Lei and Wasserman(2014)]{lei2014distribution}
Jing Lei and Larry Wasserman.
\newblock Distribution-free prediction bands for non-parametric regression.
\newblock \emph{Journal of the Royal Statistical Society: Series B},
  76\penalty0 (1):\penalty0 71--96, 2014.

\bibitem[Lei et~al.(2013)Lei, Robins, and Wasserman]{lei2013distribution}
Jing Lei, James Robins, and Larry Wasserman.
\newblock Distribution-free prediction sets.
\newblock \emph{Journal of the American Statistical Association}, 108\penalty0
  (501):\penalty0 278--287, 2013.

\bibitem[Lei et~al.(2018)Lei, G’Sell, Rinaldo, Tibshirani, and
  Wasserman]{lei2018distribution}
Jing Lei, Max G’Sell, Alessandro Rinaldo, Ryan~J. Tibshirani, and Larry
  Wasserman.
\newblock Distribution-free predictive inference for regression.
\newblock \emph{Journal of the American Statistical Association}, 113\penalty0
  (523):\penalty0 1094--1111, 2018.

\bibitem[Lei and Cand{\`e}s(2021{\natexlab{a}})]{lei2021conformal}
Lihua Lei and Emmanuel~J. Cand{\`e}s.
\newblock Conformal inference of counterfactuals and individual treatment
  effects.
\newblock \emph{Journal of the Royal Statistical Society: Series B},
  2021{\natexlab{a}}.

\bibitem[Lei and Cand{\`e}s(2021{\natexlab{b}})]{lei2021theory}
Lihua Lei and Emmanuel~J. Cand{\`e}s.
\newblock Theory of weighted conformal inference.
\newblock Technical report, Stanford University, 2021{\natexlab{b}}.

\bibitem[Leip(2020)]{Leip2020}
David Leip.
\newblock {Dave Leip's} atlas of {U.S.} presidential elections, 2020.
\newblock URL \url{http://uselectionatlas.org}.

\bibitem[Mao et~al.(2020)Mao, Martin, and Reich]{mao2020valid}
Huiying Mao, Ryan Martin, and Brian Reich.
\newblock Valid model-free spatial prediction.
\newblock \emph{arXiv preprint arXiv:2006.15640}, 2020.

\bibitem[{MIT Election Data and Science Lab}(2018)]{electionsData}
{MIT Election Data and Science Lab}.
\newblock {County Presidential Election Returns 2000-2016}, 2018.
\newblock URL \url{https://doi.org/10.7910/DVN/VOQCHQ}.

\bibitem[Podkopaev and Ramdas(2021)]{podkopaev2021distribution}
Aleksandr Podkopaev and Aaditya Ramdas.
\newblock Distribution-free uncertainty quantification for classification under
  label shift.
\newblock In \emph{Uncertainty in Artificial Intelligence}. PMLR, 2021.

\bibitem[Ramdas et~al.(2019)Ramdas, Barber, Wainwright, and
  Jordan]{ramdas2019unified}
Aaditya Ramdas, Rina~Foygel Barber, Martin~J. Wainwright, and Michael~I.
  Jordan.
\newblock A unified treatment of multiple testing with prior knowledge using
  the p-filter.
\newblock \emph{Annals of Statistics}, 47\penalty0 (5):\penalty0 2790--2821,
  2019.

\bibitem[Romano et~al.(2019)Romano, Patterson, and
  Cand{\`e}s]{romano2019conformalized}
Yaniv Romano, Evan Patterson, and Emmanuel~J. Cand{\`e}s.
\newblock Conformalized quantile regression.
\newblock \emph{Advances in Neural Information Processing Systems}, 32, 2019.

\bibitem[Ross and Pek{\"o}z(2007)]{ross2007second}
Sheldon~M. Ross and Erol~A. Pek{\"o}z.
\newblock \emph{A Second Course in Probability}.
\newblock Probability Bookstore, 2007.

\bibitem[Shafer and Vovk(2008)]{shafer2008tutorial}
Glenn Shafer and Vladimir Vovk.
\newblock A tutorial on conformal prediction.
\newblock \emph{Journal of Machine Learning Research}, 9\penalty0 (3), 2008.

\bibitem[Stankeviciute et~al.(2021)Stankeviciute, M~Alaa, and van~der
  Schaar]{stankeviciute2021conformal}
Kamile Stankeviciute, Ahmed M~Alaa, and Mihaela van~der Schaar.
\newblock Conformal time-series forecasting.
\newblock \emph{Advances in Neural Information Processing Systems}, 34, 2021.

\bibitem[Storey(2002)]{storey2002direct}
John~D Storey.
\newblock A direct approach to false discovery rates.
\newblock \emph{Journal of the Royal Statistical Society: Series B},
  64\penalty0 (3):\penalty0 479--498, 2002.

\bibitem[Storey et~al.(2004)Storey, Taylor, and Siegmund]{storey2004strong}
John~D. Storey, Jonathan~E. Taylor, and David Siegmund.
\newblock Strong control, conservative point estimation and simultaneous
  conservative consistency of false discovery rates: {A} unified approach.
\newblock \emph{Journal of the Royal Statistical Society: Series B},
  66\penalty0 (1):\penalty0 187--205, 2004.

\bibitem[Tibshirani et~al.(2019)Tibshirani, Barber, Cand{\`e}s, and
  Ramdas]{tibshirani2019conformal}
Ryan~J. Tibshirani, Rina~Foygel Barber, Emmanuel~J. Cand{\`e}s, and Aaditya
  Ramdas.
\newblock Conformal prediction under covariate shift.
\newblock \emph{Advances in Neural Information Processing Systems}, 32, 2019.

\bibitem[{United States Census Bureau}(2019{\natexlab{a}})]{censusDemos2019}
{United States Census Bureau}.
\newblock County characteristics resident population estimates,
  2019{\natexlab{a}}.
\newblock Data retrieved from
  \url{https://www.census.gov/data/tables/time-series/demo/popest/2010s-counties-detail.html}.

\bibitem[{United States Census Bureau}(2019{\natexlab{b}})]{censusEdu}
{United States Census Bureau}.
\newblock 2015-2019 american community survey 5-year average county-level
  estimates, 2019{\natexlab{b}}.
\newblock Data retrieved from
  \url{https://www.ers.usda.gov/data-products/county-level-data-sets/download-data/}.

\bibitem[{United States Census Bureau}(2019{\natexlab{c}})]{censusIncom}
{United States Census Bureau}.
\newblock Small area income and poverty estimates: 2019, 2019{\natexlab{c}}.
\newblock Data retrieved from
  \url{https://www.ers.usda.gov/data-products/county-level-data-sets/download-data/}.

\bibitem[Volkhonskiy et~al.(2017)Volkhonskiy, Burnaev, Nouretdinov, Gammerman,
  and Vovk]{volkhonskiy2017inductive}
Denis Volkhonskiy, Evgeny Burnaev, Ilia Nouretdinov, Alexander Gammerman, and
  Vladimir Vovk.
\newblock Inductive conformal martingales for change-point detection.
\newblock In \emph{Conformal and Probabilistic Prediction and Applications},
  pages 132--153. PMLR, 2017.

\bibitem[Vovk(2015)]{vovk2015cross}
Vladimir Vovk.
\newblock Cross-conformal predictors.
\newblock \emph{Annals of Mathematics and Artificial Intelligence}, 74\penalty0
  (1):\penalty0 9--28, 2015.

\bibitem[Vovk(2021)]{vovk2021testing}
Vladimir Vovk.
\newblock Testing randomness online.
\newblock \emph{Statistical Science}, 36\penalty0 (4):\penalty0 595--611, 2021.

\bibitem[Vovk et~al.(2005)Vovk, Gammerman, and Shafer]{vovk2005algorithmic}
Vladimir Vovk, Alex Gammerman, and Glenn Shafer.
\newblock \emph{Algorithmic Learning in a Random World}.
\newblock Springer Science \& Business Media, 2005.

\bibitem[Vovk et~al.(2018)Vovk, Nouretdinov, Manokhin, and
  Gammerman]{vovk2018cross}
Vladimir Vovk, Ilia Nouretdinov, Valery Manokhin, and Alexander Gammerman.
\newblock Cross-conformal predictive distributions.
\newblock In \emph{Conformal and Probabilistic Prediction and Applications},
  pages 37--51. PMLR, 2018.

\bibitem[Vovk et~al.(2021)Vovk, Petej, and Gammerman]{vovk2021protected}
Vladimir Vovk, Ivan Petej, and Alex Gammerman.
\newblock Protected probabilistic classification.
\newblock In \emph{Conformal and Probabilistic Prediction and Applications},
  pages 297--299. PMLR, 2021.

\bibitem[Xu and Xie(2021)]{xu2021conformal}
Chen Xu and Yao Xie.
\newblock Conformal prediction interval for dynamic time-series.
\newblock In \emph{International Conference on Machine Learning}. PMLR, 2021.

\bibitem[Zaffran et~al.(2022)Zaffran, F{\'e}ron, Goude, Josse, and
  Dieuleveut]{zaffran2022adaptive}
Margaux Zaffran, Olivier F{\'e}ron, Yannig Goude, Julie Josse, and Aymeric
  Dieuleveut.
\newblock Adaptive conformal predictions for time series.
\newblock In \emph{International Conference on Machine Learning}. PMLR, 2022.

\end{thebibliography}

\newpage
\appendix

\section{Extension to general nonconformity scores}
\label{app:general_scores}

In this section, we extend our new nonexchangeable inference methods for split
and full conformal to the setting of general nonconformity scores. The response
is no longer required to be real-valued, so we will consider the general setting
with data points $(X_i,Y_i)\in\Xcal\times\Ycal$.   

For split conformal, as usual, we assume that the nonconformity score function 
\smash{$ \widehat{S}:\Xcal\times\Ycal\rightarrow\R$} is pre-fitted. The
nonexchangeable split conformal set is given by  
\begin{equation}\label{eqn:def_splitCP_general_score}
\Ch_n(X_{n+1}) = \left\{y\in\Ycal : \widehat{S}(X_{n+1},y) \leq
  \quant_{1-\alpha}\left(\sum_{i=1}^n \tilde{w}_i \cdot \delta_{
   \widehat{S}(X_i,Y_i)} + \tilde{w}_{n+1} \cdot \delta_{+\infty}\right)\right\}.   
\end{equation}
For the special case \smash{$\widehat{S}(x,y) = |y - \muh(x)|$} (where
\smash{$\muh$} is a pre-fitted function), note that this reduces to the previous
definition~\eqref{eqn:def_splitCP_symm} from before.  

For full conformal, we now consider algorithms $\alg$ of the form
\begin{equation}\label{eqn:general-alg_general_score}
\alg: \ \cup_{n\geq 0} \left(\Xcal\times \Ycal\times \Tcal\right)^n \
\rightarrow \ \left\{\textnormal{measurable functions $\widehat{S}:
    \Xcal\times\Ycal\rightarrow\R$}\right\}. 
\end{equation}
(As before, the symmetric algorithm setting, with no tags $t_i$, is simply a
special case of this general formulation.) First, for any $y\in\R$ and any
$k\in[n+1]$, define 
\[
\widehat{S}^{y,k} =  \alg\left((X_{\pi_k(i)},Y^y_{\pi_k(i)},t_i) :
  i\in[n+1]\right),
\]
where the permutation $\pi_k$ is defined as before (that swaps indices $k$ and
$n+1$), and where 
\[
Y^y_i = Y_i, \  i= 1,\ldots,n, \quad Y^y_{n+1} = y,
\]
as before. Define the scores from this model,
	\[
S^{y,k}_i = 
\widehat{S}^{y,k}(X_i,Y_i), \ i=1,\dots,n, \quad S^{y,k}_{n+1} = 
 \widehat{S}^{y,k}(X_{n+1},y).
\]
Then, after drawing a random index $K$ as in~\eqref{eqn:draw_K}, the prediction
set is given by 
\begin{equation}\label{eqn:def_fullCP_general_with_general_score}
\Ch_n(X_{n+1}) = \left\{y\in\Ycal : S^{y,K}_{n+1} \leq
  \quant_{1-\alpha}\left(\sum_{i=1}^{n+1} \tilde{w}_i \cdot
    \delta_{S^{y,K}_i}\right)\right\}.
\end{equation} 
For the special case \smash{$\widehat{S}(x,y) = |y - \muh(x)|$} (where
\smash{$\muh$} is fitted on the same data), this again reduces to the previous
definition~\eqref{eqn:def_fullCP_general} from before.  

Importantly, the same theoretical result (i.e., Theorem~\ref{thm:fullCP}) holds
for these more general methods as well. The proof does not fundamentally rely on
residual scores, and the modifications required for the general case are
straightforward, so we omit the details here.

\section{Nonexchangeable jackknife+}\label{app:jack}

\subsection{Background}

The jackknife+ \citep{barber2019predictive} (closely related to
``cross-conformal prediction''~\citep{vovk2015cross}) is a method that offers a
compromise between the computational and statistical costs of the split and full conformal
methods. For each $i=1,\dots,n$, define the $i$th leave-one-out model as  
\begin{equation}\label{eq:muh-minusi}
\muh_{-i} = \alg\big((X_1,Y_1),\dots,(X_{i-1},Y_{i-1}),(X_{i+1},Y_{i+1}),
\dots,(X_n,Y_n)\big),
\end{equation}
fitted to the training data with $i$th point removed. Define also the $i$th
leave-one-out residual \smash{$R_i^{\textnormal{LOO}}=|Y_i - \muh_{-i}(X_i)|$},
which avoids overfitting since data point $(X_i,Y_i)$ is not used for training 
$\muh_{-i}$. The jackknife+ prediction interval is then given by\footnote{Abusing notation,
here $\quant_\alpha(\cdot)$ is used to denote the largest possible $\alpha$-quantile if it is not unique,
while as before, $\quant_{1-\alpha}(\cdot)$ denotes the smallest possible $(1-\alpha)$-quantile if not unique.} 
\begin{multline}\label{eqn:def_jack+}
\Bigg[ \quant_{\alpha} \left(\sum_{i=1}^n \tfrac{1}{n+1}
  \cdot\delta_{\muh_{-i}(X_{n+1}) - R^{\textnormal{LOO}}_i}
  +\tfrac{1}{n+1}\cdot\delta_{-\infty}\right), \\
\quant_{1-\alpha}\left(\sum_{i=1}^n \tfrac{1}{n+1}
  \cdot\delta_{\muh_{-i}(X_{n+1}) + R^{\textnormal{LOO}}_i}
  +\tfrac{1}{n+1}\cdot\delta_{+\infty}\right) \Bigg]. 
\end{multline}
While in practice the jackknife+ generally provides coverage close to the target
level $1-\alpha$ (and provably so under a stability assumption on $\alg$), its
theoretical guarantee only ensures $1-2\alpha$ probability of coverage in the
worst case:

\begin{theorem}[Jackknife+
  \citep{barber2019predictive}]\label{thm:background_jack+} 
If $(X_1,Y_1),\dots,(X_n,Y_n),(X_{n+1},Y_{n+1})$ are i.i.d.\ (or more generally,
exchangeable), and the algorithm $\alg$ treats the input data points
symmetrically as in~\eqref{eqn:alg_symmetric}, then the jackknife+ prediction
interval defined in~\eqref{eqn:def_jack+} satisfies 
\[
\PP{Y_{n+1}\in\Ch_n(X_{n+1})} \geq 1-2\alpha.
\]
\end{theorem}

This method can be viewed as a form of $n$-fold cross-validation; more
generally, the CV+ method \citep{barber2019predictive} uses $K$-fold
cross-validation for any desired $K$, and obtains a similar distribution-free
guarantee. 

\subsection{Methods}

We next present the nonexchangeable jackknife+ method.

\paragraph{Nonexchangeable jackknife+ with a symmetric algorithm.}
We first consider the setting where the
algorithm $\alg$ is symmetric.
To begin, we choose weights $w_i \in [0,1]$,
$i=1,\dots,n$, which are fixed ahead of time, and as before, this gives rise to
normalized weights as in~\eqref{eqn:normalized_weights}. The prediction interval
is then given by
\begin{multline}\label{eqn:def_jack+_symm}
\Bigg[ \quant_{\alpha} \left(\sum_{i=1}^n \tilde{w}_i
  \cdot\delta_{\muh_{-i}(X_{n+1}) - R^{\textnormal{LOO}}_i} +
  \tilde{w}_{n+1}\cdot\delta_{-\infty}\right), \\
\quant_{1-\alpha}\left(\sum_{i=1}^n \tilde{w}_i
  \cdot\delta_{\muh_{-i}(X_{n+1}) + R^{\textnormal{LOO}}_i} +
  \tilde{w}_{n+1}\cdot\delta_{+\infty}\right) \Bigg],
\end{multline}
where \smash{$\muh_{-i}$} is defined as in~\eqref{eq:muh-minusi},
and \smash{$R^{\textnormal{LOO}}_i = |Y_i - \muh_{-i}(X_i)|$} as before. 

\bigskip

Analogous to split and full conformal, here the original (unweighted) version of
jackknife+ is recovered by choosing weights $w_1 = \dots = w_n= 1$ in the new
algorithm.

\paragraph{Nonexchangeable jackknife+ with a nonsymmetric algorithm.}

We now extend the nonexchangeable jackknife+ to allow for a nonsymmetric
algorithm $\alg$. For any $k\in[n+1]$ and any $i\in[n]$, define the model 
\smash{$\muh^k_{-i}$} as  
\[
\muh^k_{-i} = \alg\left((X_{\pi_k(j)},Y_{\pi_k(j)},t_j) : 
    j\in[n+1],  \pi_k(j)\not\in\{i,n+1\}\right).
\] 
As before, $\pi_k$ is the permutation on $[n+1]$ that  swaps indices $k$ and
$n+1$ (or, the identity permutation in the case $k=n+1$). Equivalently, 
\[
\muh^k_{-i}
= \begin{cases}
  \alg\left((X_{j},Y_{j},t_j): j \in [n]\backslash \{i,k\}, (X_k, Y_k,
  t_{n+1})\right), 
&\textnormal{ if } k\in [n]\textnormal{ and }k\neq i, \\
  \alg\left((X_{j},Y_{j},t_j): j \in [n]\backslash \{i\}\right), 
&\textnormal{ if } k = n+1\textnormal{ or }k=i. 
\end{cases}
\]
In other words, this
model is fitted on the training data $(X_1,Y_1),\dots,(X_n,Y_n)$ but with the
$i$th point removed, and furthermore the data point $(X_k,Y_k)$ is given the tag
$t_{n+1}$ rather than $t_k$. (We note that computing the fitted model 
\smash{$\muh^k_{-i}$} does not require knowledge of the test point
$(X_{n+1},Y_{n+1})$, because $\pi_k(j)=n+1$ is excluded from the data set when
running $\alg$.) For the model \smash{$\muh^k_{-i}$}, we define its
corresponding leave-one-out residuals as  
\[
R^{k,\textnormal{LOO}}_i = |Y_i - \muh^k_{-i}(X_i)|.
\]

To run the method, we first draw a random index $K$ as in~\eqref{eqn:draw_K},
and then compute the nonexchangeable jackknife+ prediction interval as
\begin{multline}\label{eqn:def_jack+_general}
\Bigg[ \quant_{\alpha} \left(\sum_{i=1}^n \tilde{w}_i
  \cdot\delta_{\muh^K_{-i}(X_{n+1}) - R^{K,\textnormal{LOO}}_i} +
  \tilde{w}_{n+1}\cdot\delta_{-\infty}\right),
\\\quant_{1-\alpha}\left(\sum_{i=1}^n \tilde{w}_i
  \cdot\delta_{\muh^K_{-i}(X_{n+1}) + R^{K,\textnormal{LOO}}_i} +
  \tilde{w}_{n+1}\cdot\delta_{+\infty}\right) \Bigg].
\end{multline} 

\bigskip

Again, as was the case for nonexchangeable conformal, this method is a
generalization of the symmetric case for nonexchangeable jackknife+, which was
presented above.

\subsection{Theory}
As for the proof for nonexchangeable full conformal, we first need to define how we map a data sequence
$z=(z_1,\dots,z_{n+1})\in (\Xcal\times\R)^{n+1}$, with entries $z_i=(x_i,y_i)$,
to the residuals $R_{\textnormal{jack+}}(z)$. In the setting of jackknife+, however, the residuals are now a matrix rather than a vector.
Given $z$, we define ${n+1\choose 2}$ leave-two-out models: for each
$i,j\in[n+1]$ with $i\neq j$, let 
\[
\muh_{-ij} = \muh_{-ji} =
\alg\big((x_k,y_k,t_k):k\in[n+1]\backslash\{i,j\}\big).
\]
Then define the matrix of residuals 
\smash{$R_{\textnormal{jack+}}(z)\in\R^{(n+1)\times(n+1)}$} with entries 
\[
\big(R_{\textnormal{jack+}}(z)\big)_{ij} =  |y_i - \muh_{-ij}(x_i)|,
\]
for all $i\neq j$, and zeros on the diagonal.

\begin{theorem}[Nonexchangeable jackknife+]\label{thm:jack+}
Let $\alg$ be an algorithm mapping a sequence of  triplets $(X_i,Y_i,t_i)$ to a
fitted function as in~\eqref{eqn:general-alg}. Then the  nonexchangeable 
jackknife+ defined in~\eqref{eqn:def_jack+_general} satisfies
\[
\PP{Y_{n+1}\in\Ch_n(X_{n+1})}\geq 1-2\alpha - \sum_{i=1}^n\tilde{w}_i \cdot
\dtv\big(R_{\textnormal{jack+}}(Z),R_{\textnormal{jack+}}(Z^i)\big).
\] 
\end{theorem}

The proof of this result is given in~Appendix~\ref{app:proofs_jack+} below. To
summarize, we see that the coverage gap is bounded by 
\[\sum_{i=1}^n\tilde{w}_i\cdot \dtv(R_{\textnormal{jack+}}(Z),R_{\textnormal{jack+}}(Z^i)).\]
As for the full conformal guarantee, this therefore implies
the coverage gap is bounded by $\sum_i \tilde{w}_i\cdot\dtv(Z,Z^i),
$
as well. Again, while this last bound can be viewed
as more interpretable, in many settings it is substantially more loose.

While jackknife+ is defined specifically for the residual-based nonconformity
score (i.e., the score $|y - \muh(x)|$ to measure the extent to which a data
point $(x,y)$ does not conform to observed trends in the data), in other
settings we may wish to use alternative nonconformity scores. Jackknife+ is
closely related to earlier work on the cross-conformal method
\citep{vovk2015cross,vovk2018cross}.
Unlike jackknife+, the cross-conformal method can be applied to arbitrary
nonconformity scores. In Appendix~\ref{app:crossconformal}, we will present 
a nonexchangeable version of the cross-conformal algorithm.

\section{Huber-robustness of conformal prediction}
\label{app:multiplicative_bound}

In this section, we consider an alternative form of robustness, which requires 
stricter assumptions on the distribution drift but will yield a stronger
predictive coverage guarantee. First, consider a version of the classic Huber
contamination model from robust statistics, where most of the data is i.i.d.\
from the target distribution \smash{$\mathcal{D}_{\textnormal{target}}$}, but
some fraction $\epsilon$ of the data is arbitrarily corrupted. For simplicity,
to start we consider observing training data point $Z_i = (X_i,Y_i)$ from the
mixture model 
\begin{equation}\label{eq:mixture-model}
\mathcal{D}_i = (1-\epsilon) \mathcal{D}_{\textnormal{target}} + \epsilon
\mathcal{D}'_i.
\end{equation}
Here $\mathcal{D}'_i$ denotes an arbitrary adversarial distribution, that 
could potentially corrupt the $i$th training data point. However, we want to
ensure coverage with respect to the target distribution
\smash{$\mathcal{D}_{\textnormal{target}}$}---that is, the test point
$Z_{n+1}=(X_{n+1},Y_{n+1})$ will be drawn from
\smash{$\mathcal{D}_{\textnormal{target}}$}. Standard conformal prediction
assumes $\epsilon=0$. But, one may ask: how badly can such adversarial
corruptions hurt coverage? Here, we will answer that question, but do so in a
slightly more general manner. First, define a new measure of distance between
distributions,  
\begin{equation}
\label{eq:dmix}
\dmix(\mathcal{D},\mathcal{D}') = \inf\left\{ t \geq 0 : \mathcal{D} =
  (1-t)\cdot \mathcal{D}' + t\cdot \mathcal{D}'' \textnormal{ for some
    distribution $\mathcal{D}''$}\right\}. 
\end{equation}
Abusing notation, we will write \smash{$\dmix(Z,Z') =
  \dmix(\mathcal{D},\mathcal{D}')$} if $Z\sim\mathcal{D}$ and
$Z'\sim\mathcal{D}'$.  

This ``distance'' can be thought of as measuring the contamination of
$\mathcal{D}'$, in the Huber sense. Indeed, if the data did indeed come from the
mixture model in~\eqref{eq:mixture-model},  then we would have
\smash{$\dmix(Z_i,Z_{n+1})\leq \epsilon$}. (We note that \smash{$\dmix$} is not
a metric, and in particular, is not symmetric in its two arguments.) 

We now state our theory for our weighted version of split conformal, full
conformal, and jackknife+, in a more restricted setting where the data points
are independent and the algorithm is symmetric.   From this point on 
we assume $w_1 + \dots + w_n>0$ to avoid a trivial setting. Define 
\[
\bar{w}_i = \frac{w_i}{w_1+\dots+w_n}, \ i=1,\dots,n.
\]

\begin{theorem}[Multiplicative bounds]\label{thm:mult}
Suppose that $Z_1,\dots,Z_{n+1}$ are independent. 
For any symmetric algorithm $\alg$, the nonexchangeable full conformal  
method~\eqref{eqn:def_fullCP_symm} satisfies 
\[
\PP{Y_{n+1}\notin\Ch_n(X_{n+1})} \leq \frac{\alpha}{1 - \sum_{i=1}^n \bar{w}_i 
  \cdot \dmix(Z_i,Z_{n+1})}
\]
(which includes the nonexchangeable split conformal
method~\eqref{eqn:def_splitCP_symm} as a special case), and the nonexchangeable
jackknife+ method~\eqref{eqn:def_jack+_symm} satisfies   
\[
\PP{Y_{n+1}\notin\Ch_n(X_{n+1})}\leq \frac{2\alpha}{1 - \sum_{i=1}^n \bar{w}_i
  \cdot \dmix(Z_i,Z_{n+1})}.
\]
\end{theorem}

In particular, if each $Z_i$ follows an $\eps$-Huber contamination model
relative to $Z_{n+1}$ as in~\eqref{eq:mixture-model}, then the bound on the
noncoverage rate for both unweighted and weighted conformal methods inflates by
a factor of at most $1/(1-\epsilon)$, i.e., for split or full
  conformal prediction we get a noncoverage guarantee of $\alpha/(1-\epsilon)$
instead of the nominal level $\alpha$.  We note that the coverage gap here is
multiplicative---that is, $\alpha/(1-\epsilon)\approx \alpha + \alpha\epsilon$,
and so the coverage gap is proportional to $\alpha$.  If the target error level
$\alpha$ is small, then this multiplicative bound can offer much tighter error
control, as compared to the earlier additive bounds in Theorem~\ref{thm:fullCP},
if the terms \smash{$\dmix(Z_i,Z_{n+1})$} are small.

On the other hand, notice that in general we have \smash{$\dmix(Z_i,Z_{n+1})
\geq \dtv(Z_i,Z_{n+1})$}, and furthermore, it is possible to have
\smash{$\dmix(Z_i,Z_{n+1})=1$} even when \smash{$\dtv(Z_i,Z_{n+1})$} is
arbitrarily small. In a such setting the original additive bounds may give
tighter results. Of course, an additional restriction is that the multiplicative
bounds require independent data and symmetric algorithms, whereas the earlier
theorems make no such assumptions.

\section{Nonexchangeable cross-conformal}\label{app:crossconformal}

While jackknife+ is defined specifically for the residual-based nonconformity
score (i.e., the score $|y - \muh(x)|$ to measure the extent to which a data
point $(x,y)$ does not conform to observed trends in the data), in other
settings we may wish to use alternative nonconformity scores. Jackknife+ is
closely related to earlier work on the cross-conformal method
\citep{vovk2015cross,vovk2018cross}.
Unlike jackknife+, the cross-conformal method can be applied to arbitrary
nonconformity scores.

In this section, we present  a nonexchangeable version of the $n$-fold
cross-conformal algorithm \citep{vovk2015cross}, which can be implemented with
an arbitrary nonconformity score. In the case of the regression score
\smash{$\widehat{S}(x,y) = |y-\muh(x)|$}, the jackknife+ prediction interval always
contains the $n$-fold cross-conformal prediction set. (See
\cite{barber2019predictive} for a more detailed comparison of these methods in
the exchangeable setting.)  

As for the extension of nonexchangeable full conformal prediction
to the setting of general nonconformity scores (in
Appendix~\ref{app:general_scores}), the algorithm $\alg$ is now a function
mapping tagged data sets to scoring functions, as
in~\eqref{eqn:general-alg_general_score}. For any $k\in[n+1]$ and any $i\in[n]$,
define the $i$th leave-one-out scoring function \smash{$\widehat{S}^k_{-i}$} as  
\[
\widehat{S}^k_{-i} =\alg\left((X_{\pi_k(i)},Y^y_{\pi_k(i)},t_i) : 
i\in[n+1], \pi_k(j)\not\in\{i,n+1\}\right),
\]
or equivalently, 
\[
\widehat{S}^k_{-i}
= \begin{cases}
  \alg\Big((X_{j},Y_{j},t_j): j \in [n]\backslash \{i,k\}, (X_k, Y_k,
  t_{n+1})\Big), &\textnormal{ if } k\in [n]\textnormal{ and }k\neq i, \\ 
  \alg\left((X_{j},Y_{j},t_j): j \in [n]\backslash \{i\}\right), 
  &\textnormal{ if } k = n+1\textnormal{ or }k=i. 
\end{cases}\]
As before, $\pi_k$ is the permutation on $[n+1]$ that swaps indices $k$ and
$n+1$ (or, the identity permutation in the case $k=n+1$). In other words, this
scoring function is fitted on the training data $(X_1,Y_1),\dots,(X_n,Y_n)$ but
with the $i$th point removed, and furthermore the data point $(X_k,Y_k)$ is
given the tag $t_{n+1}$ rather than $t_k$. We then define the corresponding
leave-one-out scores as 
\[
S^{k,\textnormal{LOO}}_i = \widehat{S}^k_{-i}(X_i,Y_i).
\]

Finally, to define the prediction set, we first draw a random index $K$ as
in~\eqref{eqn:draw_K}, and then compute the nonexchangeable cross-conformal
prediction set as 
\[
\Ch_n(X_{n+1})
= \left\{y\in\Ycal : S^{K,\textnormal{LOO}}_{n+1} \leq
  \quant_{1-\alpha}\left(\sum_{i=1}^{n+1} \tilde{w}_i \cdot
    \delta_{S^{K,\textnormal{LOO}}_i}\right)\right\}.
\]
As for the exchangeable setting, in the special case of the standard
nonconformity score \smash{$S(x,y) = |y-\muh(x)|$}, the nonexchangeable
$n$-fold cross-conformal prediction set defined here is always contained inside
the nonexchangeable jackknife+ prediction interval defined
in~\eqref{eqn:def_jack+_general}. 

Importantly, the same guarantee that holds for jackknife+, i.e., the result of
Theorem~\ref{thm:jack+}, also holds for the $n$-fold cross-conformal method run
with an arbitrary nonconformity score. The proof of this coverage guarantee is
essentially the same as for jackknife+ and so we omit it here for brevity. (For
the exchangeable case, the connection between the proofs for these two different
methods is explained in detail in \cite{barber2019predictive}, and extends in a
straightforward way to the nonexchangeable case considered here.)

\section{Additional proofs and calculations}

\subsection{Proof of Lemma~\ref{lem:dtv_swap}}

First, by the maximal coupling theorem (e.g., \cite[Proposition
2.7]{ross2007second}), there exists a distribution $\mathcal{D}$ on
a pair of random variables
$(Z_i',Z_{n+1}')$ such that, marginally, \smash{$Z_i'\eqd Z_i$} and
\smash{$Z_{n+1}'\eqd Z_{n+1}$}, and such that 
\[ 
\PP{Z_i' = Z_{n+1}'} = 1 -\dtv(Z_i,Z_{n+1}).
\]
Now let $Z=(Z_1,\dots,Z_{n+1})$, with $Z_j$ drawn independently for each $j \in
[n+1]$, then draw \smash{$(Z_i',Z_{n+1}'),(Z_i'',Z_{n+1}'') \iidsim
  \mathcal{D}$}, independently from $Z$. Define 
\[
Z' = (Z_1,\dots,Z_{i-1},Z_i',Z_{i+1},\dots,Z_n,Z_{n+1}''),
\]
and
\[
Z'' = (Z_1,\dots,Z_{i-1},Z_i'',Z_{i+1},\dots,Z_n,Z_{n+1}').
\]
Then clearly, \smash{$Z'\eqd Z'' \eqd Z$}. In particular, recalling the swapped
indices notation~\eqref{eqn:z-swap-indices}, this implies that \smash{$(Z'')^i
  \eqd Z^i$}, and so
\[
\dtv(Z,Z^i) = \dtv(Z', (Z'')^i) .
\]
Again applying the maximal coupling theorem, we have
\begin{align*}
\dtv(Z', (Z'')^i) &\leq 1-  \PP{Z' = (Z'')^i} \\ 
&= 1 - \PP{Z_i'=Z_{n+1}', Z_i'' = Z_{n+1}''} \\
&= 1 - \PP{Z_i'=Z_{n+1}'}\cdot \PP{Z_i'' = Z_{n+1}''} \\ 
&= 1 - \left( 1 -\dtv(Z_i,Z_{n+1})\right)^2 \\ 
&= 2\dtv(Z_i,Z_{n+1}) - \dtv(Z_i,Z_{n+1})^2,
\end{align*}
completing the proof.

\subsection{Proof of Theorem~\ref{thm:overcover}}

Since nonexchangeable split conformal is simply a special case of
nonexchangeable full conformal, we only need to prove the result for full
conformal. We recall from the proof of Theorem~\ref{thm:fullCP}, found in
Section~\ref{sec:proofs_CP}, that for nonexchangeable full conformal, 
the coverage event can be characterized as
\[
Y_{n+1}\in\Ch_n(X_{n+1}) \ \iff \ R^{Y_{n+1},K}_{n+1} \leq
\quant_{1-\alpha}\left(\sum_{i=1}^{n+1} \tilde{w}_i \cdot
  \delta_{R^{Y_{n+1},K}_i}\right),
\]
or equivalently,
\[
Y_{n+1}\in\Ch_n(X_{n+1}) \ \iff \ R(Z^K)_K\leq
\quant_{1-\alpha}\left(\sum_{i=1}^{n+1} \tilde{w}_{\pi_K(i)} \cdot
  \delta_{R(Z^K)_i}\right).
\]
Therefore,
\begin{align*}
&\PP{Y_{n+1}\in\Ch_n(X_{n+1}) }\\
&=\PP{R(Z^K)_K\leq \quant_{1-\alpha}\left(\sum_{i=1}^{n+1}
\tilde{w}_{\pi_K(i)} \cdot \delta_{R(Z^K)_i}\right)} \\
&=\sum_{k=1}^{n+1}\PP{K=k\textnormal{ and }R(Z^k)_k\leq
\quant_{1-\alpha}\left(\sum_{i=1}^{n+1} \tilde{w}_{\pi_k(i)} \cdot
\delta_{R(Z^k)_i}\right)} \\
&=\sum_{k=1}^{n+1}\tilde{w}_k\cdot \PP{R(Z^k)_k\leq 
\quant_{1-\alpha}\left(\sum_{i=1}^{n+1} \tilde{w}_{\pi_k(i)} \cdot 
\delta_{R(Z^k)_i}\right)} \\
&\leq \sum_{k=1}^{n+1}\tilde{w}_k\cdot \PP{R(Z)_k\leq
\quant_{1-\alpha}\left(\sum_{i=1}^{n+1} \tilde{w}_{\pi_k(i)} \cdot 
\delta_{R(Z)_i}\right)} \\
&\hspace{150pt} +\sum_{k=1}^{n+1}\tilde{w}_k\cdot 
\dtv(R(Z),R(Z^k)) \\
&=\EE{ \sum_{k=1}^{n+1}\tilde{w}_k\cdot \One{R(Z)_k\leq
 \quant_{1-\alpha}\left(\sum_{i=1}^{n+1} \tilde{w}_{\pi_k(i)} \cdot 
\delta_{R(Z)_i}\right)}} \\
&\hspace{150pt} +\sum_{k=1}^{n+1}\tilde{w}_k\cdot
\dtv(R(Z),R(Z^k)).
\end{align*}
Here, as in the proof of  Theorem~\ref{thm:fullCP}, the third equality holds
as $K$ is drawn independently from $Z$. Below, we will show that, for any 
\emph{distinct} and fixed $r_1,\dots,r_{n+1}\in\R$, it holds that  
\begin{equation}\label{eqn:overcover_to_show}
\sum_{k=1}^{n+1}\tilde{w}_k\cdot \One{r_k\leq
\quant_{1-\alpha}\left(\sum_{i=1}^{n+1} \tilde{w}_{\pi_k(i)} \cdot
\delta_{r_i}\right)}<1-\alpha+\tilde{w}_{n+1}.
\end{equation} 
Applying this inequality with \smash{$r_i = R(Z)_i$} (and
recalling, by assumption in the theorem, the values 
\smash{$R(Z)_1,\dots,R(Z)_{n+1}$}
are distinct with probability 1), we obtain 
\[
\PP{Y_{n+1}\in\Ch_n(X_{n+1})} \leq 1-\alpha+\tilde{w}_{n+1} + 
\sum_{k=1}^{n+1}\tilde{w}_k \cdot
\dtv(R(Z) ,R(Z^k)),
\]
which would complete the proof of the theorem.

Now we need to verify~\eqref{eqn:overcover_to_show}. Define
\[
\mathcal{K} = \left\{ k\in[n+1] : r_k\leq 
\quant_{1-\alpha}\left(\sum_{i=1}^{n+1} \tilde{w}_{\pi_k(i)} \cdot 
\delta_{r_i}\right) \right\},
\] 
so that proving~\eqref{eqn:overcover_to_show} is equivalent to proving that
\smash{$\sum_{k\in\mathcal{K}}\tilde{w}_k< 1-\alpha + \tilde{w}_{n+1}$}.  
Let 
\[
k_* = \arg\max_k \left\{ r_k : k\in\mathcal{K} \right\},
\]
indexing the largest value $r_k$ over indices $k\in\mathcal{K}$. Since
$k_*\in\mathcal{K}$ by definition, 
\[
r_{k_*}\leq \quant_{1-\alpha}\left(\sum_{i=1}^{n+1} \tilde{w}_{\pi_{k_*}(i)}
  \cdot \delta_{r_i}\right) \ \implies \ \sum_{i=1}^{n+1}
\tilde{w}_{\pi_{k_*}(i)} \cdot \One{r_i < r_{k_*}} < 1-\alpha.
\]
As $k_*$ is defined to attain the maximum, we also have \smash{$\mathcal{K}
  \subseteq \{k\in[n+1] : r_k \leq r_{k_*}\}$}. Therefore, 
\begin{align*}
\sum_{k\in\mathcal{K}}\tilde{w}_k
& \leq\sum_{k=1}^{n+1}\tilde{w}_k\cdot\One{r_k \leq r_{k_*}} \\
& =\tilde{w}_{k_*} + \sum_{k=1}^{n+1}\tilde{w}_k\cdot\One{r_k < r_{k_*}} \\
& =\tilde{w}_{k_*} + \sum_{k=1}^{n+1} 
\left(\tilde{w}_k - \tilde{w}_{\pi_{k_*}(k)}\right) \cdot \One{r_k < r_{k_*}}  
+ \sum_{k=1}^{n+1}\tilde{w}_{\pi_{k_*}(k)}\cdot \One{r_k < r_{k_*}} \\ 
& <\tilde{w}_{k_*} + \sum_{k=1}^{n+1}
\left(\tilde{w}_k - \tilde{w}_{\pi_{k_*}(k)}\right) \cdot 
\One{r_k < r_{k_*}}+ (1-\alpha). 
\end{align*}
The second line holds because $r_1,\ldots,r_{n+1}$ are distinct, and the last
line holds by the calculations above. 

Finally, consider the term \smash{$(\tilde{w}_k - \tilde{w}_{\pi_{k_*}(k)})
  \cdot \One{r_k < r_{k_*}}$} in the remaining sum. If $k=k_*$, then 
\smash{$\One{r_k < r_{k_*}} = 0$}. If $k = n+1$, then 
\[
\left(\tilde{w}_k - \tilde{w}_{\pi_{k_*}(k)}\right)\cdot\One{r_k < r_{k_*}}
= \left(\tilde{w}_{n+1} - \tilde{w}_{k_*}\right)\cdot\One{r_k < r_{k_*}} 
\leq \tilde{w}_{n+1} - \tilde{w}_{k_*}.
\]
If $k\not\in\{k_*,n+1\}$, then \smash{$\pi_{k_*}(k)=k$} and so the term is again
zero. Therefore, we have 
\[
\sum_{k=1}^{n+1}\left(\tilde{w}_k - \tilde{w}_{\pi_{k_*}(k)}\right) \cdot
\One{r_k < r_{k_*}} \leq \tilde{w}_{n+1}-\tilde{w}_{k_*},
\]
and combining this with the work above, we have shown that 
\[
\sum_{k\in\mathcal{K}}\tilde{w}_k < 1-\alpha+ \tilde{w}_{n+1}.
\]
This verifies~\eqref{eqn:overcover_to_show}, and therefore we have proved the
theorem.   

\subsection{Calculation for~\eqref{eqn:example_covariate_timeseries}} 
\label{app:calculations_eqn:example_covariate_timeseries}

Define \smash{$R = R(Z) = \mathcal{P}^\perp_X(\eps)$},
\smash{$U = R/ \|R\|_2$} and \smash{$L = \|R\|_2$.}  Then $R=U\cdot L$. Let
\smash{$R^i = R(Z^i)$}, and write \smash{$U^i =
R^i/\|R^i\|_2$}. Note that $R^i$ and $U^i$ are obtained from $R$ and $U$,
respectively, by swapping the $i$th and $(n+1)$st entries, and note also that
$\|R\|_2 = \|R^i\|_2$ and so $R^i = U^i\cdot L$.  We then have
\[
\dtv(R,R^i) = \dtv(U\cdot L, U^i\cdot L)
\leq \dtv((U,L), (U^i,L)) = \dtv(U,U^i),
\]
where the last step holds because $U\independent L$, and consequently,
$U^i\independent L$ also. (To see why $U\independent L$ holds, we note that
$R\mid X \sim \mathcal{N}(0, \sigma^2\mathcal{P}^\perp_X)$, and thus
$U\independent L\mid X$ by properties of the normal distribution; moreover,
$L\mid X \sim \sigma\cdot\chi_{n+1-p}$, meaning that $L\independent X$.)  On the
other hand, we have
\[
\dtv(U,U^i) = \dtv(R/\|R\|_2,R^i/\|R^i\|_2)\leq \dtv(R,R^i),
\]
and so we see that $\dtv(R,R^i) = \dtv(U,U^i)$. From this point on, we only need
to bound \smash{$\dtv(U,U^i)$}. 

Conditional on the subspace \smash{$\textnormal{span}(X)^\perp$}, the unit
vector $U$ is drawn uniformly from this subspace intersected with the unit
sphere, and therefore the joint density of $(X,U)$ is given by
\[
f_{(X,U)}(x,u) \propto \frac{1}{(2\pi)^{(n+1)p/2}|\Sigma|^{1/2}}
e^{-\textnormal{vec}(x)^\top\Sigma^{-1}\textnormal{vec}(x)/2}
\]
with respect to Lebesgue measure on the manifold 
\smash{$\left\{(x,u)\in\R^{(n+1)\times p}\times \mathbb{S}^n:
x\perp u\right\}$}. 
Therefore the marginal density of $u$ is given by
\[
g_U(u)\propto \int_{x\in\mathbb{R}^{(n+1)\times p}; x\perp u}
\frac{1}{(2\pi)^{(n+1)p/2}|\Sigma|^{1/2}}
e^{-\textnormal{vec}(x)^\top\Sigma^{-1}\textnormal{vec}(x)/2}\;\mathsf{d}x,
\]
where the integral is taken over the $np$-dimensional subspace of matrices $x$ 
where all columns are orthogonal to $u$. Equivalently we can take $x = W_uy$
where $W_u\in\R^{(n+1)p\times np}$ is an orthonormal basis for the subspace
orthogonal to $u$, and so 
\begin{multline*}
g_U(u)\propto \int_{y\in\mathbb{R}^{n\times p}}\frac{1}{(2\pi)^{(n+1)p/2} 
|\Sigma|^{1/2}}e^{-(W_u\textnormal{vec}(y))^\top
\Sigma^{-1}(W_u\textnormal{vec}(y))/2} \;\mathsf{d}y \\
=\frac{(2\pi)^{np/2}|(W_u^\top \Sigma^{-1}W_u)^{-1}|^{1/2}}
{(2\pi)^{(n+1)p/2}|\Sigma|^{1/2}}  
\propto |(W_u^\top \Sigma^{-1}W_u)^{-1}|^{1/2} =  
|W_u^\top \Sigma^{-1}W_u|^{-1/2}. 
\end{multline*}
Since $[W_u \ | \ u\otimes \mathbf{I}_p]$ is an orthogonal matrix, we can verify
through matrix identities that 
\[ 
|W_u^\top \Sigma^{-1}W_u| =
|(u\otimes \mathbf{I}_p)^\top\Sigma (u\otimes \mathbf{I}_p)| 
\cdot |\Sigma|^{-1},
\]
and therefore,
\[
g_U(u) = g(u) \propto \frac{1}{\sqrt{|(u\otimes \mathbf{I}_p)^\top 
\Sigma (u\otimes \mathbf{I}_p)|}}.
\]
We can also calculate the marginal density of $U^i$,
\[g_{U^i}(u) = g(u^i) \propto \frac{1}{\sqrt{|(u^i\otimes \mathbf{I}_p)^\top 
\Sigma (u^i\otimes \mathbf{I}_p)|}},
\]
and note that these two densities have the same normalizing constant, so we have 
\[
\frac{g_{U^i}(u)}{g_U(u)} = \frac{g(u^i)}{g(u)} 
= \sqrt{\frac{|(u\otimes \mathbf{I}_p)^\top \Sigma (u\otimes \mathbf{I}_p)|}
{|(u^i\otimes \mathbf{I}_p)^\top \Sigma (u^i\otimes \mathbf{I}_p)|}}.
\]
Next, the multiplicative property of the determinant yields
\begin{multline*}
|(u^i\otimes \mathbf{I}_p)^\top \Sigma (u^i\otimes \mathbf{I}_p)| 
=|(u\otimes \mathbf{I}_p)^\top \Sigma (u\otimes \mathbf{I}_p)| \\ 
{}\cdot \left|\left((u\otimes \mathbf{I}_p)^\top \Sigma (u\otimes
 \mathbf{I}_p)\right)^{-1/2}\cdot (u^i\otimes \mathbf{I}_p)^\top \Sigma
 (u^i\otimes \mathbf{I}_p)\cdot \left((u\otimes \mathbf{I}_p)^\top \Sigma
 (u\otimes \mathbf{I}_p)\right)^{-1/2}\right|,
\end{multline*} 
and so
\begin{multline*}
\frac{g_{U^i}(u)}{g_U(u)} = \Bigg|\left((u\otimes \mathbf{I}_p)^\top 
\Sigma (u\otimes \mathbf{I}_p)\right)^{-1/2}\cdot
(u^i\otimes \mathbf{I}_p)^\top \Sigma (u^i\otimes \mathbf{I}_p) \\
{}\cdot \left((u\otimes \mathbf{I}_p)^\top 
\Sigma (u\otimes \mathbf{I}_p)\right)^{-1/2}\Bigg|^{-1/2}.
\end{multline*}
Next, for any positive definite matrices $A,B\in\R^{p\times p}$, we calculate 
\begin{multline*}
|A^{-1/2}\cdot B \cdot A^{-1/2}|
\leq \|A^{-1/2}\cdot B \cdot A^{-1/2}\|^p
= \|\mathbf{I}_p + A^{-1/2}\cdot (B-A) \cdot A^{-1/2}\|^p\\
\leq \left(1 +  \|A^{-1/2}\cdot (B-A) \cdot A^{-1/2}\|\right)^p
\leq \left(1 +  \|A^{-1}\| \cdot\|B-A\|\right)^p,
\end{multline*}
and so applying this with $A=(u\otimes \mathbf{I}_p)^\top \Sigma (u\otimes
\mathbf{I}_p)$ and $B=(u^i\otimes \mathbf{I}_p)^\top \Sigma (u^i\otimes
\mathbf{I}_p)$, we have
\[
\frac{g_{U^i}(u)}{g_U(u)} \geq \left(1 +  \|\Sigma^{-1}\|\cdot
\left\|(u^i\otimes \mathbf{I}_p)^\top \Sigma (u^i\otimes \mathbf{I}_p) 
- (u\otimes \mathbf{I}_p)^\top \Sigma (u\otimes \mathbf{I}_p)\right\|
\right)^{-p/2}.
\]

Now we calculate the remaining matrix norm. Fix any unit vector $v\in\R^p$. We
have 
\begin{align*}
&\left|v^\top\left((u^i\otimes \mathbf{I}_p)^\top \Sigma 
(u^i\otimes\mathbf{I}_p) - (u\otimes \mathbf{I}_p)^\top 
\Sigma (u\otimes \mathbf{I}_p)\right)v\right|\\
&=\left|(u^i\otimes v)^\top\Sigma 
(u^i\otimes v) - (u\otimes v)^\top \Sigma (u\otimes v)\right|\\
&=\left|(u^i\otimes v)^\top\Sigma 
((u^i-u)\otimes v) + ((u^i-u)\otimes v)^\top \Sigma (u\otimes v)\right|\\
&\leq \|\Sigma\|\cdot \left(\|u^i\otimes v\|_2\cdot 
\|(u^i-u)\otimes v\|_2 + \|(u^i-u)\otimes v\|_2\cdot
\|u\otimes v\|_2\right)\\
&= \|\Sigma\|\cdot \left(\|u^i\|_2\cdot \|v\|_2\cdot \|u^i-u\|_2
\cdot \|v\|_2 + \|u^i-u\|_2\cdot\|v\|_2\cdot\|u\|_2\cdot\|v\|_2\right)\\
&= 2\|\Sigma\|\|u^i-u\|_2\\
&= \sqrt{8}\|\Sigma\||u_i-u_{n+1}|.
\end{align*}
Combining everything so far, then, we have
\[
\frac{g_{U^i}(u)}{g_U(u)} 
\geq \left(1 +  \sqrt{8}\|\Sigma\| \|\Sigma^{-1}\||u_i-u_{n+1}|\right)^{-p/2}
= \left(1 +  \sqrt{8}\kappa_\Sigma|u_i-u_{n+1}|\right)^{-p/2}.
\]
In particular, this implies that
\[
1 - \frac{g_{U^i}(u)}{g_U(u)} 
\leq 1 - \left(1 +  \sqrt{8}\kappa_\Sigma|u_i-u_{n+1}|\right)^{-p/2} 
\leq p\sqrt{2}\kappa_\Sigma\cdot |u_i-u_{n+1}|.
\]
Next we have
\begin{align*}
\dtv(U,U^i) 
&=\int_{u\in\mathbb{S}^n} \left(g_U(u) - g_{U^i}(u)\right)_+\;\mathsf{d}u\\
&=\int_{u\in\mathbb{S}^n}g_U(u) 
\left(1 - \frac{g_{U^i}(u)}{g_U(u)}\right)_+\;\mathsf{d}u\\
&\leq \int_{u\in\mathbb{S}^n}g_U(u) \cdot  p\sqrt{2}\kappa_\Sigma\cdot 
|u_i-u_{n+1}|\;\mathsf{d}u\\
&=\EE{p\sqrt{2}\kappa_\Sigma\cdot |U_i-U_{n+1}|}\\
&\leq p\sqrt{2}\kappa_\Sigma\cdot\left(\EE{ |U_i|} + \EE{|U_{n+1}|}\right).
\end{align*}
Now we need to bound $\EE{|U_i|}$. Recall that $R=U\cdot L$, with $U\independent
L \mid X$. We can therefore calculate 
\[ 
\EEst{R_i^2}{X}
=\EEst{U_i^2\cdot L^2}{X}
=\EEst{U_i^2}{X} \cdot \EEst{L^2}{X}
= \EEst{U_i^2}{X} \cdot \sigma^2(n+1-p),
\]
since $L\mid X \sim \sigma\cdot\chi_{n+1-p}$. Therefore,
\[
\EEst{U_i^2}{X} = \frac{\EEst{R_i^2}{X}}{\sigma^2(n+1-p)} = 
\frac{ \sigma^2 (\mathcal{P}^\perp_X)_{ii} }{\sigma^2(n+1-p)} 
\leq \frac{1}{n+1-p},
\]
where the last equality holds because $
R\mid X \sim \mathcal{N}(0,\sigma^2 \mathcal{P}^\perp_X)$.
Therefore,
\[
\EE{|U_i|} \leq \sqrt{\EE{U_i^2}} \leq \frac{1}{\sqrt{n+1-p}}.
\]
Since this also holds for $U_{n+1}$ in place of $U_i$, we therefore have 
\[
\dtv(U,U^i)  \leq   \kappa_\Sigma\sqrt{8}\cdot \frac{p}{\sqrt{n+1-p}},
\]
which completes the proof.

\subsection{Proofs for the jackknife+}\label{app:proofs_jack+}

\subsubsection{Background: proof of Theorem~\ref{thm:background_jack+}}

Before proving our new results for nonexchangeable jackknife+, we first recall
the proof of Theorem~\ref{thm:background_jack+}, from
\cite{barber2019predictive},  for the exchangeable case. Denote by
\smash{$\muh_{-ij}$} the model fitted by running the symmetric algorithm
$\alg$ on the $n-1$ data points \smash{$\big\{(X_k,Y_k) : k\in[n+1]
  \backslash\{i,j\}\big\}$}. Let $R_{\textnormal{jack+}}\in\R^{(n+1)\times(n+1)}$ be the matrix with
entries   
\[
(R_{\textnormal{jack+}})_{ij} = |Y_i - \muh_{-ij}(X_i)|,
\]
for each $i\neq j$, and zeros on the diagonal. By exchangeability of the $n+1$
data points, the matrix $R_{\textnormal{jack+}}$ also satisfies an exchangeability property, namely, 
\smash{$\Pi\cdot R_{\textnormal{jack+}} \cdot \Pi^\top \eqd R_{\textnormal{jack+}}$} for any fixed permutation matrix
$\Pi$. Moreover, for each $i\in[n]$, we have
\[
\muh_{-i,(n+1)}=\muh_{-(n+1),i} = \muh_{-i},
\] 
where \smash{$\muh_{-i}$} is the usual leave-one-out model defined earlier, and
so also
\[ 
(R_{\textnormal{jack+}})_{n+1,i} = |Y_{n+1} - \muh_{-i}(X_{n+1})|
\textnormal{ and }
(R_{\textnormal{jack+}})_{i,n+1} = R^{\textnormal{LOO}}_i = |Y_i - \muh_{-i}(X_i)|.
\]
Next, define the set of ``strange'' points 
\[
\Scal(R_{\textnormal{jack+}}) = \left\{i\in [n+1] \ : \ \sum_{j=1}^{n+1} \One{(R_{\textnormal{jack+}})_{ij}>(R_{\textnormal{jack+}})_{ji}} \geq
  (1-\alpha)(n+1)\right\}.
\]
In \cite[Proof of Theorem 1]{barber2019predictive} it is shown that the bound
\[
|\Scal(R_{\textnormal{jack+}})| \leq 2\alpha(n+1)
\]
must hold deterministically as a consequence of Landau's theorem for tournaments 
\citep{landau1953dominance}. Furthermore, by definition of the jackknife+
prediction interval, \citet[Proof of Theorem 1]{barber2019predictive} verify
that failure of coverage, i.e., the event 
\smash{$Y_{n+1}\not\in\Ch_n(X_{n+1})$}, implies that $n+1 \in \Scal(R_{\textnormal{jack+}})$. 
Thus, we have
\[
\PP{Y_{n+1}\not\in\Ch_n(X_{n+1})} \leq \PP{n+1\in\Scal(R_{\textnormal{jack+}})} = 
\frac{1}{n+1}\sum_{i=1}^{n+1}\PP{i\in\Scal(R)} \leq 2\alpha,
\]
where the  equality holds due to the exchangeability property of the
matrix $R_{\textnormal{jack+}}$, while the last step follows the bound on the size of the set of
``strange'' points.

\subsubsection{Proof of Theorem~\ref{thm:jack+}}

Observe that, for $i \in [n]$,
\[
\big(R_{\textnormal{jack+}}(Z^K)\big)_{\pi_K(i),K} 
= |Y_i - \muh^K_{-i}(X_i)| = R^{K,\textnormal{LOO}}_i,
\]
where $\pi_K$ is the permutation swapping indices $K$ and $n+1$ as before, and
also 
\[
\big(R_{\textnormal{jack+}}(Z^K)\big)_{K,\pi_K(i)} 
= |Y_{n+1} - \muh^K_{-i}(X_{n+1})|. 
\]
Therefore, 
\begin{align*}
&\sum_{i=1}^n \tilde{w}_i\cdot \One{|Y_{n+1} - \muh^K_{-i}(X_{n+1})| >
R^{K,\textnormal{LOO}}_i} \\ 
&=\sum_{i=1}^n \tilde{w}_i\cdot 
\One{\big(R_{\textnormal{jack+}}(Z^K)\big)_{K,\pi_K(i)} >
\big(R_{\textnormal{jack+}}(Z^K)\big)_{\pi_K(i),K}} \\
&=\sum_{i\in[n+1]\backslash\{n+1\}} \tilde{w}_i\cdot 
\One{ \big(R_{\textnormal{jack+}}(Z^K)\big)_{K,\pi_K(i)} >  
\big(R_{\textnormal{jack+}}(Z^K)\big)_{\pi_K(i),K}} \\
&=\sum_{i\in[n+1]\backslash\{K\}} \tilde{w}_{\pi_K(i)}\cdot 
\One{ \big(R_{\textnormal{jack+}}(Z^K)\big)_{Ki} >  
\big(R_{\textnormal{jack+}}(Z^K)\big)_{iK}} \\
&\leq \sum_{i\in[n+1]\backslash\{K\}} \tilde{w}_i\cdot 
\One{ \big(R_{\textnormal{jack+}}(Z^K)\big)_{Ki} > 
\big(R_{\textnormal{jack+}}(Z^K)\big)_{iK}} \\
&= \sum_{i=1}^{n+1} \tilde{w}_i\cdot 
\One{ \big(R_{\textnormal{jack+}}(Z^K)\big)_{Ki} >  
\big(R_{\textnormal{jack+}}(Z^K)\big)_{iK}},
\end{align*}
where the third step holds by simply substituting $i$ with $\pi_K(i)$ in the sum
indexing, and the next step (the inequality) holds since
\smash{$\tilde{w}_{\pi_K(i)} \leq \tilde{w}_i$} for all
$i\in[n+1]\backslash\{K\}$, as we either have $i=\pi_K(i)$, or $i=n+1$ in which
case we have \smash{$\tilde{w}_{\pi_K(n+1)} = \tilde{w}_K\leq \tilde{w}_{n+1}$},
as $w_K\in[0,1]$ by assumption.

Next, by its construction, we can verify that the noncoverage event satisfies  
\[
Y_{n+1}\not\in\Ch_n(X_{n+1}) \ \implies \
\sum_{i=1}^n \tilde{w}_i\cdot \One{|Y_{n+1} - \muh^K_{-i}(X_{n+1})| > 
  R^{K,\textnormal{LOO}}_i} \geq 1-\alpha.
\]
The proof of this claim in the unweighted case is given in \citet[Proof of
Theorem 1]{barber2019predictive}; the proof for the weighted case is similar.  
Combined with the above, this gives
\begin{equation}\label{eqn:quantile_noncoverage_jack+}
Y_{n+1}\not\in\Ch_n(X_{n+1}) \ \implies \
\sum_{i=1}^{n+1}\tilde{w}_i\cdot \One{
  \big(R_{\textnormal{jack+}}(Z^K)\big)_{Ki} >
  \big(R_{\textnormal{jack+}}(Z^K)\big)_{iK}} \geq 1-\alpha.
\end{equation}
Now for any $r\in\R^{(n+1)\times(n+1)}$, define
\begin{equation}\label{eqn:define_S_r_matrix}
\Scal(r) = \left\{i\in[n+1] : \sum_{j=1}^{n+1} \tilde{w}_j\cdot \One{r_{ij} >
    r_{ji}} \geq 1-\alpha\right\}, 
\end{equation}
a weighted set of ``strange'' points. The following lemma (proved in
Appendix~\ref{app:proof-jack-strange} for completeness) verifies that
\smash{$\sum_{i\in \Scal(r)} \tilde{w}_i \leq 2\alpha$} for any $r$.

\begin{lemma}[\cite{lei2021theory}]\label{lem:jack_strange} 
Fix \smash{$\tilde{w}_1,\dots,\tilde{w}_{n+1} \geq 0$}, with 
\smash{$\sum_{i=1}^{n+1}\tilde{w}_{n+1}=1$}. Let $\Scal(r)$ be defined as 
  in~\eqref{eqn:define_S_r_matrix}. Then 
\[
\sum_{i\in\Scal(r)}\tilde{w}_i \leq 2\alpha\textnormal{ for all
  $r\in\R^{(n+1)\times(n+1)}$.}
\]
\end{lemma}

In words, the above lemma shows that the (weighted) fraction of ``strange''
points cannot exceed $2\alpha$. From~\eqref{eqn:quantile_noncoverage_jack+}, we
see that noncoverage of $Y_{n+1}$ implies strangeness of point $K$:   
\begin{equation}\label{eq:noncoverage-implies-strangeness2}
Y_{n+1}\not\in\Ch_n(X_{n+1}) \ \implies \ 
K \in\Scal\big(R_{\textnormal{jack+}}(Z^K)\big), 
\end{equation}
and finally, following the exact same steps as in the proof of
Theorem~\ref{thm:fullCP}, we have 
\[
\PP{K\in\Scal\big(R_{\textnormal{jack+}}(Z^K)\big)} \leq 
2\alpha + \sum_{i=1}^n\tilde{w}_i \cdot
\dtv\big(R_{\textnormal{jack+}}(Z),R_{\textnormal{jack+}}(Z^i)\big),
\]
which completes the proof.

\subsubsection{Proof of Lemma~\ref{lem:jack_strange}}
\label{app:proof-jack-strange} 

Lemma~\ref{lem:jack_strange} is stated and proved in \cite{lei2021theory}; we
reproduce the proof here for completeness since that paper is currently an
unpublished manuscript. For each $i\in\Scal$, by definition of $\Scal$, we have 
\begin{multline*}
1-\alpha
\leq \sum_{j=1}^{n+1} \tilde{w}_j\One{r_{ij}>r_{ji}}
\leq \sum_{j\in\Scal(r)} \tilde{w}_j\One{r_{ij}>r_{ji}} \;+
\sum_{j\in[n+1]\backslash \Scal(r)} \tilde{w}_j \\
= \sum_{j\in\Scal(r)} \tilde{w}_j\One{r_{ij}>r_{ji}} +
1 - \sum_{j\in \Scal(r)} \tilde{w}_j,
\end{multline*}
where the last step holds since $\sum_{i=1}^{n+1}\tilde{w}_i=1$ by definition. 
Taking a weighted sum over $i\in\Scal(r)$,
\[
(1-\alpha)\sum_{i\in\Scal(r)} \tilde{w}_i \leq \sum_{i\in\Scal(r)} \tilde{w}_i
\cdot \left[\sum_{j\in\Scal(r)} \tilde{w}_j\One{r_{ij}>r_{ji}} +1 - \sum_{j\in
    \Scal(r)} \tilde{w}_j\right].
\]
Rearranging terms, we have
\begin{align*}
\left(\sum_{i\in\Scal(r)} \tilde{w}_i\right)^2 
&\leq \sum_{i,j\in\Scal(r)} \tilde{w}_i  \tilde{w}_j\One{r_{ij}>r_{ji}} + \alpha
\sum_{i\in\Scal(r)} \tilde{w}_i \\
&=\frac{1}{2} \sum_{i,j\in\Scal(r)} \tilde{w}_i
\tilde{w}_j\left(\One{r_{ij}>r_{ji}}+\One{r_{ji}>r_{ij}}\right)
+ \alpha \sum_{i\in\Scal(r)} \tilde{w}_i \\
&\leq \frac{1}{2} \sum_{i,j\in\Scal(r)} \tilde{w}_i  \tilde{w}_j + 
\alpha \sum_{i\in\Scal(r)} \tilde{w}_i
= \frac{1}{2} \left(\sum_{i\in\Scal(r)} \tilde{w}_i\right)^2+ 
\alpha \sum_{i\in\Scal(r)} \tilde{w}_i.
\end{align*}
Rearranging terms again we have
\[ 
\frac{1}{2} \left(\sum_{i\in\Scal(r)} \tilde{w}_i\right)^2 \leq
\alpha \left(\sum_{i\in\Scal(r)} \tilde{w}_i\right) \ \implies \ 
\sum_{i\in\Scal(r)} \tilde{w}_i\leq 2\alpha,
\]
which proves the lemma.

\subsection{Proof of Theorem~\ref{thm:mult}}

To prove the theorem, we first need a lemma on weighted sums of Bernoulli random
variables. Its proof will follow shortly.

\begin{lemma}\label{lem:inverse_Bernoulli}
Fix $p_1,\dots,p_n,a_1,\dots,a_n\in[0,1]$, with $a_1 + \dots + a_n>0$. Let
$B_1,\dots,B_n$ be independent, with $B_i \sim\textnormal{Bernoulli}(p_i)$. Then 
\[
\frac{a_1 + \dots + a_n+1}{a_1 p_1 + \dots + a_n p_n+1} 
\leq \EE{\frac{a_1 + \dots + a_n + 1}{a_1B_1 + \dots + a_nB_n + 1}} 
\leq \frac{a_1 + \dots + a_n}{a_1 p_1 + \dots + a_n p_n}. 
\]
\end{lemma}

The lower bound clearly holds by Jensen's inequality, but the upper bound is
more challenging to prove. Several special cases of this upper bound are
well-known in the multiple testing literature. For example, the case where $a_1
= \dots = a_n=1$ and $p_1 = \dots = p_n$ is proved in \cite[Theorem 
3]{storey2004strong} and used for proving FDR control of Storey's modification
of the Benjamini-Hochberg procedure \citep{storey2002direct}. 
The case where $p_1=\dots = p_n$ (and $a_1,\ldots,a_n$ are arbitrary) can be
found in \cite[Lemma 3]{ramdas2019unified} and is used for proving FDR control
for a hierarchical multiple testing procedure (the p-filter).  

We are now ready to prove the theorem. By definition of \smash{$\dmix$}~\eqref{eq:dmix}, 
note that we can view $Z_1,\dots,Z_{n+1}$ as being generated by
the following procedure.   
\begin{itemize}
\item Draw $C_1,\dots,C_n$ independently, with
  \smash{$C_i\sim\textnormal{Bernoulli}(\dmix(Z_i,Z_{n+1}))$}.
\item For each $i \in[n]$ with $C_i=0$, and for $i=n+1$, draw $Z_i$ i.i.d.\ from the
  distribution of $Z_{n+1}$.   
\item For each $i\in[n]$ with $C_i=1$, draw $Z_i$ from the contamination 
  distribution, i.e., the distribution $\mathcal{D}''$ achieving the infimum in
  \eqref{eq:dmix} (applied with $\mathcal{D}$ and $\mathcal{D}'$ equal to the distribution
  of $Z_i$ and of $Z_{n+1}$, respectively).  
\end{itemize}

Below, we will show that for nonexchangeable   conformal,  
\begin{equation}\label{eqn:mult_to_show_CP}
\PPst{Y_{n+1}\not\in\Ch_n(X_{n+1})}{C_1,\dots,C_n} 
\leq \frac{\alpha}{\sum_{i=1}^n \tilde{w}_i \One{C_i=0} + \tilde{w}_{n+1}}, 
\end{equation}
while for nonexchangeable jackknife+,
\begin{equation}\label{eqn:mult_to_show_jack+}
\PPst{Y_{n+1}\not\in\Ch_n(X_{n+1})}{C_1,\dots,C_n} 
\leq \frac{2\alpha}{\sum_{i=1}^n \tilde{w}_i \One{C_i=0} + \tilde{w}_{n+1}}.
\end{equation}
Having shown this, observe that
\begin{multline*}
\EE{\frac{1}{\sum_{i=1}^n \tilde{w}_i \One{C_i=0} + \tilde{w}_{n+1}}}
= \EE{\frac{\sum_{i=1}^n w_i + 1}{\sum_{i=1}^n w_i \One{C_i=0} + 1}} \\
\leq \frac{\sum_{i=1}^n w_i}{\sum_{i=1}^n w_i (1-\dmix(Z_i,Z_{n+1}))}
= \frac{1}{\sum_{i=1}^n \bar{w}_i (1-\dmix(Z_i,Z_{n+1}))} \\
=\frac{1}{ 1 - \sum_{i=1}^n \bar{w}_i \dmix(Z_i,Z_{n+1})},
\end{multline*}
where the inequality holds by Lemma~\ref{lem:inverse_Bernoulli}. This implies that
\begin{multline*}
\PP{Y_{n+1}\not\in\Ch_n(X_{n+1})}
= \EE{\PPst{Y_{n+1}\not\in\Ch_n(X_{n+1})}{C_1,\dots,C_n}} \\
\leq \frac{\alpha}{ 1 - \sum_{i=1}^n \bar{w}_i \dmix(Z_i,Z_{n+1})},
\end{multline*}
for nonexchangeable   conformal prediction, and 
\begin{multline*}
\PP{Y_{n+1}\not\in\Ch_n(X_{n+1})} 
= \EE{\PPst{Y_{n+1}\not\in\Ch_n(X_{n+1})}{C_1,\dots,C_n}} \\
\leq \frac{2\alpha}{ 1 - \sum_{i=1}^n \bar{w}_i \dmix(Z_i,Z_{n+1})},
\end{multline*}
for nonexchangeable jackknife+, as desired.

To complete the proof, we now need to verify the
bounds~\eqref{eqn:mult_to_show_CP} and~\eqref{eqn:mult_to_show_jack+}.
For the bound~\eqref{eqn:mult_to_show_CP} for conformal prediction, we have  
\begin{multline*}
Y_{n+1}\not\in\Ch_n(X_{n+1}) \ \iff \ R^{Y_{n+1},K}_{n+1} 
> \quant_{1-\alpha}\left(\sum_{i=1}^{n+1}\tilde{w}_i 
\delta_{R^{Y_{n+1},K}_i} \right) \\ 
\iff \ \sum_{i=1}^{n+1} \tilde{w}_i \One{R^{Y_{n+1},K}_{n+1}
\leq R^{Y_{n+1},K}_i} \leq \alpha.
\end{multline*}
Now let $w'_i = w_i\One{C_i=0}$ and let 
\[
\tilde{w}'_i = \frac{w'_i}{w'_1+\dots+w'_n+1}, \ i=1,\dots,n; 
\ \tilde{w}'_{n+1}=\frac{1}{w'_1+\dots+w'_n+1}.
\]
Then, deterministically,
\[
\sum_{i=1}^{n+1} \tilde{w}_i \One{R^{Y_{n+1},K}_{n+1}
\leq R^{Y_{n+1},K}_i} \geq \sum_{i=1}^{n+1} \tilde{w}_i \cdot
\One{C_i=0}\cdot \One{R^{Y_{n+1},K}_{n+1}\leq R^{Y_{n+1},K}_i},
\]
and so we can write
\[
Y_{n+1}\not\in\Ch_n(X_{n+1}) \ \implies \ 
\sum_{i=1}^{n+1} \tilde{w}'_i \One{R^{Y_{n+1},K}_{n+1}\leq  R^{Y_{n+1},K}_i} 
\leq \alpha \cdot \frac{w_1+\dots+w_n+1}{w'_1+\dots+w'_n+1}.
\]
Now suppose we had instead conditioned on $C_1,\dots,C_n$ and we ran
nonexchangeable full conformal on the same data set $Z$ but with weights $w'_i$
in place of $w_i$, and with a level \smash{$\alpha \cdot
  \frac{w_1+\dots+w_n+1}{w'_1+\dots+w'_n+1}$} in place of $\alpha$. Let 
\smash{$\Ch'_n(X_{n+1})$} be the resulting prediction interval. Then by the same 
arguments as above, we have
\[
Y_{n+1}\not\in\Ch_n'(X_{n+1}) 
\ \iff \ \sum_{i=1}^{n+1} \tilde{w}'_i \One{R^{Y_{n+1},K}_{n+1}\leq
  R^{Y_{n+1},K}_i} \leq \alpha \cdot 
\frac{w_1+\dots+w_n+1}{w'_1+\dots+w'_n+1}.
\]
and combining this with the work above, we obtain
\[
Y_{n+1}\not\in\Ch_n(X_{n+1}) \ \implies \ Y_{n+1}\not\in\Ch_n'(X_{n+1}) .
\]
Moreover, Theorem~\ref{thm:fullCP} ensures that
\begin{multline*}
\PPst{Y_{n+1}\not\in\Ch_n'(X_{n+1})}{C_1,\dots,C_n}
\leq \alpha \cdot \frac{w_1+\dots+w_n+1}{w'_1+\dots+w'_n+1} \\
+ \sum_{i=1}^n \tilde{w}'_i \cdot \dtv(Z,Z^i\mid C_1,\dots,C_n),
\end{multline*}
where \smash{$\dtv(Z,Z^i\mid C_1,\dots,C_n)$} is the total variation distance
between the conditional distributions of $Z$ and of $Z^i$ conditional on
$C_1,\dots,C_n$. Furthermore, we can see that \smash{$\dtv(Z,Z^i\mid
  C_1,\dots,C_n) =0$} for all $i\in[n]$ with $C_i=0$. Since
\smash{$\tilde{w}'_i$} is nonzero only for $i$ with $C_i=0$, we thus have 
\begin{multline*}
\PPst{Y_{n+1}\not\in\Ch_n'(X_{n+1})}{C_1,\dots,C_n} \\ 
\leq \alpha \cdot \frac{w_1+\dots+w_n+1}{w'_1+\dots+w'_n+1}  
= \frac{\alpha}{\sum_{i=1}^n\tilde{w}_i\One{C_i=0} + \tilde{w}_{n+1}},
\end{multline*}
where the last step applies the definitions of \smash{$\tilde{w}_i$}~\eqref{eqn:normalized_weights} and
of $w'_i$.
 Therefore, 
\begin{multline*}
\PPst{Y_{n+1}\not\in\Ch_n(X_{n+1})}{C_1,\dots,C_n}\\
\leq \PPst{Y_{n+1}\not\in\Ch_n'(X_{n+1})}{C_1,\dots,C_n}\leq  
\frac{\alpha}{\sum_{i=1}^n\tilde{w}_i\One{C_i=0} + \tilde{w}_{n+1}},
\end{multline*}
which verifies~\eqref{eqn:mult_to_show_CP}.

Finally, the proof of the bound~\eqref{eqn:mult_to_show_jack+} for the
jackknife+ is nearly identical. As calculated before, we have 
\begin{multline*}
Y_{n+1}\not\in\Ch_n(X_{n+1}) 
\ \implies \ \sum_{i=1}^n\tilde{w}_i
\One{|Y_{n+1}-\muh_{-i}(X_{n+1})|>R^{\textnormal{LOO}}_i} \geq 1-\alpha \\  
\ \iff \ \sum_{i=1}^n\tilde{w}_i \One{|Y_{n+1}-\muh_{-i}(X_{n+1})|
\leq  R^{\textnormal{LOO}}_i}\leq \alpha.
\end{multline*}
Define $w'_i$ and \smash{$\tilde{w}'_i$} as above. Then, deterministically, 
\begin{multline*}
\sum_{i=1}^n\tilde{w}_i \One{|Y_{n+1}-\muh_{-i}(X_{n+1})|
\leq R^{\textnormal{LOO}}_i} \\ \geq \sum_{i=1}^n\tilde{w}_i\cdot\One{C_i=0} 
\cdot  \One{|Y_{n+1}-\muh_{-i}(X_{n+1})|\leq R^{\textnormal{LOO}}_i},
\end{multline*}
and so we can write
\begin{multline*}
Y_{n+1}\not\in\Ch_n(X_{n+1}) \\
\implies \ \sum_{i=1}^{n+1} \tilde{w}'_i \One{|Y_{n+1}-\muh_{-i}(X_{n+1})|
\leq R^{\textnormal{LOO}}_i} \leq \alpha \cdot
\frac{w_1+\dots+w_n+1}{w'_1+\dots+w'_n+1}.
\end{multline*}
As argued before, suppose that we had instead conditioned on $C_1,\dots,C_n$, 
then ran nonexchangeable jackknife+ on the same data set $Z$ but with weights 
$w'_i$ in place of $w_i$, and with a level \smash{$\alpha \cdot
  \frac{w_1+\dots+w_n+1}{w'_1+\dots+w'_n+1}$} in place of $\alpha$.   
Then by the same arguments as in the proof of Theorem~\ref{thm:jack+}, 
we have
\begin{multline*}\PPst{Y_{n+1}\not\in\Ch_n(X_{n+1})}{C_1,\dots,C_n}\\
=\PPst{\sum_{i=1}^{n+1} \tilde{w}'_i \One{|Y_{n+1}-\muh_{-i}(X_{n+1})|\leq
    R^{\textnormal{LOO}}_i} \leq \alpha \cdot
  \frac{w_1+\dots+w_n+1}{w'_1+\dots+w'_n+1}}{C_1,\dots,C_n} \\ 
\leq 2 \alpha \cdot \frac{w_1+\dots+w_n+1}{w'_1+\dots+w'_n+1} + 
\sum_{i=1}^n \tilde{w}'_i \cdot \dtv(Z^i,Z\mid C_1,\dots,C_n) \\
=2 \alpha \cdot \frac{w_1+\dots+w_n+1}{w'_1+\dots+w'_n+1}
= \frac{2\alpha}{\sum_{i=1}^n\tilde{w}_i\One{C_i=0} + \tilde{w}_{n+1}},\end{multline*}
where the next-to-last step is shown exactly as for full conformal. This
verifies~\eqref{eqn:mult_to_show_jack+}. 

\subsubsection{Proof of Lemma~\ref{lem:inverse_Bernoulli}}

The lower bound holds by Jensen's inequality. For the upper bound, we will
instead prove the claim 
\begin{equation}\label{eqn:inverse_Bernoulli_inductive_step}
\EE{\frac{a^\top\mathbf{1} + c + 1}{a^\top B+c+1}} 
\leq \frac{a^\top\mathbf{1}+c}{a^\top p+ c},
\end{equation} 
for any $c\geq 0$ and any $p_1,\dots,p_n,a_1,\dots,a_n\in[0,1]$ with $a_1 +
\dots + a_n+c>0$, where $a=(a_1,\dots,a_n)$, $p=(p_1,\dots,p_n)$,
$B=(B_1,\dots,B_n)$, and as before, the expectation is taken with respect to
independent Bernoulli random variables
$B_i\sim\textnormal{Bernoulli}(p_i)$. Initially this appears stronger than the 
claim in the lemma (i.e., the lemma claims this bound only for the case $c=0$),
but in fact these claims are equivalent. 

To see why, suppose that the lemma holds and now we want to
prove~\eqref{eqn:inverse_Bernoulli_inductive_step} for some $p,a\in[0,1]^n$ and 
some $c>0$. Let $m \geq c$ be any integer, and let
\smash{$\tilde{p},\tilde{a}\in[0,1]^{n+m}$} be defined as   
\[
\tilde{p} = (p_1,\dots,p_n,1\dots,1), \ 
\tilde{a} = (a_1,\dots,a_n, c/m,\dots, c/m). 
\]
Then writing \smash{$\tilde{B} = (\tilde{B}_1,\dots,\tilde{B}_{n+m})$} for
independent Bernoullis\smash{$\tilde{B}_i \sim
  \textnormal{Bernoulli}(\tilde{p}_i)$}, the
claim~\eqref{eqn:inverse_Bernoulli_inductive_step} is equivalent to  
\[
\EE{\frac{\tilde{a}^\top\mathbf{1}+1}{\tilde{a}^\top \tilde{B} + 1}}
\leq \frac{\tilde{a}^\top\mathbf{1}}{\tilde{a}^\top \tilde{p}},
\]
which holds by applying the lemma with \smash{$\tilde{p},\tilde{a},n+m$} in
place of $p,a,n$. 

\paragraph{Case 1: $a_1 = \dots = a_n=1$ and $p_1 = \dots = p_n$.} 

Proving that~\eqref{eqn:inverse_Bernoulli_inductive_step} holds for this case is
equivalent to proving that   
\[
\EE{\frac{n+c+1}{A + c + 1}} \leq \frac{n+c}{np_1+c}
\]
for $A\sim\textnormal{Binomial}(n,p_1)$.
In particular, if $c=0$, then this is the well-known bound
\smash{$\EE{\frac{n+1}{A+1}}\leq
  \frac{1}{p_1}$} (e.g., \cite[Theorem 3]{storey2004strong}), while if $p_1=0$
then the result is trivial. If instead $c>0$ and $p_1>0$, then we calculate 
\begin{align*}
\EE{\frac{n+c+1}{A + c + 1}}
&=1 + \EE{\frac{n - A}{A + c + 1}} \\
&=1 + \sum_{k=0}^{n-1} \PP{A=k}\cdot \frac{n-k}{k+c+1} \\ 
&=1 +  \frac{1-p_1}{p_1}\cdot \sum_{k=0}^{n-1} \PP{A=k+1} \cdot 
\frac{k+1}{k+c+1} \\
&=1 + \frac{1-p_1}{p_1}\cdot \EE{\frac{A}{A+c}} \\
&\leq 1 + \frac{1-p_1}{p_1}\cdot \frac{\EE{A}}{\EE{A}+c}
=\frac{n+c}{np_1+c},
\end{align*}
where the  inequality holds by Jensen's inequality.

\paragraph{Case 2: $a_1,\ldots,a_n$ arbitrary and $p_1 = \dots = p_n$.}  

Proving that~\eqref{eqn:inverse_Bernoulli_inductive_step} holds for this 
next case is equivalent to proving that  
\[
\EE{\frac{a^\top \mathbf{1} + c+1}{a^\top B + c+ 1} }
\leq \frac{a^\top \mathbf{1} + c}{a^\top \mathbf{1}\cdot p_1 + c}.
\]
For the special case $c=0$, this result is shown in \cite[Lemma
3]{ramdas2019unified}. Let $A=(A_1,\dots,A_n)$,
where $A_i\sim \textnormal{Bernoulli}(a_i)$ are drawn
independently for $i=1,\ldots,n$, and $A\independent B$.
Note that, conditional on $B$, it holds that $A^\top B \independent A^\top
(\mathbf{1}-B)$. Therefore,
\begin{align*}
\EEst{\frac{A^\top\mathbf{1}+c+1}{A^\top B + c + 1}}{B}
&=1 + \EEst{\frac{A^\top (\mathbf{1}-B)}{A^\top B + c + 1}}{B}\\ 
&=1 +  \EEst{A^\top (\mathbf{1}-B)}{B} \cdot 
\EEst{\frac{1}{A^\top B + c + 1}}{B}\\
&\geq 1 +  \frac{\EEst{A^\top (\mathbf{1}-B)}{B}}
{\EEst{A^\top B + c + 1}{B}}
\textnormal{ (by Jensen's inequality)}\\
&= 1 +  \frac{a^\top (\mathbf{1}-B)}{a^\top B + c + 1}
= \frac{a^\top \mathbf{1}+c+1}{a^\top B + c + 1},
\end{align*}
and therefore after marginalizing over $B$, we obtain
\[
\EE{\frac{a^\top \mathbf{1} + c+1}{a^\top B + c+ 1}}
\leq \EE{\frac{A^\top\mathbf{1}+c+1}{A^\top B + c + 1}}.
\]

Next, writing $S=A^\top\mathbf{1}$, we see that $A^\top B$ follows a
Binomial$(S,p_1)$ distribution conditional on $S$, and therefore, 
\[ 
\EEst{\frac{A^\top \mathbf{1} +c+1}{A^\top B+c+1}}{S}
= \EEst{\frac{S+c+1}{\textnormal{Binomial}(S,p_1)+c+1}}{S}
\leq \frac{S+c}{Sp_1 +c},
\]
where the last step holds by case 1. We can also observe that
\smash{$s \mapsto \frac{s+c}{sp_1 +c}$} is a concave function, and so   
\begin{multline*}\EE{\frac{a^\top \mathbf{1} + c+1}{a^\top B + c+ 1}}\leq
\EE{\frac{A^\top \mathbf{1} +c+1}{A^\top B+c+1}}\\
= \EE{\EEst{\frac{A^\top \mathbf{1} +c+1}{A^\top B+c+1}}{S}} 
\leq\EE{ \frac{S+c}{Sp_1 +c}}\leq \frac{\EE{S}+c}{\EE{S}\cdot p_1 + c} 
= \frac{a^\top \mathbf{1}+c}{a^\top\mathbf{1}\cdot p_1 + c},
\end{multline*}
as desired.

\paragraph{Case 3:  $a_1,\ldots,a_n$ and $p_1,\ldots,p_n$ arbitrary.} 

In this final case, we will prove~\eqref{eqn:inverse_Bernoulli_inductive_step},
proceeding by induction on $n$.  For $n=1$, this reduces to case 1, so we can
proceed to the case $n\geq 2$. Without loss of generality, assume $p_1\leq
\dots\leq p_n$. If $p_n=0$ then the claim is trivial. Otherwise, let $A_i \sim
\textnormal{Bernoulli}(p_i/p_n)$ for $i=1,\dots,n-1$, and let $C_i \sim
\textnormal{Bernoulli}(p_n)$ for $i=1,\dots,n$, with
$A_1,\dots,A_{n-1},C_1,\dots,C_n$ all drawn independently. Then
\smash{$a_1B_1+\dots+a_nB_n \eqd a_1A_1C_1 + \dots + a_{n-1}A_{n-1}C_{n-1} 
+ a_n C_n$}. 

Next, define random weights $W=(W_1,\dots,W_n)$, where $W_i = a_i A_i$ for each
$i=1,\dots,n-1$ and $W_n=a_n$. Then by case 2, we have  
\[
\EEst{\frac{W_1 + \dots + W_n + c+1}
{W_1 C_1 + \dots + W_n C_n + c+1}}{W} 
\leq \frac{W^\top \mathbf{1}+c}{W^\top\mathbf{1}\cdot p_n + c}.
\]  
Equivalently, we have shown that
\[
\EEst{\frac{a_{-n}^\top A+a_n + c+1}{a^\top B + c+1}}{A}
\leq \frac{a_{-n}^\top A + a_n + c}{(a_{-n}^\top A + a_n)\cdot p_n + c} 
= \frac{1}{p_n} - \frac{c\left(\frac{1}{p_n}-1\right)}{(a_{-n}^\top A + a_n)
\cdot p_n + c}.
\] 
Thus,
\begin{align}
\notag&\EE{\frac{a^\top\mathbf{1}+c+1}{a^\top B + c + 1}}
= \EE{\frac{a^\top\mathbf{1}+c+1}{a_{-n}^\top A+a_n + c+1}\cdot 
\frac{a_{-n}^\top A+a_n + c+1}{a^\top B + c + 1}}\\
\notag&= \EE{\frac{a^\top\mathbf{1}+c+1}{a_{-n}^\top A+a_n + c+1}\cdot 
\EEst{\frac{a_{-n}^\top A+a_n + c+1}{a^\top B + c + 1}}{A}}\\
\notag&\leq \EE{\frac{a^\top\mathbf{1}+c+1}{a_{-n}^\top A+a_n + c+1}\cdot 
\left(\frac{1}{p_n} - \frac{c\left(\frac{1}{p_n}-1\right)}{(a_{-n}^\top A + a_n)
\cdot p_n + c}\right)}\\
\label{eqn:Bern_lemma_last_step}&\leq \EE{\frac{a^\top\mathbf{1}+c+1}{a_{-n}^\top A+a_n + c+1}}\cdot 
\EE{\frac{1}{p_n} - \frac{c\left(\frac{1}{p_n}-1\right)}{(a_{-n}^\top A + a_n)
\cdot p_n + c}},
\end{align}
where the last step holds since the first quantity is a monotone decreasing
function of $a_{-n}^\top A$, and the second quantity is a monotone increasing
function of $a_{-n}^\top A$. By induction, we can
apply~\eqref{eqn:inverse_Bernoulli_inductive_step} at size $n-1$ in place of $n$ 
to see that the first expected value is bounded as
\[
\EE{\frac{a^\top\mathbf{1}+c+1}{a_{-n}^\top A+a_n + c+1}}
\leq \frac{a^\top\mathbf{1}+c}{a_{-n}^\top (p_n^{-1}p_{-n}) + a_n + c} 
= p_n \cdot  \frac{a^\top\mathbf{1}+c}{a^\top p + cp_n}.
\]
Moreover, applying Jensen's inequality, we calculate
\begin{multline*}
\EE{\frac{1}{p_n} - \frac{c\left(\frac{1}{p_n}-1\right)}
{(a_{-n}^\top A + a_n)\cdot p_n + c}}
\leq \left(\frac{1}{p_n} - \frac{c\left(\frac{1}{p_n}-1\right)}
{(a_{-n}^\top \EE{A} + a_n)\cdot p_n + c}\right)\\
=\left(\frac{1}{p_n} - \frac{c\left(\frac{1}{p_n}-1\right)}
{a^\top p + c}\right)
=\frac{a^\top p + cp_n}{p_n(a^\top p + c)}.\end{multline*}
Combining these calculations with~\eqref{eqn:Bern_lemma_last_step} above, we have
\[
\EE{\frac{a^\top\mathbf{1}+c+1}{a^\top B + c + 1}}
\leq p_n \cdot  \frac{a^\top\mathbf{1}+c}{a^\top p + cp_n}\cdot 
\frac{a^\top p + cp_n}{p_n(a^\top p + c)}
=\frac{a^\top \mathbf{1}+c}{a^\top p + c},
\]
which proves that~\eqref{eqn:inverse_Bernoulli_inductive_step} holds as
desired. 

\section{Simulations for split conformal and jackknife+}
\label{app:additional_simulations}  

The simulations presented in Section~\ref{sec:simulations}
used only full conformal methods. In this section, we repeat these simulated
data experiments with split conformal prediction and jackknife+. The same data
is used in these experiments as was generated for the results in 
Section~\ref{sec:simulations}.  (Code for reproducing these additional 
experiments is available at 
\url{https://rinafb.github.io/code/nonexchangeable_conformal.zip}.)

For split conformal prediction, we split the training data indices $[n]$ by
assigning odd indices to the training set and even indices to the holdout
set. The methods compared for split conformal are SplitCP+LS, NexSplitCP+LS,
NexSplitCP+WLS, defined exactly as the full conformal experiments but with
split conformal in place of full conformal. For jackknife+, the methods compare
are Jack+LS, NexJack+LS, NexJack+WLS, again defined analogously.  

For both split conformal and jackknife+, the details for defining \smash{$\muh$}
for choosing the weights $w_i$ and tags $t_i$ are exactly the same as for the
full conformal experiments given in Section~\ref{sec:simulations}. Also as 
in the full conformal experiments, after a burn-in period of the first 100 time
points,  at each time $n=100,\dots,N-1$ we run the inference methods with
training data $i=1,\dots,n$ and test point $n+1$. The results shown are averaged
over 200 independent replications of the simulation.   

{\begin{table}[htb]\small\centering
\begin{tabular}{l|cc|cc|cc}
&\multicolumn{2}{c|}{Setting 1 (i.i.d.\ data)} 
&\multicolumn{2}{c|}{Setting 2 (changepts)}
&\multicolumn{2}{c}{Setting 3 (drift)}\\
&Coverage& Width &Coverage& Width&Coverage& Width\\\hline
SplitCP+LS & 0.902 &3.34&0.836 &6.04 &0.839 &3.76\\
NexSplitCP+LS &0.915 &3.51&0.893 &7.09&0.896&4.43\\
NexSplitCP+WLS &0.915 &3.56& 0.914&4.33&0.914&3.59
\end{tabular}
\caption{Simulation results showing mean prediction interval coverage and width,
  averaged over all time points and over 200 trials, for split conformal
  methods.}
\label{tab:simulation_splitCP}

\bigskip
\begin{tabular}{l|cc|cc|cc}
&\multicolumn{2}{c|}{Setting 1 (i.i.d.\ data)} 
&\multicolumn{2}{c|}{Setting 2 (changepoints)}
&\multicolumn{2}{c}{Setting 3 (drift)}\\
&Coverage& Width &Coverage& Width&Coverage& Width\\\hline
Jack+LS & 0.899 &3.30&0.834 &5.98 &0.837 &3.72\\
NexJack+LS &0.906 &3.38&0.881 &6.79&0.887&4.27\\
NexJack+WLS &0.906 &3.40& 0.905&4.11&0.905&3.44
\end{tabular}
\caption{Simulation results showing mean prediction interval coverage and width,
  averaged over all time points and over 200 trials, for jackknife+ methods.} 
\label{tab:simulation_jack+}
\end{table}}

\begin{figure}[p]
\includegraphics[width=\textwidth]{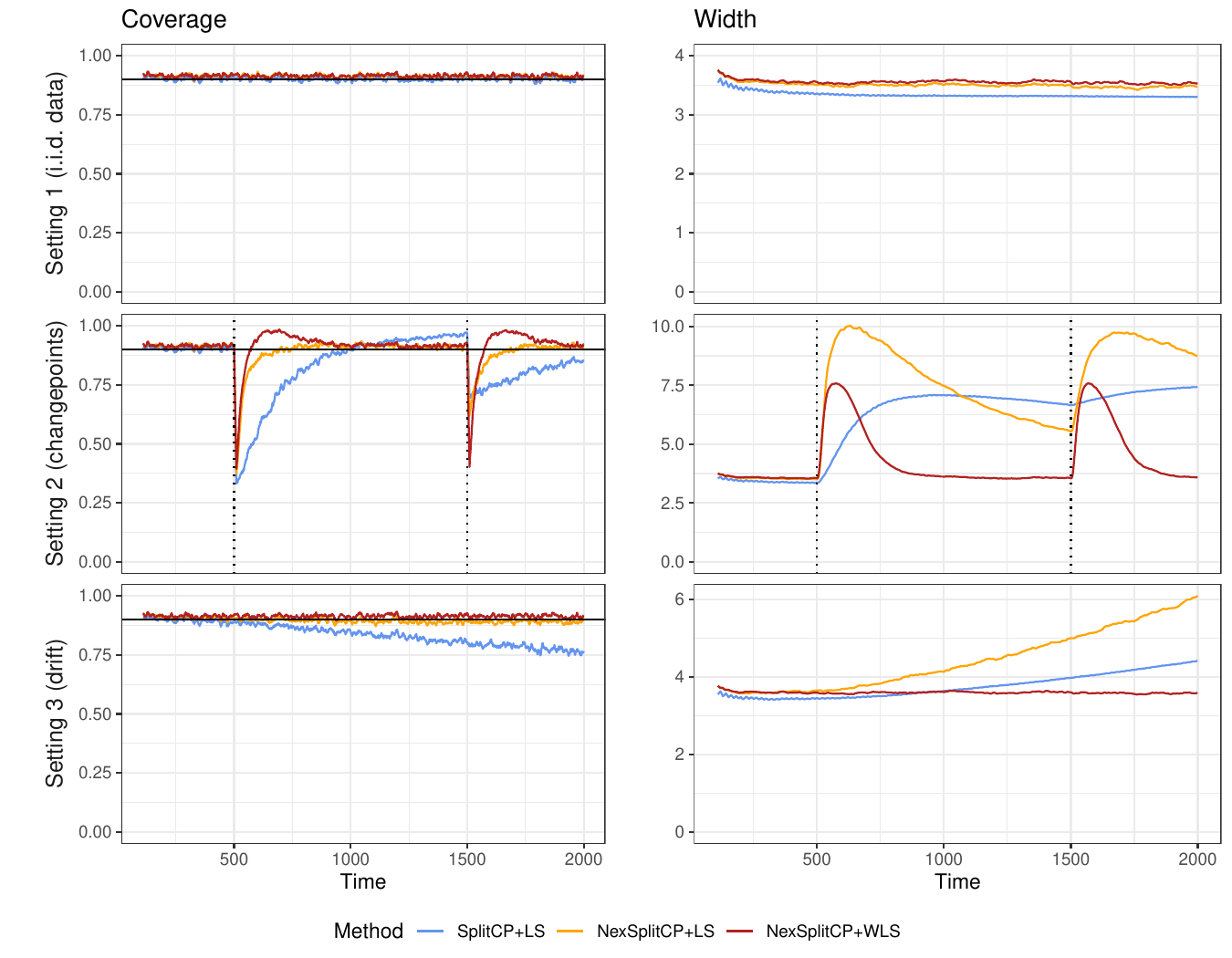}
\caption{Simulation results showing mean prediction interval coverage and width
  for split conformal methods, averaged over 200 independent trials. The
  curves are smoothed by taking a rolling average with a window of 10 time
  points.} 
\label{fig:simulation_splitCP}
\end{figure}
\begin{figure}[p]
\includegraphics[width=\textwidth]{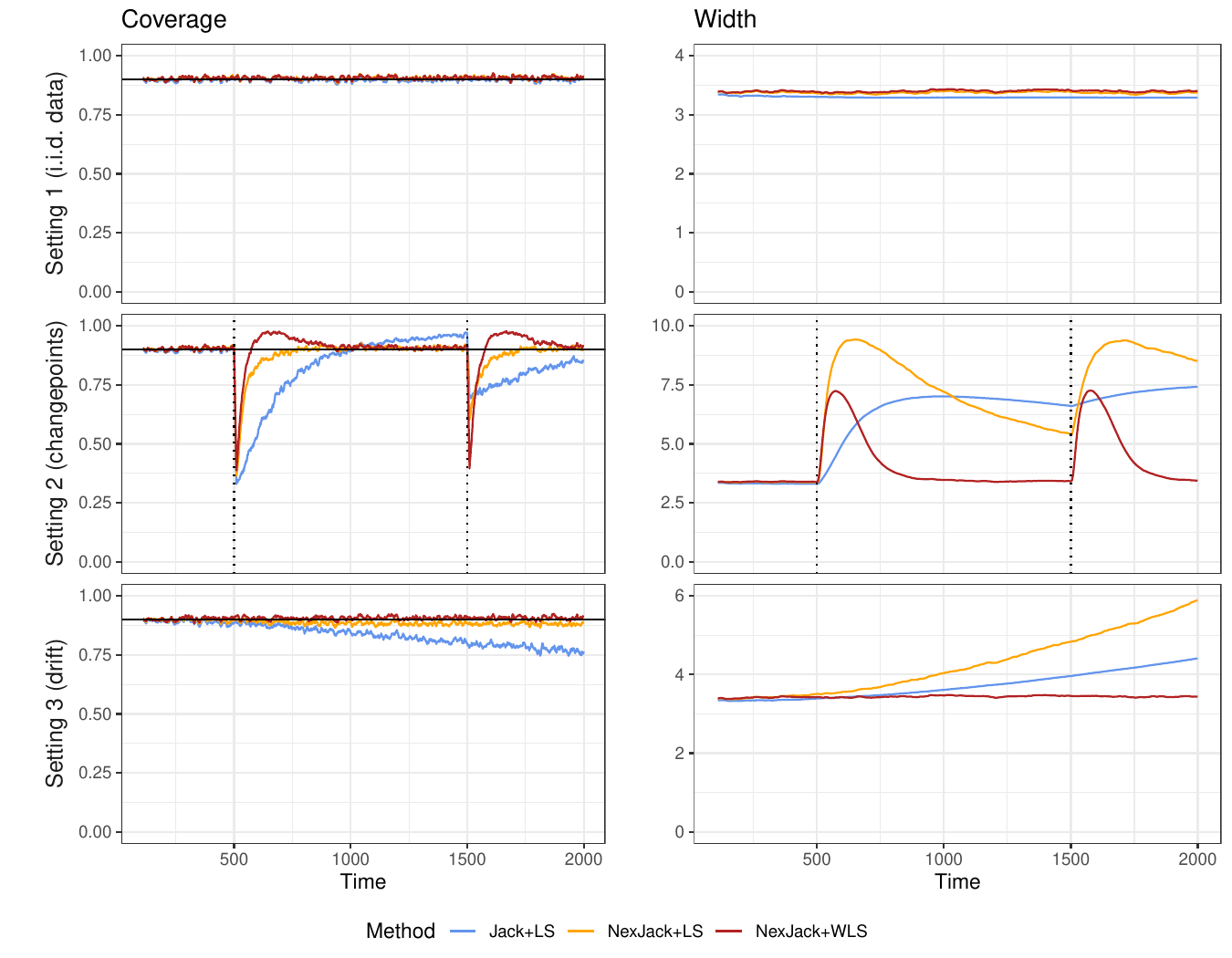}
\caption{Simulation results showing mean prediction interval coverage and width
  for jacknife+ methods, averaged over 200 independent trials. The curves are
  smoothed by taking a rolling average with a window of 10 time points.} 
\label{fig:simulation_jack+}
\end{figure}

Results for split conformal and for jackknife+ are summarized in
Tables~\ref{tab:simulation_splitCP} and~\ref{tab:simulation_jack+},
respectively, while Figures~\ref{fig:simulation_splitCP}
and~\ref{fig:simulation_jack+} display the average coverage and the prediction
interval width over the time range of the simulation.  Overall, we see similar
trends as for the full conformal prediction experiments in
Section~\ref{sec:simulations}, where for the i.i.d.\ data in Setting 1 the
performance of all three versions of each method are comparable, while for the
changepoint data in Setting 2 and the distribution drift data in Setting 3, the
original methods lose coverage substantially, while nonexchangeable versions of
split conformal and of jackknife+ remain closer to the target coverage level,
and the nonsymmetric algorithm (weighted least squares) allows for a narrower
prediction interval.

\section{Election data set description}
\label{app:election}

To prepare the data, we followed the protocol from \cite{gibbs2021adaptive} and,
therefore, quote from the above reference:
\begin{quote}
  \em The county-level
  demographic characteristics used for prediction were the proportion of the
  total population that fell into each of the following race categories (either
  alone or in combination): black or African American,  American Indian or
  Alaska Native, Asian, Native Hawaiian or other Pacific islander. In addition
  to this, we also used the proportion of the total population that was male, of
  Hispanic origin and that fell within each of the age ranges 20-29, 30-44,
  45-64, and 65+.  Demographic information was obtained from 2019 estimates
  published by the United States Census Bureau and available at
  \citep{censusDemos2019}. In addition to these demographic features we also used
  the median household income and the percentage of individuals with a bachelors
  degree or higher as covariates. Data on county-level median household incomes
  was based on 2019 estimates obtained from \citep{censusIncom}. The percentage
  of individuals with a bachelors degree or higher was computed based on data
  collected during the years 2015-2019 and published at \citep{censusEdu}. As an
  aside, we remark that we used 2019 estimates because this was the most recent
  year for which data was available.
\end{quote} 

For 2016 covariate data, we used the same data sources, subject to the important
distinction that we---almost exclusively---used published figures available by 
2016. (The U.S.\ Census Bureau sometimes updates its figures so we cannot rule
out the possibility that a few entries were changed post 2016.) Finally, vote
counts for the 2016 election were obtained from \cite{electionsData}, while 2020
election data was taken from \cite{Leip2020}. In total, matching covariate and
election vote count data were obtained for 3111 counties. Merging 2016 and 2020
data left us with 3076 counties (1119 in the training set and 1957 in the test
set).
\end{document}